# AAPPeC: Agent-based Architecture for Privacy Payoff in eCommerce

by
Abdulsalam Yassine

A thesis submitted to the
Faculty of Graduate and Postdoctoral Studies
In partial fulfillment of the requirements for the degree of

Doctor of Philosophy
in
Electrical and Computer Engineering

Ottawa-Carleton Institute for Electrical and Computer Engineering
School of Information Technology and Engineering

Faculty of Engineering
University of Ottawa

# Acknowledgements

Writing this thesis marks the happy conclusion of the journey that I started a few years ago. Throughout the journey, I greatly and sincerely benefited from the support and companionship of many people.

First and foremost, I had the great fortune of being supervised by a very supportive and helpful supervisor. His integrity, dedication, insightful advice, and encouragement have greatly contributed in fulfilling this degree, and so I give my heartfelt gratitude to Dr. Shervin Shirmohammadi. Throughout my thesis-study period, he was always there for me with prompt guidance and feedback and lots of good ideas. I would have been lost without his support. I will never find the words to thank him for believing in me and giving me the chance to pursue this degree, and I much regret the fact that I will have had little chance to pay him back for his kindness.

I have enormous gratitude for Dr. Thomas T. Tran, my co-supervisor, for his help, rich discussions, and sound advice. His valuable comments on my work and experienced guidance were significant in shaping the quality of this research. For that reason, it is difficult to put into words the debt that I owe to Dr. Thomas T. Tran.

During the last two years, I have had the pleasure of working with Dr. Ali Nazari, for whom I am greatly thankful.  His technical expertise combined with his friendship has been very helpful to me during the critical stages of this work. I would especially thank him for his excellent work on helping me build the experimental tests.

I also wish to dedicate my thanks to my parents from the bottom of my heart. Although they are physically far away from me, I feel they are with me all the time. It is their unconditional support and love that have made all my accomplishments possible.

Last, but certainly not least, I would like to express my sincere gratitude to my family, especially my beloved wife Reem and my two beautiful kids Mostafa and Yasmine for all the sacrifices and patience, and, most importantly, for their understanding that sometimes "I am extremely busy."

# Abstract

With the rapid development of applications in open distributed environments such as eCommerce, privacy of information is becoming a critical issue. Today, many online companies are gathering information and have assembled sophisticated databases that know a great deal about many people, generally without the knowledge of those people. Such information changes hands or ownership as a normal part of eCommerce transactions, or through strategic decisions that often includes the sale of users' information to other firms. The key commercial value of users' personal information derives from the ability of firms to identify consumers and charge them personalized prices for goods and services they have previously used or may wish to use in the future. A look at present-day practices reveals that consumers' profile data is now considered as one of the most valuable assets owned by online businesses. In this thesis, we argue the following: if consumers' private data is such a valuable asset, should they not be entitled to commercially benefit from their asset as well? The scope of this thesis is on developing architecture for privacy payoff as a means of rewarding consumers for sharing their personal information with online businesses. The architecture is a multi-agent system in which several agents employ various requirements for personal information valuation and interaction capabilities that most users cannot do on their own. The agents in the system bear the responsibility of working on behalf of consumers to categorize their personal data objects, report to consumers on online businesses' trustworthiness and reputation, determine the value of their compensation using risk-based financial models, and, finally, negotiate for a payoff value in return for the dissemination of users' information.

# Table of Contents

## Chapter 5. AAPPeC design specification       79

## Chapter 6. Proof of concept implementation       105

## Chapter 7. Evaluation       130

# List of Figures

# List of Tables

# List of Definitions

| | |
|---|---|
| Agent | A tool that a user might use to accomplish some goal. |
| Autonomous | Being self-contained and capable of making independent decisions |
| Consumer Agent | An agent working on behalf of the consumer |
| Electronic Commerce | A collection of people and services that interact through an electronic medium |
| Infomediary | An entity that collects data and performs service matching between consumers and service providers |
| Service Provider Agent | An agent who works on behalf of a service provider; an infomediary who also works on behalf of a service provider based on an agreed-upon contract |
| Privacy | The legal and ethical principle that individuals control data that can identify or be identified with themselves |
| Privacy risk | The perceived risk of revealing private data |
| Private data | The data that defines a single identity and its preferences |
| Property right to privacy | The right to own private data |
| Privacy credential | Reliability measure of obtaining privacy trust attributes |
| Reputation | The public opinion towards a person |
| Ticket | An electronic certificate that can be used to prove that a client has certain permissions. |
| Trust attributes | Third party credentials obtained by the service provider |
| TTP | Trusted Third Party |

# List of Acronyms

| | |
|---|---|
| AAPPeC | Agent-based Architecture for Privacy Payoff in eCommerce |
| JADE | Java Agent Development Environment |
| HTTP | Hyper Text Transfer Protocol |
| P3P | Platform for Privacy Preferences |
| FIPA | Foundation of Intelligent Physical Agents |
| PCR | Privacy Credential Rating |
| W3C | World Wide Web Consortium |
| XML | Extensible Markup Language |
| TTP | Trusted Third Party |

# List of Symbols

| Symbol | Description |
|---|---|
| $\beta_{ij}$ | Privacy risk of private data in category i under context j |
| $\Psi(\Lambda, \beta)$ | Magnitude of privacy risk of private data set |
| $\varphi$ | Set of credential attributes |
| $\Phi$ | Online service provider reliability factor |
| $E(U_q)$ | Expected return U of revealing private data record q |
| $\phi_{(t)}{}^A$ | Time-dependent function for an agent using time-dependent tactics |

# Chapter 1. Introduction

Collection and analysis of personal information are among the most far-reaching developments in online retail practices (Chellappa and Sin 2005). Like gold nuggets, data nuggets about consumers are random pieces of raw information that have little value until they are brought together and modeled into something with a desirable structure and form (Taylor 2004). This is the key to successful online marketing strategies, which rely on data warehousing technologies to bring structure to data, businesses intelligence tools to give the information context and meaning, and customer relationship management applications to provide consistent and personalized knowledge about consumers (Grover and Teng 2001). Only then do online retailers truly get to know the person behind the customer. Such endeavor, however, has resulted in the unprecedented attrition of individual privacy and individual's right and the desire for informational self-determination (Cavoukian 2009).

Today, many online companies are gathering information and assembling sophisticated databases that contain a great deal of information about many people, generally without the knowledge of those people (Cavoukian 2009; Prins 2006). Such information then changes hands or ownership as part of normal eCommerce transactions or during a firm's strategic decisions which often include selling consumers' information lists to other firms (Prins 2006). With the increasing economic importance of services based on the processing of personal data, it is clear that firms, especially online businesses, have high incentives to acquire consumers' personal information (Taylor 2004). Collection of personal information about customers brings the following two important benefits for online businesses: First, data about individual (potential) customers allows for targeted advertisements and therefore materializes as a higher return on investment (Cavoukian 2009; Prins 2006). Second, information about an individual buyers' willingness to pay enables online merchants with market power to impose pricing strategies that increase sales and revenues (Taylor 2004). A look at present-day practices reveals that consumers' profile data is now considered one of the most valuable assets owned by online businesses (Rendelman 2001). As a result, a flourishing market in the sale of personal information has been created (Taylor 2004). Although private data is not traded separately, when aggregated together, the tiniest nuggets of personal information have value (Prins 2006; Taylor 2004; Cavoukian 2009).

In such an information market, however, consumers are currently unable to participate for two reasons. Firstly, they have no means by which they can benefit from the profitable commercial exchange of their own asset (i.e., their personal information). Although it is known (Acquisti 2004; Chellappa and Sin 2005; Harn et al. 2002; Han et al. 2003) that sometimes consumers make choices in which they surrender a certain degree of privacy in exchange for non monetary benefit (such as better quality, customized offers, specials, coupons, etc.), according to Taylor (2004) these choices are far less than what the consumers deserve given the risk of privacy exposure.

Secondly, they are powerless to prevent the unauthorized dissemination of their private data. In this sense then, the market is "unfair." This is the main motivation behind this work, as explained next.

## 1.1   Research Motivation

Based on the above imbalance, not only is consumers' privacy becoming a critical issue, but so is the risk at which they are put from any damages arising from the loss of their privacy, such as revenue loss or increased cost. The economic consequences of the misuse of personal information can be significant for an individual and society as a whole (Roznowski 2001; Tang and Smith 2008). As it currently stands, privacy risk is a business risk resulting from the collection, use, retention, and disclosure of users' personal information. As in all business risk, privacy risk could result in revenue loss.

For many people, there is not much difference between the collection and use of their personal information and the shooting of their portrait to be published in a magazine (Cavoukian and Hamilton 2002). Both are representatives of the individual. According to Cavoukian and Hamilton (2002), "if someone takes a picture of another person and wants to reproduce it in a book, the normal practice is for the photographer to obtain consent from the person to whom the picture belongs. This person might be in a position to collect royalties from its use. But if someone gathers information about an individual—where he lives, what he does, what his habits and lifestyle are—and then reproduces it electronically in a list that is sold over and over again to a variety of organizations, chances are this individual would not even know it is happening. That is, not until this individual starts to receive such annoyances as junk mail, telemarketing calls, spam e-mail, and pop-up Web ads". Unlike the photographer who may negotiate with the individual a royalty fee against using the picture, the "electronic intruders" collect all the gains for themselves and leave the consumers with nothing except moral and financial hardship (Bergelson 2003). Consumers need mechanisms that help them valuate their personal data and give them the ability to negotiate a payoff or compensation at least in part from the damages that might occur to their privacy.

As will be explained in more detail in subsection 1.3, the scope of this thesis is on developing an agent-based architecture for privacy payoff as a means of rewarding consumers for sharing their personal information.

### 1.1.1      Motivating Example

Although the Internet has brought unquestionable benefits to people around the globe and helped make our society connected like never before, it has also made it possible for unknown characters to surreptitiously watch users in every click they make when they visit their favorite websites. Online tracking technologies such as HTTP cookies, Flash cookies, Web bugs, bots, etc. allow service providers to monitor users' behavior as well as scrutinize their lifestyle and personal habits and preferences. Not only that, but also the data the users reveal to complete a transaction (such as name, address, email, telephone numbers, etc.) is often sold to third parties for marketing and advertisement purposes. The following real-life example describes the vast extent of personal information that can be collected in the online world.

**Example 1: By Saul Hansell. The New York Times October 24, 2008**

"*It was summer time when I decided to buy a plastic watch before going to the beach, so I ordered one from Swatch.com. What I didn't know was that Swatch sent information about my purchase to a company called Acerno, which runs an advertising network that reaches 80 of the top 100 Web sites. Since then, some of the ads I have seen on those sites could well have been placed by a company that specifically wanted to reach watch buyers. Acerno, which has operated for three years with almost no publicity, says it now has files on 140 million people in the United States, nearly all the online shoppers. These are gathered from 375 online stores, including Spiegel, eHarmony, Video Professor, Michelin and*

*American Girl, where it tracks not only what Internet users buy but also what products they read about. It uses this information to place ads on more than 400 sites on behalf of marketers. Now every time I check my email I find at least one advertisement email about watches. When I visit some other websites, I also find online advertisements about watches.*"

The above example conveys some of the privacy invasions that the average person can experience in the online world. It is a typical scenario of how online service providers could aggregate the users' information and then sell it or share it with other advertisement firms. It shows that users' personal information is very valuable for online businesses. Consumers' behavior, taste, preferences, lifestyle, etc. means money. While online businesses use consumers' information to rack up great amounts of money, consumers are excluded from the profitable commercial exchange of their own assets. Therefore, a mechanism is needed to allow consumers to gain control over the dissemination of their private data and to be far-sighted about the value of their own personal information.

The proposed architecture, described in Chapter 4, is a multi-agent system in which several agents play different roles and coordinate their activities to provide a user-centric system. The agents in the system bear the responsibility of working on behalf of consumers to categorize their personal data objects, report to consumers on online businesses' trustworthiness and reputation, determine the value of their compensation using known financial models, and finally negotiate for a payoff value in return for the dissemination of users' information. To perform their tasks, the agents in the system leverage computational and artificial intelligent methods, such as fuzzy logic, data categorization based on attributes ontology, and strategic negotiation in order to maximize the benefit to the consumers.

The viability of the system is demonstrated through detailed analysis and a prototype implementation, called AAPPeC, which is based on the JADE (Java Agent Development Environment) multi-agent framework. Thus, the proposed architecture would be of scientific value and a contribution to the field.

## 1.2    Research Problem

Collection and processing of personal information for commercial use without the consent of the original owner (i.e., the user to whom the information refers) has created a privacy challenge to users, to the growth of the Internet, and to service providers as well (Bartow 2000; Bergelson 2003; Gopal et al. 2006). According to Cavoukian (2009), violation of consumers' personal information is "an external cost or negative externality, often created by online businesses, therefore the cost of resolving the problem should be borne by online businesses." While many approaches have been proposed (the details of these approaches are provided in Chapter 3 – Related Work) to assure consumers that their personal information in online transactions is protected, none offers the necessary guarantees of fair information practices. Consumers need mechanisms that guarantee their participation in the decision making of their own asset (i.e., their personal information) and furthering their interests in environments where personal data is commercially exchanged among Web-based service providers. The research problem of this thesis focuses on a number of issues related to the users' privacy and the commercial use of personal information. These issues can be summarized in the following way:

**Control**: An individual must have full control over the dissemination of personal information deciding who may or may not have access to his personal information. In particular, determining reliable decision criteria for sharing personal information based on interest (payoff), trust, risk, and protection level in the context of data usage.

**Payoff Value**: Determining the value of a given private data object by means of credit assigning and weighting concept given the risk of potential damages resulting from the misuse of personal information.

**Personal information usage**: Determining a mechanism based on trust that examines service providers' promises regarding privacy and private data handling. The mechanism must be able to provide a reliability metrics reflecting the service provider's competency towards personal information usage.

**Negotiation**: Facilitating a dynamically automated negotiation process between agents representing consumers and agents representing service providers. In particular, leverage a negotiation strategy for the goal of maximizing the users' benefit.

As will be demonstrated later, these issues are not completely addressed by existing solutions. Here we briefly present two examples, and refer the reader to Chapter 3 for a detailed analysis of related work. One example is the Platform of Privacy Preference (P3P) developed by the World Wide Web Consortium (W3C), which is emerging as the standard protocol that gives consumers control over their personal information and is considered a fundamental model for handling privacy concerns in the online world. However, mechanisms for automatic determination of the value of private data when exchanging rules in P3P format do not exist yet. Another problem with P3P is that there is no way for P3P to test or track if a website is actually following its advertised privacy policies thereby giving users a false sense of security about the site (EPIC 2002; Lioudakis et al. 2007). Another example is the PLUTO protocol which extends the P3P specification (Desmarais et al.

2007). The PLUTO protocol incorporates government privacy laws so as to enable the controlled sharing of private information. Despite the possible benefits of such extension, specifically the ability to technically describe what is legal and what is not, the protocol does not consider the value of private information in the presence of potential benefits from sharing personally identifiable information.

Today, there is evidence to suggest that the market is evolving a new industry in which agents negotiate for consumers' information (Dighton 2003; Hagel, J., III and Armstrong, A. (1997); Hann et al. 2003; Hui 2006, Alshammri 2009). Companies called "infomediaries" will act as custodians and brokers of customer information. In such an information market, consumers need a reliable mechanism that guarantees their participation in deals related to their personal information. In particular, they need a mechanism that is convenient for them to use, allows them to participate in the information market and ultimately control who has access to their personal information. The creation of such a user-centric architecture is the primary subject of this thesis.

## 1.3    Research Goal and Objectives

According to the above discussion, the general goal of this thesis is as follows:

> *To develop an architecture that allows users to participate in the information market, benefit from sharing their personal information, and ultimately decide who may or may not have access to their private data.*

In particular, the following objectives are extracted from this goal:

**Objective 1**

An absolute requirement for the architecture is granularity awareness of personal information. Hence, we aim to define modular attribute ontology that ensures the classification of a well-defined private data according to its usage context and semantic. While the usage context may vary as well as the varying importance a user may assign to one piece of specific data over another, context-dependent risk weights will be used to weigh the importance of different situations. Furthermore, the mechanism aims at determining the privacy risk value which will be used to determine the user's payoff once it has been decided to reveal the personal data.

**Objective 2**

One of the challenges that users face in the online world is how service providers use their personal information. Such knowledge is very important for them in order to foresee the possible privacy risks. It also allows them to assign privacy risk weights to their private data objects. Hence, there is urgency for mechanisms that allow consumers to make assessments regarding the potential privacy risks resulting from the usage of their private data in transactions involving risk. For this reason, we aim to develop a privacy credential rating mechanism that assesses the competency of the service provider with respect to privacy and private data handling, yet this mechanism allows service providers to increase and correctly represent their trustworthiness.

**Objective 3**

Personal information is a valuable marketing asset and a source of revenue for service providers. Consumers need to be far-sighted about the value of their personal information. Hence they need a mechanism that associates the value of the information to the individual's behavior taking into consideration his or her privacy concerns. Our aim is to determine the reward amount which the consumer should receive based on the perceived privacy risk of each data object as well as the risk resulting from the composition of private data objects.

**Objective 4**

Since it is a daunting prospect for an individual consumer to negotiate a return value with an online service provider against the dissemination of personal information, we aim at developing a negotiation strategy to ensure that the agent which works on behalf of the consumer is maximizing the consumer's benefit during the negotiation process. For this reason, we have extended a negotiation protocol developed by Skylogiannis et al. (2007) for the purpose of our architecture.

It should also be mentioned that it was decided to use software agents for the design and implementation of the system, because they have a high degree of self-determination—they decide for themselves what, when, and under what conditions their actions should be performed (Sierra 2004). They are becoming a choice technology for carrying out complex tasks that many users cannot do on their own (Wooldridge, M. 2002; Skylogiannis et al. 2007; Alshammri 2009). For example, in order for a user to valuate his or her personal information in an open environment such as eCommerce, the user needs to fulfill certain requirements of personal data categorization and data composition based on attribute ontology, risk assessment, and quantification, automatic negotiation, etc. These tasks are very tedious and sometimes difficult and require certain expertise. Software agents, on the other hand, are capable of autonomously performing complex tasks and meeting their design objectives in the environment where they reside (Wooldridge and Jennings 1995).

## 1.4    Main Contributions

The contributions of this thesis are:

- The identification of failures and shortcomings of related work to privacy-preserving technologies, agent-based reputation and trust systems that support privacy protection, and agent negotiation models for automated privacy negotiation, with a discussion of the reasons for these failures or the need for their extension as part of a general solution that will satisfy our architecture.

- The design and implementation of a novel agent-based architecture to overcome the current shortcomings of systems that do not consider private data as a set of assets that have value in use as well as in exchange. The new architecture is based on a detailed analysis of the emerging information market and the privacy problem that faces online users.

- A proposal of a new technique to categorize and evaluate privacy risks derived from users' private data set based on attribute ontology of the semantic equivalence of personal information as well as the rate of substitution. Such a technique is very important for the process of privacy risk quantification.

- The employment, for the first time, of a well-known financial model to determine the payoff value that the consumers should receive against the revelation of their personal information. This model allows us to determine the return value based on the privacy risk quantification.

- An advancement of the state of the art by proposing a new assessment scheme to determine the trustworthiness of service providers based on privacy credentials. This new assessment is targeted to determine the competency of the service provider regarding the handling of personal information.

- The extension of a known negotiation scheme for the purpose of privacy payoff negotiation that allows the agent representing consumers to be strategically placed during the negotiation process.

- A proof-of-concept implementation based on the JADE multi-agent framework.

- An empirical evaluation of the privacy credential rating model. The evaluation study was designed to test the effect of the rating on the participants' attitude, awareness, and perceived trust.

### 1.4.1   Peer-reviewed Publications

**Refereed Journal and conference Papers**

Yassine, Abdulsalam, Ali Asghar Nazari Shirehjini, Shervin Shirmohammadi, and Thomas T. Tran. "Knowledge-empowered agent information system for privacy payoff in

Thomas T. Tran. "An intelligent agent-based model for future personal information markets." In Web Intelligence and Intelligent Agent Technology (WI-IAT), 2010 IEEE/WIC/ACM International Conference on, vol. 2, pp. 457-460. IEEE, 2010

## 1.5   Road Map

The road map for the rest of this thesis is outlined below:

**Chapter 2 Privacy – Background and Issues** provides background information on privacy issues; it presents a brief background on the debate with respect to people's right to privacy and their property rights to private data**.**

**Chapter 3 – Related Work** examines a sample of systems that we believe to be representative and specifically related to the work in our topic. This chapter offers a panoramic view of current models and systems. It also provides a set of relevant aspects associated to the presented models that are used to establish a classification of related work.

**Chapter 4 – Proposed Architecture** presents the proposed system architecture and explains its components.

**Chapter 5 – AAPPeC Design Specification** presents a view of the AAPPeC software engineering design specification using GAIA and JADE.

**Chapter 6 – Proof of Concept Implementation** presents the JADE implementation of AAPPeC and provides experimental scenarios that affect the privacy payoff of consumers as well as service providers.

**Chapter 7 – Evaluation** focuses in the field study which was designed to examine the effect of the proposed privacy credential model on the consumers' attitude and perceived trust.

**Chapter 8 – Conclusion and Future Work** concludes our work and discusses our future research plan given the direction and the open questions resulting from this thesis.

# Chapter 2. Privacy – Background and Issues

## 2.1    Introduction

This chapter provides background information on privacy; in particular, it presents a brief background on the debate with respect to people's right to privacy and their property rights over private data. Although the nature of this discussion is not technical, its inclusion is essential and will enrich the understanding of the privacy problem and hence provide a better view to the general theme of the proposal.

What is privacy?

Firstly, according to (Warren and Brandies 1980), it is "the right to be left alone," which means the protection against intrusion by unwanted information. Secondly, it is the ability to control information about oneself and one's activities; this is related to information about a specific identifiable person and protection of that information from tampering by others (Samuelson 2000).

There is a common consensus among privacy advocates regarding the above definition. However, this consensus breaks down when they discuss the ownership rights to private data. Specifically, some advocates believe that individuals have property rights over their private data, and hence have the right to commercially benefit from their own personal information. Others object to the idea of having property rights over private data. They argue that "the property rights" approach is insufficient to protect privacy because of the nature of the privacy concept itself (Prins 2006; Swire 1997). By this they mean that the legal system values privacy as a human right (Samuelson 2000; Schwartz 2004) which has no relation to property rights, and therefore a "property rights" approach is irrelevant.

The next two subsections provide an in-depth overview of this debate.

## 2.2   Property, Privacy, and the Right to Private Data

The suggestion that privacy should be enforced by an ownership right has sparked a debate about the opportunities and risks of the "propertization" of personal data (Prins, C. 2006; Bartow 2000; Vera Bergelsom 2003). Proponents of "propertization" believe that the concept of privacy protects personal information from the unauthorized use and sharing with others. Generally, individuals have a legal right to be left alone and thus to prevent others from accessing their personal data. As a result, the legislative system that supports the privacy principle should make it clear that people have an exclusive right to their private data and thereby provide them with instruments guaranteeing them an exclusive right to their personal data (Laudon, K.C. 1996; Vera Bergelsom 2003).

However, proponents of property rights for private data argue that despite the enormous legal armaments that have evolved over the past 100 years, most people feel their privacy has not improved (Laudon, K.C. 1996). This is due to the vast changes that have occurred within society during this period. A closer look at our society reveals that knowledge of data about individuals is crucial for a variety of important purposes: education, medical care, commerce, marketing, terrorism detection, etc. (Prins 2006). Moreover, advances in information technology have made it very simple to collect and use information. The technical infrastructure of the Internet combined with profiling techniques and other advanced processing applications make it easy and cheap to collect, combine, and use enormous amounts of data. Nowadays, many online companies have access to personal information and can sell it to third parties for lucrative amounts of money (Vera Bergelsom 2003).

Moreover, the present-day practice of selling personal information in the online world has evolved in a way such that individuals make choices in which they surrender a certain degree of privacy in exchange for deals (such as price discounts, money, improved qualities of service, customized offers, specials, etc.) that are perceived to be worth the risk of information disclosure (Chellappa and Sin 2005; Gopal et al. 2006; Grover and Teng 2001; Harn et al. 2002; Metzger and Docter 2003; Miyazaki and Fernandez 2001). In this respect, private information is an asset with a market price and opportunity cost.

Nevertheless, advocates of property rights stress that the solution to the privacy problem is not stronger privacy laws or new legislation, but making the existing information market fairer by giving individuals the right to own their own private data asset.

## 2.3    Reflections on Property Rights over Private Data

As mentioned earlier, not all privacy advocates are in favor of the idea of a separate legal acknowledgment of the propertization of personal data. Some even argue that the whole discussion is out of proportion and particularly unhelpful because such a discussion improperly privileges form over substance (Kang and Buchner 2004).

In reaction to specific suggestions that grant people a property interest in their personal data, Schwartz (Schwartz 2004) has pointed to structural and fundamental difficulties with such a propertization approach. For example, lack of collective action and the fact of bounded rationality may benefit the parties who process and share our information but not those who help us place limits on this processing. Under these circumstances, where the power dynamics are not in favor of ordinary individuals, people will be offered unvarying terms with respect to their privacy property, and are likely to accept whatever data processors offer them (Prins 2006). Today's practices on the Internet mean that e-businesses and other users of personal data apply a "take it or leave it" approach which leaves the consumer no choice but to accept or reject a service. Individuals thus appear to consent "gladly" to certain uses of their personal data. In other words, vesting a property right in private data would not make any difference because it would be a daunting prospect for any individual consumer to bargain with a huge online service provider. Therefore, individuals in effect would have no effective choice in the matter of the use of their "propertized" data (Schwartz 2004; Prins 2006).

Another issue that comes with vesting a property right in personal data is the question of what kinds of private data property rights should be vested? A natural question that arises is, if vesting a property right in personal data entails "someone" defining precisely what is worth being called a property right, then, according to (Prins 2006), "who will make the paternalistic choice between data that are and are not within the ambit of an individual's personal property? Would it be legislators, the courts, or individuals themselves?"

## 2.4    Summary

This chapter presented a background on the debate regarding people's right to privacy and their property rights over private data. Two opinions from the research community were discussed. One side of the debate is against the propertization of the private data while the other side sees that the changes that happen to our society dictate that users should be the sole owners of their private data and therefore they are free to benefit from sharing it. Based on the presented discussion the following observations are worth noting: first, there is a difference between privacy as a right and private data as an asset. While the former draws its authority from established laws and humanistic principles, the latter is a property which can be traded for a value. Second, the day-to-day practices of online companies with respect to personal information dictates that people need tools and practical mechanisms to be established as an integral part of an online society that is using their own personal information to shape its existence. Third, any solution to the privacy problem must guarantee individuals' participation in the decision-making process of how their own assets, i.e., their personal information, are used. While such participation is a basic market requirement, and its morality is protected by law, yet in reality neither the participation nor the enforcement of its morality is occurring. In the simplest form, individuals may argue that in the absence of ownership rights to private data assets, should they not be entitled to benefit from the sale of their asset as well?

# Chapter 3. Related Work

## 3.1 Introduction

An enormous amount of work has been conducted in the area of agent-based eCommerce systems and information privacy in eCommerce. It is well beyond the scope of this thesis to examine all of it. Instead, this chapter examines a sample of systems that we believe to be representative and specifically related to the work in our topic. This chapter tries to offer a panoramic view on current models and systems. Although far from being exhaustive, it gives a rather complete idea of the current state of the art.

We also provide in this chapter a set of relevant aspects associated to the presented models that are used to establish a classification of related work. The aim is to identify shortcomings of work related with a discussion of the reasons for these failures or the need for their extension as part of a general solution that will satisfy our architecture.

The organization of this chapter is as follows: work related on agent-based eCommerce systems is presented in subsection 3.2, privacy systems in eCommerce in 3.3 followed by classification of related work in subsection 3.4. We then close the chapter by a summary in subsection 3.5.

## 3.2 Agent-based eCommerce Systems

Traditionally, eCommerce is considered one of the main applications of agent-based systems (Sierra 2004, Wooldridge 2002). The rich content and the dynamic nature of eCommerce have made shopping activity a process that requires large effort. A human buyer, for example, requires collecting and interpreting information on merchants, products and services, making optimal decisions, and finally entering appropriate purchase and payment information. Software agents help automate a variety of activities, mostly time-consuming ones, and thus lower the transaction costs. Examples of agent-based studies that were conducted to automate such activities are described in the following paragraphs.

One of the most studied activities in eCommerce is negotiation. In eCommerce agent-based negotiation systems, an agent-buyer and an agent-seller apply strategies and negotiation protocols so the negotiation process between buyers and sellers is simplified. The work of Bagnal and Toft (2006) and Aknine et al. (2004) are examples of such studies. Bagnal and Toft (2006) describe an autonomous adaptive agent for single seller sealed-bid auctions. The aim of their work is "to determine the best agent learning algorithm for bidding in a variety of single seller auction structures in both static environments where a known optimal strategy exists and in complex environments where the optimal strategy may be constantly changing." Aknine et al. (2004) propose a solution that enables an agent to manage several simultaneous negotiation processes with the ability to detect failures.

Agent-based systems are also applied in the formation process of consumer-driven eCommerce societies, such as the work of Sensoy and Yolum (2009). The study proposes a multi-agent system of consumers that represent their service needs semantically using ontologies. The aim of the system is to allow consumer agents to create new service descriptions and share them with other consumer agents thus creating an eCommerce society of consumers. According to Sensoy and Yolum (2009), such a system leads to better offerings from service providers as they compete among each other to provide attractive promotions and target service to consumers more effectively.

Parallel to the above studies, agent-based systems were also applied in eCommerce privacy negotiation as in the work of El-khatib (2003) and Blažič et al. (2007). El-khatib (2003) presented an agent-based privacy policy negotiation. In the model, a user agent negotiates on the user's behalf with another agent that represents the service provider. The interaction diagram is shown in Figure 1. The negotiation protocol defines the syntax and the semantic of the exchanged messages among negotiating agents. It makes use of the existing P3P protocol to express the terms and conditions of the privacy policy carried in these messages. The architecture has four major components: a policy generator, a rule evaluator, a consumer's agent, and service provider's agent.

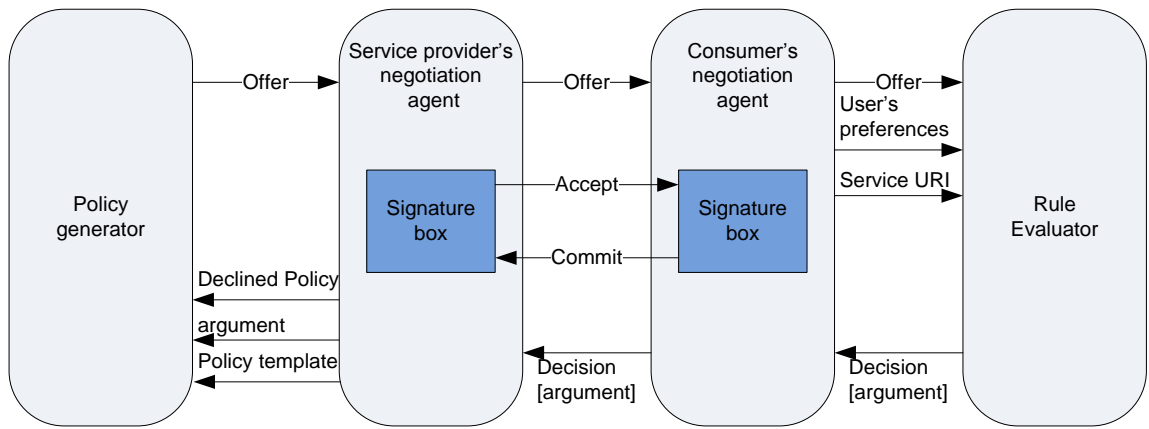

**Figure 1: Agent-Based Privacy Negotiation (El-khatib 2003)**

The policy generator uses a policy template to form all possible policy offers, while the rule evaluator running on the consumer's side compares the privacy policy offer received from the service provider with the consumer's privacy preferences. The service provider's agent and the consumer's agent negotiate a privacy policy using a protocol developed based on P3P protocol. Despite the advantages of having a negotiation agent working on behalf of consumers, users in the system lack a context-aware mechanism that can help them decide on the privacy practices that they prefer. In our system, users can assign privacy risk weights to private data based on the context of the transaction and the competency of the service provider to whom the information is transferred.

Blažič et al. (2007) developed a conceptual privacy-enabling architecture of infrastructural pervasive middleware that uses trust management, privacy negotiation, and identity management during the inter-entity communication lifecycle. The aim of the architecture is to balance between the conflicting interest of pervasiveness on the one hand and protecting privacy on the other. At the user side there are two agents, one for security and one for privacy. Similar agents are also present at the service provider's side. When both sides negotiate with each other through a negotiation manager, the privacy agents in both sides exchange privacy policy that describes the privacy preferences of the users. The identity manager (a component of the privacy agent) creates and maintains virtual identities, maintains a database on history of disclosures of personal identifying information, and estimates the threats of privacy exposures.

While our system employs agents to automate activities similar to the above-mentioned work, it tackles a novel problem (privacy payoff) in eCommerce that has not been addressed by any previous work. Each agent in our system autonomously assumes a specific activity such as consumers' personal data categorization, trust and reputation, payoff computation, and negotiation. Furthermore, our study, like that of Sensoy and Yolum (2009), uses agents which utilize attribute ontologies to aggregate consumers' information records according to a specific interest, thus forming a community of eCommerce consumers. The formation of such a community is beneficial for an individual consumer, because the agent negotiating on behalf of a consumers' community would be in a better position to bargain over the revelation of the personal information and get something of value in return.

To give a better understanding of our work, the following subsections describe some of the existing work in privacy systems for eCommerce, although they are not agent-based.

## 3.3 Privacy Systems in eCommerce

In this subsection, we provide a view of privacy systems in eCommerce. We focus in this review on systems that are designed to protect consumers' privacy in electronic transactions that require revealing sensitive data to online service providers. This subsection presents three types of systems, namely Access control privacy systems, Privacy negotiation systems and Privacy through trust and reputation.

### 3.3.1 Access Control Privacy Systems

The World Wide Web Consortium (W3C) has proposed the Platform for Privacy Preferences (P3P) W3C specification to enable service providers to post machine-readable privacy policies. Built into Internet Explorer, P3P is widely deployed as a privacy language. P3P provides a way for a website to encode its relevant practices and to communicate them to the users that visit the site. Also, P3P solves the problem of how does a user's web client decide when it can send personal information to a website. In essence, P3P is a method to provide privacy promises to the users for fair information practices; nevertheless, once users reveal their personal information to a service provider, there is no guarantee about the fate of the users' private data.

Desmarais et al. (2007) introduced an interesting approach to integrating legal requirements into the mechanism of data revelation. The mechanism proposed by Desmarais et al. (2007) is called PLUTO. It suggests that in order for electronic communities to share privacy sensitive information it is important to integrate the legal issues surrounding privacy into the tools that will manage the release of private information. PLUTO was designed to work with existing privacy legislations to create a solution to manage privacy information for electronic communities; specifically, the PLUTO protocol was designed as a solution to the Canadian privacy guidelines detailed in "The Personal Information Protection and Electronic Documents Act." PLUTO extends the Kerberos authentication protocol, which has been developed at the Massachusetts Institute of Technology (MIT), to prove the identity of the data requester on the server (and to prove the server's identity to the client). In order to adapt Kerberos for privacy work, a privacy context field was added to Kerberos. That context information can then be used to describe the privacy restrictions on the session. A sample Kerberos ticket with the added privacy context field is shown in Figure 2.

| Kerberos Ticket | | | | | |
|---|---|---|---|---|---|
| Client Portion (encrypted with Client Key) | | Server Portion (encrypted with Server Key) | | | |
| Session Key | Authentication Details | Session Key | Time To Live | Authentication Details | Privacy Context |
| | | | Timestamp / Lifespan | | |

**Figure 2: Kerberos Ticket extended for Pluto (Desmarais et al. 2007)**

Palen and Dourish (2003) argue that privacy is not simply a problem of setting rules and enforcing them, but rather an ongoing process of negotiating boundaries of disclosure, identity, and time. They suggest *genres of disclosure* for managing interpersonal privacy, which are "socially-constructed patterns of privacy management," as a sort of design pattern approach to protect users' privacy. Examples might include creating and managing accounts at shopping websites, taking appropriate photographs at social events, exchanging contact information with a new acquaintance, and the kinds of information one reveals to strangers.

A person fulfills a role under a genre of disclosure through her performance of her expected role in that genre, and the degree to which a system does *not* align with that genre is the degree to which it fails to support the user's and the genre's privacy regulation process. Although there are many lessons learned from this work, the two most salient here are that strict rule-based user interfaces may not always match peoples' needs in managing their privacy, and that social, organizational, and contextual settings are important factors to consider when designing and deploying privacy preserving applications.

Kenney and Borking (2002) introduced the concept of privacy engineering and accordingly proposed architecture to manage personal data help in Digital Rights Management (DRM) systems. The architecture defines a language to represent and describe DRM rights in terms of permissions, constraints, and obligations between users and contents. It emphasizes privacy as a data protection aspect and combines encryption techniques with access mechanisms with implicit trust in the controller entity which manages the coordination activity. The main advantage of this architecture is the access control unit, which incorporates policy management for access rights and permission. The architecture proposes a solution for eCommerce providers to ensure that they are ethically compliant to users' data protection. Such a practice will increase the trust in service providers and increase the tangible value of the service offered.

Lauden (1996) proposes the National Information Accounts, a market-based negotiation system in which information about individuals is traded at a market-clearing price, to the level where supply satisfies demand. The use of information would be limited to a specific period of time, and, maybe, for specific purposes. This system entails a strict control of the information transfer in the society, possibly imposed and maintained by the government. This model, although interesting and ingenious, does not suit the multiple possibilities to collect, store, and process information in a networked society, centered on the Internet, as a global, unregulated communication channel.

Lioudakis et al. (2007) propose a middleware architecture that incorporates personal data repositories and enforces regulations for personal data disclosure. A high-level description of this architecture is shown in Figure 3.

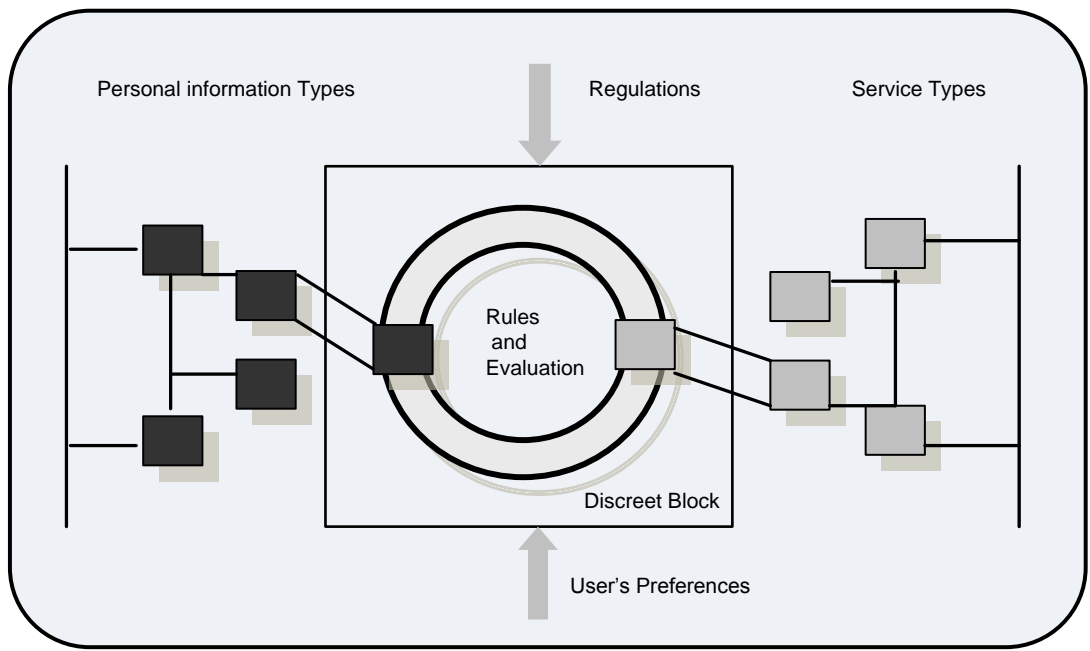

**Figure 3: Middleware Privacy Architecture (Lioudakis et al. 2007)**

The proposed architecture mediates between the users and the service providers while constituting a middleware shield for individuals' personal data based on rules and legislations. The proposed approach is to use a policy framework that will incorporate a large set of rules. All privacy-critical functionalities are integrated into a privacy-proof entity, namely the Discreet Box, which is assigned with the mission of enforcing the privacy legislation principles, on the basis on which it is conceived. The Discreet Box is installed at the service provider's premises but totally controlled by the Privacy Authority which decides on the data to be released. The Discreet Box incorporates the personal data repository that caches personal data and enforces retention periods, a policy framework that takes the decisions for personal data disclosure and interfaces to the service provider for the request/response cycles. The main shortcoming of this system is that in eCommerce applications where thousands of service providers do business directly with users, it is difficult to convince each provider to install a Discrete Box similar to the one described in the work of Lioudakis et al. Our approach in this thesis is that data control should be given to the users directly; they are the sole owner of the data and they decide who may or may not have access to it.

IBM Tivoli Privacy Manager for e-business (Bucker et al. 2003), shown below in Figure 4, is a well-known example of industry frameworks for privacy management in e-business.

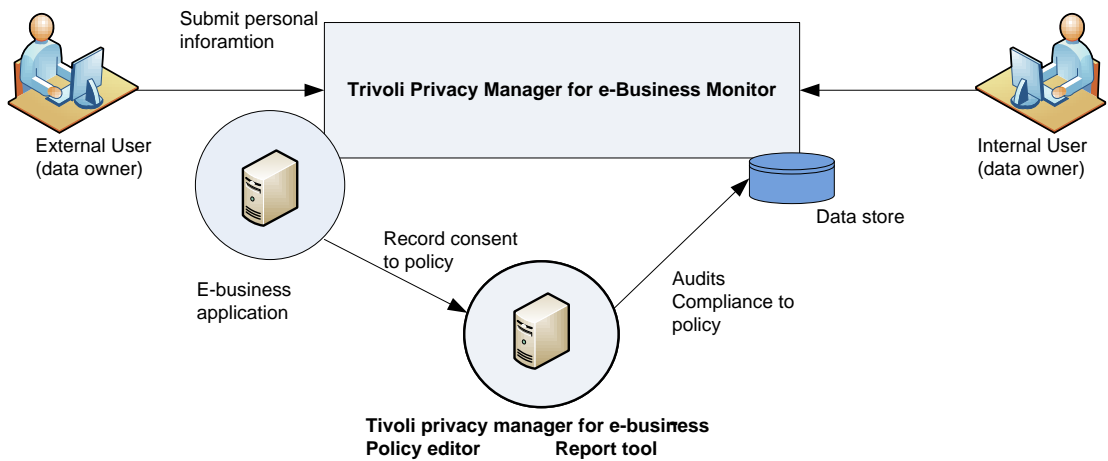

**Figure 4: IBM Tivoli privacy management solution components (Bücker et al. 2003)**

The main goal of the system is to ensure that actual privacy practice within the enterprise meets the advertised privacy promises. For example, if an organization has a mixture of 50 legacy and web-based applications and 32 of those applications come into contact with personal information kept in customer databases, then Privacy Manager consistently applies an organization's rules to all 32 applications. If the policy changes, either because of new legal rules or modifications to internal policies, Privacy Manager can make the changes centrally. However, a challenge arises when dealing with off-the-shelf applications, as it is difficult to get inside these applications to enforce privacy rules. Therefore, the requests from these applications are either accepted as is or rejected entirely.

All the abovementioned solutions have their weak points. First, although they manage to address the issue of privacy to a great extent or to enable respectful privacy management for users' profiles and identities, they fail in providing the necessary guarantees for fair information practices to the users. In fact, once the user reveals his private data to the organization, its use or abuse by means of processing and disclosure are still based on good intents. Second, the privacy policies specified in the context of these frameworks cannot be efficiently audited or verified as far as their regulatory compliance and consistency are concerned. Even an organization with the best intentions may specify a privacy policy that is not legislation-proof. Third, none of these solutions consider the value of private data in the presence of potential benefits from sharing it.

### 3.3.2        *Privacy Negotiation Systems*

Privacy negotiation is one approach to leveling the ground by allowing a user to negotiate with a service provider to determine how that service provider will collect and use the user's data.

Preibush (2005) proposes a framework for negotiating privacy versus personalization. During the negotiation process the service provider starts with a basic offer, consisting of a small discount and a few items of personal data to be asked. The framework examines how service providers may resolve the trade-off between their personalization efforts and users' individual privacy concerns through negotiations. Based on a formalization of the user's privacy revelation, a Bayesian game is considered for the negotiation where the service provider faces different types of users. The framework proposes two extensions to the P3P protocol to facilitate the negotiation process. It does not, however, consider the value of private data in question and therefore the service provider has the upper hand in the negotiation process.

He and Jutla (2006) develop a context-based negotiation model for the handling of private information. The idea of context is essential as users may have different rules about their data depending on the organization with which they are interacting. Moreover, users have different degrees of emotion towards various types of data (Acquisti and Grossklags 2005). For instance, financial and health data are highly emotional items while demographics such as postal code information may be less so. For this purpose they use OWL ontological representation to specify the revelation rules and to provide relevant substitute data for counteroffers within a negotiation session. In our work, however, the context does not rely on the nature of the data being revealed only, but also we include the trustworthiness of the service provider to be part of the context. And for this reason, we provide a trustworthiness assessment about the service provider who is going to receive the data. By so doing, we empower users with information that helps them make the right decision when they decide to reveal their data.

### 3.3.3 Privacy Enhancement through Trust and Reputation Systems

Recently, trust and reputation systems have been widely used in the context of eCommerce. With the aid of a reputation system, the buyer, for example, can distinguish between sellers based on their reputations, and trust only those sellers who have a reputation for being trustworthy in the current transaction. Thus, a reputation system can help traders achieve efficient useful outcomes, where it is equilibrium-enhancing behavior for the buyer to trust and a sufficiently patient seller to be trustworthy as long as the seller expects to remain in the market.

One example of such an approach is presented in the work of Rezgui et al. (2003). The proposed reputation system develops an automated process for privacy enforcement in the online environment. The reputation system monitors Web services and collects information related to their reputation for the purpose of privacy protection. The architecture, shown in Figure 5, has three main components: Reputation Manager, Probing Agents, and Service Wrappers.

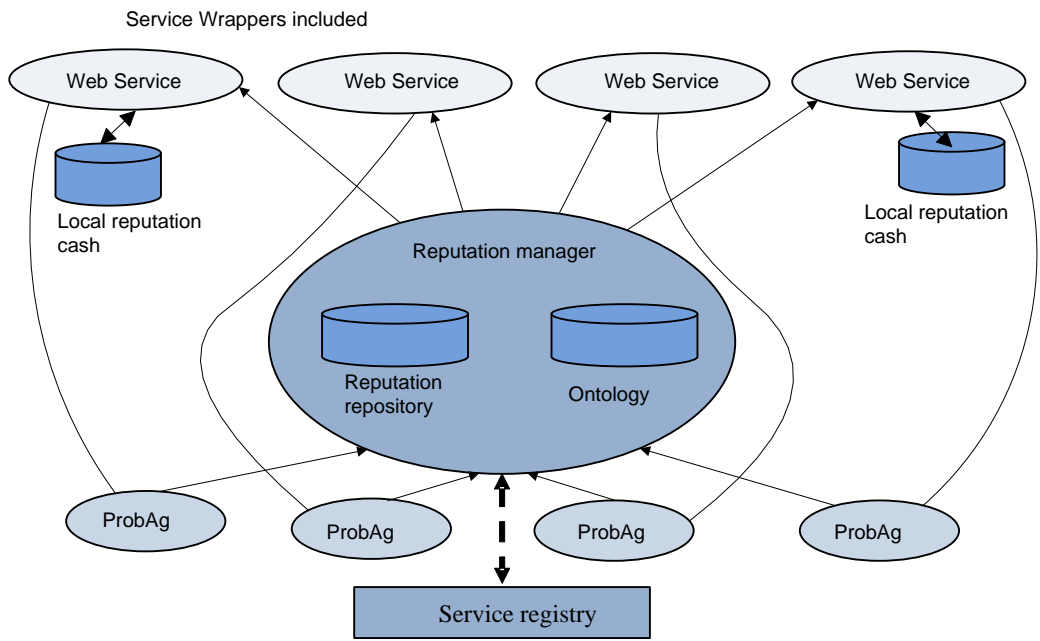

**Figure 5: Architecture of the reputation management system (Rezgui et al. 2003)**

When a user submits a request to the Reputation Manager, it collects reputation information stored in the reputation repository and sends the result back to the user. The collected reputation values are privacy-based attributes used to measure the privacy concerns of the selected service. The proposed solution also defines a set of attribute ontology that defines the semantic similarities of private data objects, which we found very useful and hence have adopted in our architecture when defining private data objects.

Wu Z. and Weaver (2007) propose a mechanism whereby the client discovers the service provider's requirements from a web service privacy policy document, then formulates a trust primitive by associating a set of client attributes in a pre-packaged credential with a semantic name, signed with the client's digital signature, to negotiate a trust relationship. The trust primitive is defined as "the minimal subset of attributes in a pre-packaged digital credential, which has a complete semantic meaning according to a set of policy requirements. A trust primitive is signed by the credential holder, which is either an individual user or a security domain." The clients' privacy is preserved because only those attributes required to build the trust relationship are revealed. For example, each time there is a request to retrieve a trust primitive from an outside domain it will be verified by the attribute service to selectively disclose the subset of attributes associated with the trust primitive. Additionally, trust primitives prevent initiation of selective disclosure from anyone except the credential holder. The system assumes that the users have a certain level of understanding of privacy concerns in the environment, which we find to be not realistic. This is because in general users have a hard time determining the trust attributes for disclosure unless they know the environment and the recipient of the data. Besides, each time there are requirements for additional personal information disclosure, a new round of trust negotiation is initiated to satisfy the additional privileges.

Ba et al. (2003) describe an economic incentive mechanism to encourage trustworthy behavior in markets. The authors model individual transactions as instances of the prisoner's dilemma where defectors might be punished in the repeated game, making cooperation the more profitable strategy. The proposed system imposes this punishment by using a trusted third party (TTP) and a cryptographic system to allow such punishment. This scheme makes the TTP an active participant in the execution of transactions. TTPs are currently used to verify identities on the Internet via digital certificates. The authors suggest the augmentation of this scheme by having the TTP track trustworthiness as well as identity. When engaging in a transaction, the agent offers its certificate to the other, who then verifies it with the TTP. If the TTP verifies the certificate, it confirms not only the identity, but the trustworthiness of the agent. The idea presented in this framework is very interesting; however, each time a transaction occurs between a client and a service provider it needs to be recorded at the TTP premises, which could be troublesome given the complexity of managing the enormous amount of transactions that take place.

## 3.4    Classification of Related Work

In this subsection, we present a short summary of the requirements criteria that is needed to develop an architecture that allows consumers to participate in the market of their personal information. The criteria here are developed based on the analysis of work related and the aim of the proposed architecture as defined by section 1.3. Following are the classification criteria: granularity awareness, sense of control, ownership of information, value of information, and tracking of information usage. The goal is to discuss how such criteria are addressed by existing models and for those that have not been addressed we state whether this research is concerned with tackling them and/or justify the reasons for not examining them in more detail. This discussion is presented next and at the end we provider a comparison summary of our findings in Table 1.

***Granularity awareness:*** Privacy of information is context dependent. Revealing information about sport activities is completely different from revealing information about financial earnings. Users are very sensitive to certain types of information depending on the context of its usage and to whom it is going to be revealed. A single-context model is designed to associate a single rule to private data revelation without taking into account the context. A multi-context model has the mechanisms to deal with different contexts associated with the revelation of personal information. Granularity of personal information has been addressed by studies such as those of Demarais et al. (2007), Lioudakis et al. (2007), He and Jutla (2006), Rezgui et al. (2003), Wu and Weaver (2007), and Ba et al. (2003). In our architecture, however, granularity of personal information is further addressed by associating the contextual usage risk to the personal data objects. By so doing, we are able to derive privacy risks that are associated with a transaction from the composition of different data objects.

***Sense of control:*** One method to assure the users that their private information will not fall into the hands of un-trusted parties is to provide them with a mechanism by which they can make decisions about who may or may not have access to their personal information. Only then will users feel that they have control over the dissemination of their personal information. Palen and Dourish (2003), Kenney and Borking (2002), Lioudakis et al. (2007), Lauden (1996), Preibush (2005), and Blazic et al. (2007) have studied mechanisms that provide such functionality either through negotiation or through mechanisms installed at the providers' premises to monitor the access to personal information. Earlier in this chapter we have explained the limitation of mechanisms installed at the providers' premises, therefore we do not consider such an approach. Our work is somewhat similar to Lauden (1996) and Preibush (2005). Although both of these allow access to information after negotiation, our work is different because we provide users with information about the competency of the party to whom the information will be revealed. By so doing, we help users decide who is trustworthy and who is not, and then it is up to the user to grant access to their private data. We recognize that this is critical to users before they grant access to their personal information, and, therefore, our mechanism provides such functionality.

*Ownership of information:* Privacy solutions that allow or enable ownership to private data mark information in a manner that allows association with an individual, behavior, or action. As we have explained in Chapter 2, the debate about the ownership rights to personal information is far from being over and is beyond the scope of this thesis. Our stand, however, is similar to the one taken by Lauden (1996). While Lauden suggested a national information account to store data about users and preserve their ownership rights, we believe that this would be very restrictive. We consider that users are the real owners of their personal information and in order to exercise this right they should be given the ability to control its fate. This right also includes their entitlement to benefit from the commercial use of their personal information as an asset.

*Personal Information value:* There is no doubt that personal information has value in use as well as in exchange. Personal information means money. By knowing the users' behavior, age, contacts, income, family size, hobbies, etc., service providers will be able to tailor their prices, target their advertisement, and enhance their services and products, etc. We believe this is a significant gap in the research and decided to address this limitation in our architecture. We have developed a risk-based mechanism to determine the value of personal information based on contextual risks. The contextual risks are based on the privacy concerns of the user to whom the information refers.

*Personal information usage:* The most challenging aspect of protecting users' privacy is tracking the usage of their personal information. The work presented by Lioudakis et al. (2007), IBM Trivoli, Blazic et al. (2007), Rezgui et al. (2003), and Ba et al. (2003) address this issue by defining usage policies versus user's preferences. In this thesis, we took a different approach. We decided to track the usage of personal information by analyzing the credentials of the party receiving the information. While there is no technical way for abuse prevention once the user reveals his or her personal information, knowing the privacy credentials of the party receiving the information will give an indication how they are going to use it. Therefore, providing users with mechanisms that enable them to know who can be trusted to handle their information and who cannot is very crucial in this matter.

Table 1 provides a summary of the systems analyzed in this chapter from the point of view of the classification points presented in this subsection.

We have to make some consideration about this table:

- We have described classification criteria that allow a comparison between privacy-aware models. However, due to the diversity of such models, the classification aspects do not always fit exactly with the characteristics of the models and in some circumstances the classification for a specific model in one category or another is subjective according to our interpretation.

- We have considered only the features explicitly presented by the authors; this is done without making suppositions on possible extensions.

Table 1: Summary of related work classification

| • | Granularity awareness | Sense of control | Ownership of information | Personal Information value | Personal information usage |
|---|---|---|---|---|---|
| Palen and Dourish (2003) | | √ | | | |
| Kenney and Borking (2002) | | √ | | | |
| Demarais et al. (2007) | √ | | | | |
| Lioudakis et al. (2007) | √ | √ | | | √ |
| Lauden (1996) | | √ | √ | √ | |
| IBM Trivoli | | | | | √ |
| Preibush (2005) | | √ | | √ | |
| El-Khatib (2003) | | | | | |
| He and Jutla (2006) | √ | | | | |
| Blazic et al. (2007) | | √ | | | √ |
| Rezgui et al. (2003) | √ | | | | √ |
| Wu and Weaver (2007) | √ | | √ | | |
| Ba et al. (2003) | √ | | | | √ |
| AAPPeC | √ | √ | √ | √ | √ |

## 3.5     Summary

In this chapter, we examined a sample of systems that we believe to be representative and specifically related to the work on our topic. In particular, this chapter reviewed work related to agent-based eCommerce systems and information privacy in eCommerce.

We presented classification criteria that allow us to study existing privacy-aware models and compare them to our proposed system. Based on this study, we have identified some of the shortcomings and provided discussion of the reasons for these failures and the need for their extension/solution.

The next chapter addresses these shortcomings by providing architecture for privacy payoff. The architecture addresses the granularity of personal information through the consideration of privacy risk weight assignment to individual private data objects. The architecture also addresses the issue of sense of control by allowing users to participate in the decision making on the dissemination of their personal information. Personal information valuation is also considered through the valuation of the payoff that the consumer should receive as a reward for sharing their personal information. The architecture allows consumers to be informed about the usage of their personal information through privacy reports that detail how service providers are expected to protect and handle private information.

# Chapter 4. AAPPeC Architecture

## 4.1    Introduction

In this chapter, the proposed architecture is presented. AAPPeC is an agent-based system in which each agent autonomously assumes a specific activity such as consumers' personal data categorization, trust and reputation, payoff computation, and negotiation. A high-level view that describes the architecture is made explicit next. This is followed by detailed descriptions of each agent in the system. Specifically, subsection 4.3 presents the architecture components of the database agent with emphasis on the attribute ontology details, data categorization, and privacy risk quantification. Then in subsection 4.4 we present the trust and reputation agent which implements a fuzzy logic mechanism to determine the reliability of the service provider with respect to privacy and private data handling. In subsection 4.5, we present the architecture and the analytical components of the payoff agent. The negotiation agent architecture is presented in subsection 4.6, where we discuss the negotiation strategy, offer construction, and negotiation protocols. Finally, the facilitator agent architectural components are presented in subsection 4.7.

## 4.2    High-level View

Figure 6 depicts the high-level architecture of the proposed system. Consumers open their accounts in the system and record their taste and preferences while online service providers negotiate, through their engaged agents, with the system to acquire the data. The multi-agent system takes on the responsibility of helping consumers valuate their personal data based on the perceived privacy risk of each private data object. Each agent is an independent entity with its own role and information. However, in order for an individual agent to satisfy its goal, it must interact with other agents. For example, the payoff agent's goal is to determine the value that the consumer should receive against the revelation of her personal information. However, the payoff valuation is based on the privacy risk that is determined by the database agent. Thus, in order for the payoff agent to satisfy its goal it must interact with the database agent. In the next subsections, we will explain the agent goals and their interactions in more detail.

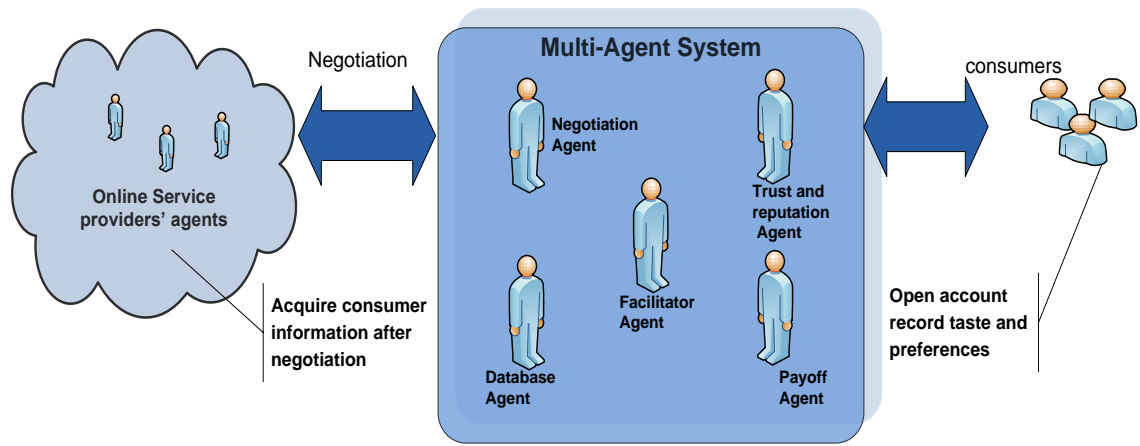

**Figure 6: High-level architecture of the proposed system**

Figure 7 shows the internal components of the multi-agent system. A database agent classifies private data into categories based on attribute ontology that captures the privacy sensitivity of the personal data (explained in subsection 4.3). A trust and reputation agent determines the trustworthiness of the service provider based on the competency of the online service provider with respect to privacy and private data handling (explained in subsection 4.4). A payoff agent employs risk-based financial models to compute the compensation value. A negotiation agent strategically negotiates on behalf of the consumers with the online service provider's agent to maximize their benefit in return for information dissemination. Finally, a facilitator agent manages the interaction of the agents and orchestrates the order of task execution.

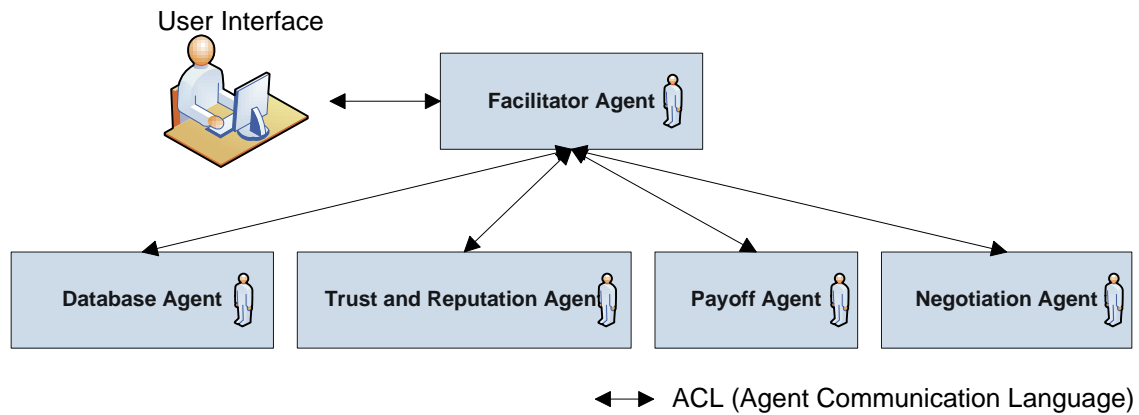

**Figure 7: Architecture components of the multi-agent system**

To explain how the system works, consider the following scenario: Alice is a privacy pragmatic person. According to Spiekermann (2001), privacy pragmatic person is defined as a person who is sensitive to data profiling but generally is willing to provide data. Alice is willing to share her personal data preferences with certain online service providers for a discount value or a reward fee. But Alice has certain requirements before she consents to complete the transaction:

(1) She wants complete information about the service provider's privacy practices and its trustworthiness.

(2) She wants to determine the level of risk involved in the transaction based on information from (1).

(3) She wants the system to valuate her privacy risk of revealing each personal data object as well as the privacy risk of potential private data combination.

(4) She wants her reward or discount value to be valuated based on the involved risk from (3).

(5) She wants the party that negotiates on her behalf to be strategic during the negotiation process so she can get the maximum benefit.

(6) She wants to have the privilege of accepting or rejecting the final offer.

To satisfy Alice's requirements, the system employs several agents where each one performs a specific task. The order of task execution is captured in Figure 8 and explained in the following paragraphs.

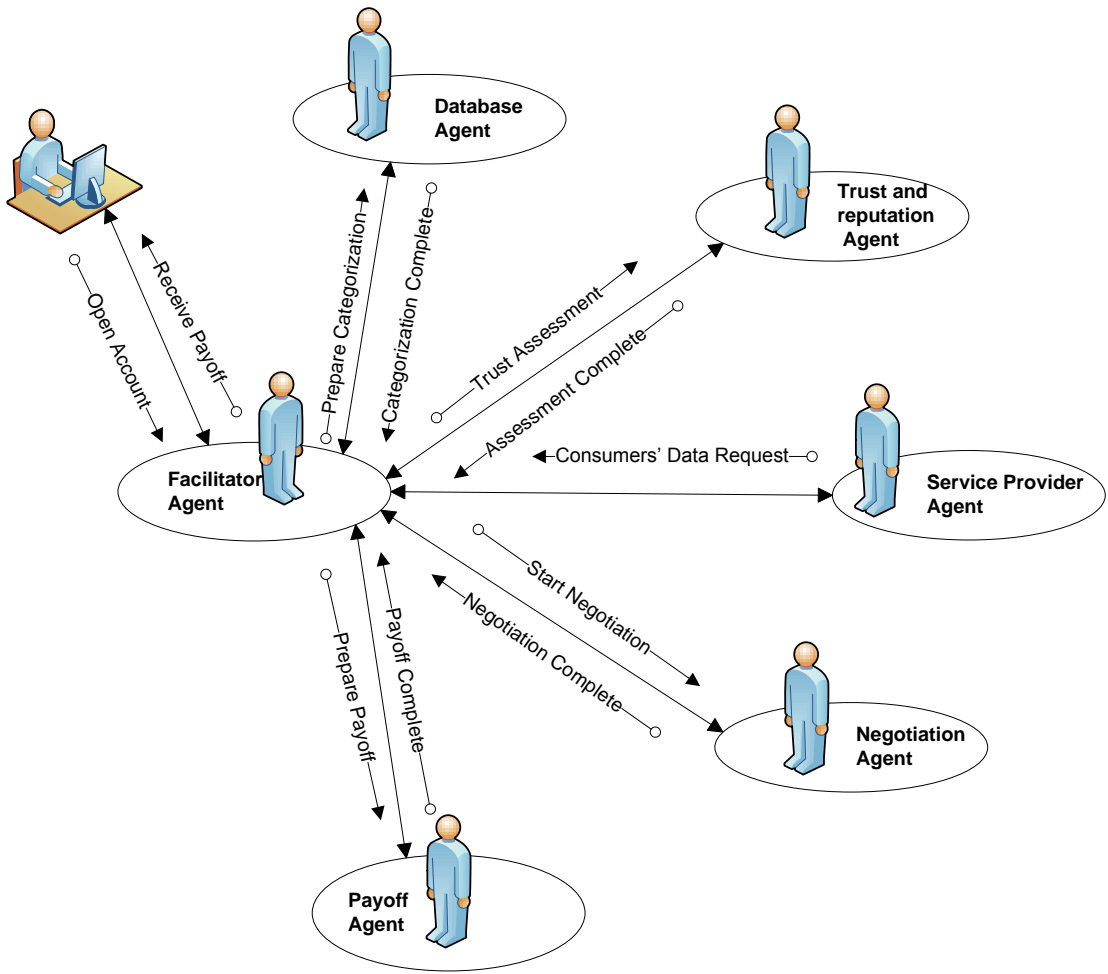

**Figure 8: Interaction diagram of the proposed system**

(1) Alice opens her account that records her taste, preferences, and personal data (essentially a detailed subscription form that needs to be filled once, but can be updated as desired; an example of such a form is provided in Chapter 6). A database agent automatically classifies the data into different categories, such as contact data, personal ID data, and hobby data. The decision about data categorization is performed based on attribute ontology formulation to reason about the composition of private data.

(2) When the service provider submits a request to obtain personal data (for example, data about consumers who like sports), the trust and reputation agent assesses the trustworthiness of the service provider. The trustworthiness assessment is mainly based on the competency of the service provider with respect to privacy and private data handling. The trust and reputation agent employs fuzzy logic techniques (details of these technique are discussed in subsection 4.4) to rate the privacy credentials of the online business. Privacy credentials (such as privacy seal, membership to privacy auditing services, security seals, authentication mechanisms, contents of privacy statement, etc.) are attributes which can be thought of as the means by which one can judge the competency of the online business with respect to privacy and private data protection (thus satisfying Alice's requirement 1).

(3) The trustworthiness rating is communicated back to Alice so she can make an informed decision when assigning the level of privacy risk to the categories of her private data objects (thus satisfying Alice's requirement 2).

(4) After Alice assigns her privacy risk weights to her private data objects, the database agent computes the total privacy risk value for her (thus satisfying Alice's requirement 3).

(5) The payoff agent uses the quantified privacy risk value to compute the payoff which Alice should receive if she decides to share her personal information with the online business. The payoff or the compensation is seen as a risk premium valuated in response to the amount of potential damages that might occur (cost of risk) with respect to the risk exposure (thus satisfying Alice's requirement 4). The payoff agent uses a computational model similar to the models used by the financial and insurance institutions (explained in subsection 4.5).

(6) Once the payoff is determined, the negotiation agent negotiates with the online business on Alice's behalf. The intelligence of the negotiation agent is realized by the negotiation strategy; i.e., the strategy of making concessions and evaluating the incoming offers and making decisions to interact with the negotiation opponent (thus satisfying Alice's requirement 5). The outcome of the negotiation is communicated with Alice. Alice either accepts the offer if it is equal or greater than the expected payoff and in this case shares her data with the online business, or denies the offer and in this case her personal data will not be shared with the online business (thus satisfying Alice's requirement 6).

The order of task execution described above is coordinated by the facilitator agent as seen in Figure 8. All incoming and outgoing tasks are managed by the facilitator agent. Once it dispatches the work order, the execution of the task is completely handled by the responsible agent. The facilitator agent implements a communication module to handle all the interaction with the agents in the system. Each agent in the system implements a similar communication module for its interaction. The discussion of such a module is saved until we discuss the architecture of the facilitator agent in subsection 4.7.

In the proposed system, enforcement is an important aspect for consumers to feel safe when using the system. Such enforcement requires some form of security assurances to the consumers so that their data is well protected in the system. The discussion about the integration of these security assurances with AAPPeC design is beyond the scope of this thesis. However, there are two possible security mechanisms that are worth mentioning. The first possible mechanism is the use of digital certifications where data stored in the system are encrypted with private and public keys. The entity that runs the system would not be able to use the data unless a certification of authentication is provided by a Trusted Third Party such as Comodo's Certification Authority (CA). Trusted third parties allow the consumers to approve or disapprove any operation that might occur on their data before issuing the authentication certificate. More discussion about information security management in user-centered multi-agent systems can be found in Piolle et al. (2007).

Another possible way is the use of technology such as "U Prove" privacy technology (U-Prove Documentations 2010). The U-Prove technology is built around the concept of U-Prove token, a binary string containing cryptographically protected information. There are three parties involved in the U-Prove token: the Issuer, who provides the tokens, the Prover, who needs a token, and the Verifier, the third party interested in authenticating the user. Each issued U-Prove token contains an unforgeable digital signature of its Issuer on the entire contents, created by the Issuer by applying its private key. The Issuer's U-Prove signature serves as its authenticity mark on the U-Prove token; it enables anyone to verify that the U-Prove token was issued by the Issuer and that its contents have not been altered. According to Microsoft, U-Prove technology can be used in virtually any communication and transaction system, examples being: "digital rights management, electronic records, online auctions, loyalty schemes, and e-gaming."

Next, the details of each agent's architecture are provided.

## 4.3    Database Agent

This subsection describes the details of the database agent. Figure 9 describes the components of the database agent. It consists of a communication module responsible for sending and receiving messages, while internally interacting with the data manager.

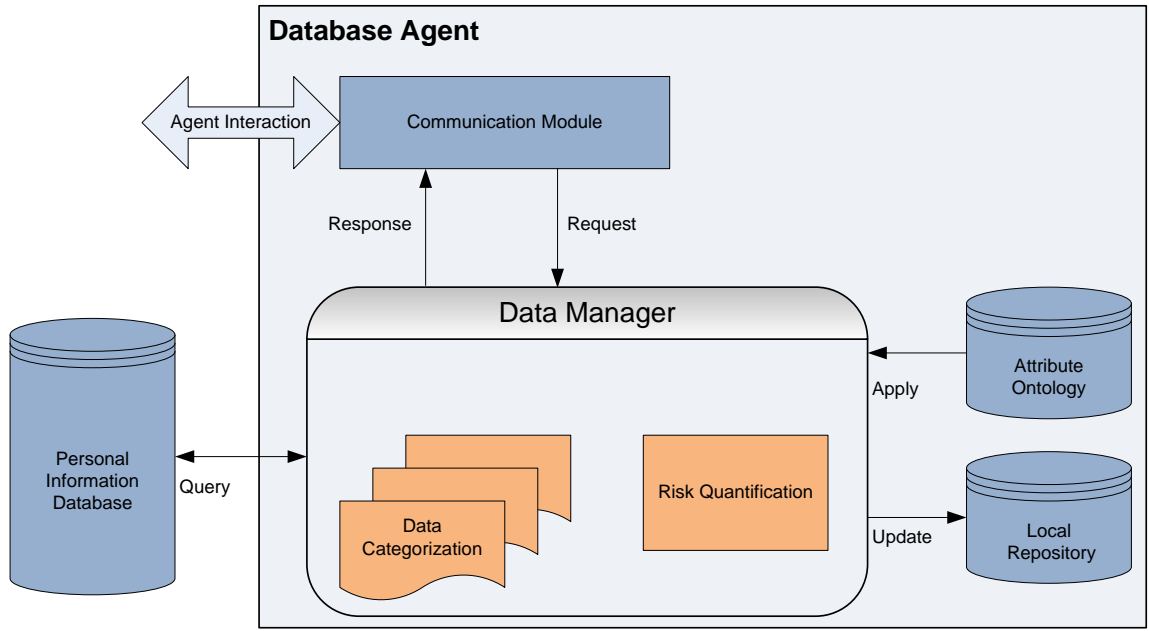

**Figure 9: Architecture of the database agent**

When a message is received at the communication module, it is then relayed to the data manager using a message-queuing mechanism. The data manager is responsible for categorizing personal information and quantifying the privacy risk value. In our system, the database agent uses attribute ontology that fulfils various requirements of privacy concerns related to private data composition. Attribute ontology (discussed in subsection 4.3.1) provides a single configuration source to the data manager to reason about the categorization of the private data and the quantification of privacy risk. The attribute ontology is a compact description of private data synonyms and composition rules. When the data manager finishes its task, it stores the results in the local repository stores for future uses.

The operation of the database agent is as follows: a user opens an account that details his or her personal information, taste, and preferences. The facilitator agent sends a request to the database agent to prepare the categorization of the private data. The database agent has access to the data specification of the user as shown in Figure 9. It applies the attribute ontology rules on the data specification and performs the data categorization. The structure of the data categorization is maintained and accessed by the database agent only. When the database agent receives another request to determine the privacy risk value, it uses the categorization structure to quantify the privacy risk of data composition.

The next subsection discusses the attribute ontology details.

### 4.3.1     Attribute Ontology

In order to determine how the agent will perform the categorization of personal information, it is quite essential to clarify different characteristics of personal information objects. We identify two attributes of data ontology, which pose certain design requirements for the proposed categorization mechanism.

Any operation that involves private data may be viewed as a process that may potentially result in privacy risk depending on the sensitivity level of the private attributes involved in the operation. The database agent categorizes the private data in a way that captures two important characteristics, namely, semantic equivalency and substitution rate given in the definitions that follow below.

**Semantic equivalency**: Consider two transactions that anticipate as their input a person's home phone number and the same person's family name. One transaction describes the parameters: PhoneNumber and FamilyName while the second transaction describes the parameters: PhoneNumber and LastName. From a privacy perspective, both transactions are equivalent (Rezgui et al. 2003). This is due to the semantic equivalence of FamilyName and LastName. To capture this equivalency among attributes, the agent uses ontology sets of semantically defined attributes. Below are examples of such sets (Rezgui et al. 2003):

Set 1 = {FamilyName, LastName, Surname}

Set 2 = {Address, HomeAddress, Location}

**Substitution rate**: The substitution rate of private data captures the level of risk in relation to private data revelation. Private data attributes that are considered substitutable have a constant substitution rate (Preibush, 2005); i.e., the level of exposure risk stays the same. On the other hand, private data attributes that are not substitutable may result in an increase in the privacy risk. An example of substitution rate is explained next in subsection 4.3.2, data classification.

### 4.3.2 Data Classification

The agent has access to data specification which reflects the consumer's true personal information and then classifies it into $M$ different categories $(C_1, C_2,..,C_M)$, such as personal identification, contact information, address, hobbies, tastes, etc. In every category $C_i$, private data are further divided into subsets; i.e., $C_i = \{S_{ik}$ for $k=1, 2,...\}$ based on the attribute ontology that applies to the data set; an example is shown in Figure 10.

In Figure 10, we consider category Contact, which is divided into two subsets, Telephone and E-mail. The database agent uses the attribute ontology to reason about the classification of the data as follows:

First, data in different categories may have different context-dependent weight. The value of the private data may differ from one context to another (personal data value in context sports and personal data value in context health); therefore, its composition may have different implications on the level of revelation.

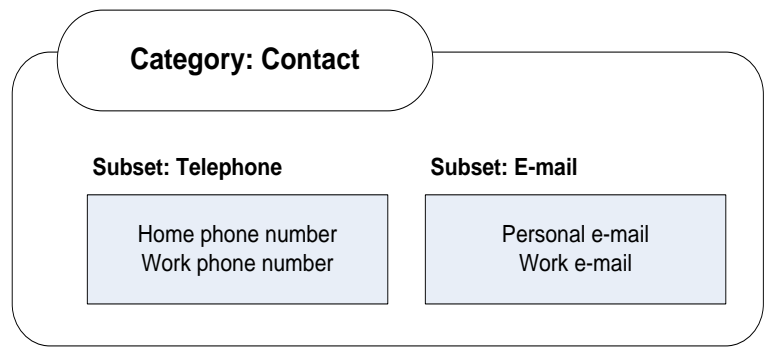

**Figure 10: Example of private data categorization**

Second, the substitution rate of private data in the same subset is constant and independent from the current level of revealed data, i.e., assuming that one of the private data items has been revealed, revealing the rest of the data in the same subset will not increase the privacy disclosure risk. For instance, a consumer's contact information can be expressed by work phone number or home phone number. Knowing both of them at the same time allows only marginal improvements. This will allow considering each private data subset as one unit.

Third, data in different subsets are not substitutable; revealing any one of them will increase the privacy risk. For example, the consumer's telephone number and her email address constitute two possible ways to contact the consumer but they are not completely interchangeable.

Next, we explain the quantification process of the privacy risk.

### *4.3.3          Risk Quantification*

After the agent classifies the personal data into different categories, as explained in the previous subsection, the agent now computes the disclosure privacy risk of private data. But first, it needs to capture the consumers' preferences about revealing different attributes of personal data for each category $i$ under context $j$. The context here refers to the situation and the nature of the private data involved in the situation. For example, an individual may have different perception for privacy risk when in country "A" versus when the same individual is in country "B". Such difference may have implication on the way this individual would be assigning weights to his private data categories.

**Furthermore,** privacy cost (cost resulting from privacy risk) valuation differs from one type of information to another. For example, privacy violation of an individual's medical information can be extremely costly for its owner, but the violation of the person's past shopping history would hopefully not be as detrimental. The context under which the private data is revealed could affect the level of privacy cost. For example, providing medical information to an online drug store is not the same as providing medical information to a government health agency even if it is the same information. In the former, the risk of leaking information about the health condition of the consumer to an insurance company could result in service denial from the insurance company; however, in the latter, the risk of leaking health information to a third party is less likely to happen. We use the parameter $\beta_{ij} \in [0, 1]$ to capture the privacy risk of each private data in category $i$ under context *j*. $\beta_{ij}$ represents the consumer's valuation of her private data under each context (we will explain how the consumer will specify this value in Chapter 6). More specifically, it represents the consumer's type (for example, Alice is privacy pragmatic), where consumers with a higher $\beta$ incur higher privacy costs as a result of privacy violation. Each consumer is assumed to be stochastically equivalent and has independent distributed valuation for each context. According to Preibush (2005) and Yu et al. (2006), individuals have various global valuation levels for each transaction that involves private data.

Let us consider a consumer with identity *I*, and several private data attributes $\Lambda = \{\Lambda_1, \Lambda_2, ..., \Lambda_n\}$. We let the privacy risk to vary with $\Lambda$ and $\beta$ according to the functional form $\Psi(\Lambda, \beta)$, where $\Psi(\Lambda, \beta)$ characterizes the magnitude of privacy risk resulting from the composition of different private data attributes $\Lambda$ and the cost $\beta$ (privacy risk cost) of revealing them. The calculation of $\Psi(\Lambda, \beta)$ is as follows (from now on, and for the sake of clarity, we will drop the symbols $\Lambda$ and $\beta$):

Let $X_i$ be the cardinality of $C_i$ (i.e., $X_i = /C_i/$). We perform normalization over the whole private data set. The normalized data size $NX_i$ reflects the ratio of each category in a consumer's privacy.

$$NX_i = \frac{X_i}{\sum_{k=1}^{M} X_k} \qquad (4.1)$$

where *M* is the number of the private data categories.

After the consumer assigns the value $\beta_{ij}$ for each category *i* under context *j*, the agent computes the weighted privacy risk, as in (4.2) and then normalizes it, as in (4.3). The value $N\Psi_{ij}$ in (4.3) reflects the risk of revealing all data in category *i* under context *j*. The purpose of this normalization is to put the privacy risk of each category into the range of [0,1].

$$\Psi_{ij} = NX_i * \beta_{ij} \qquad (4.2)$$

$$N\Psi_{ij} = \frac{\Psi_{ij}}{\displaystyle\sum_{n=1}^{M} \Psi_{nj}} \qquad (4.3)$$

The privacy risk weight of revealing $\alpha_i$ subsets from category *i* is calculated as in (4.4).

$$\Psi_i = \frac{\alpha_i}{X_i} * N\Psi_{ij} \qquad (4.4)$$

**Example**: Consider that we have two consumers, Alice and Bob. Alice is privacy pragmatic and Bob is privacy fundamentalist (i.e. very concerned about the use of his personal data Spiekermann [2001]). For the sake of simplicity, consider that both of them have similar categories as shown in Figure 11 and explained below:

*Category* $C_1$ (Contact) has two subsets $S_{11}$ (telephone) and $S_{12}$ (e-mail). Subset (telephone) has private data $d_1$ (home phone number), $d_2$ (work phone number), while subset (e-mail) has private data $d_3$ (personal e-mail) and $d_4$ (work e-mail).

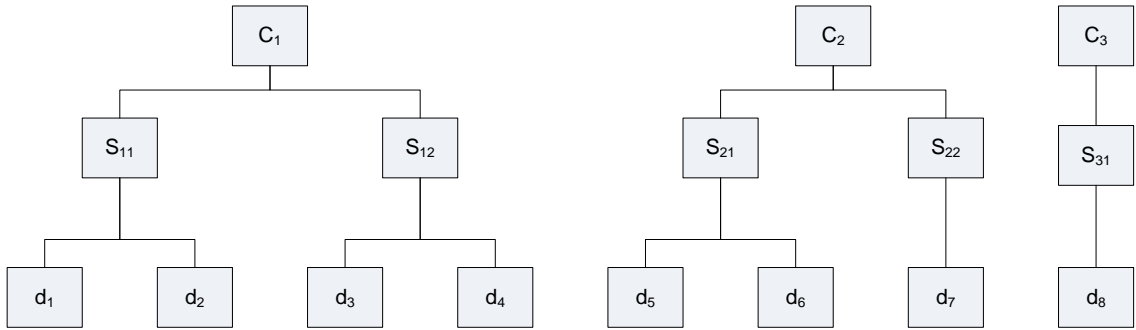

**Figure 11: Private data categorization**

*Category* $C_2$ (Hobbies) has two subsets, $S_{21}$ (online games) and $S_{22}$ (sports), subset (online games) has two private data, $d_5$ (Ogame) and $d_6$ (Hattrick), while subset (sports) has private data $d_7$ (soccer).

*Category* $C_3$ (Income) has one subset, $S_{31}$ (salary), with private data $d_8$ ($100K).

Consider that an online business called sportsworld.com is interested in acquiring information about individuals who like sports, and the provider is interested in getting the following information: *home phone number*, *personal email*, *soccer*, and the *salary value*. Assume that Alice and Bob have obtained the trustworthiness assessment of this provider (i.e., its reputation and its privacy credential ratings) from the trust and reputation agent. Both of them assign the privacy risk weights to their private data categories as shown in Table 2.

Applying equations (4.2), (4.3), and (4.4), the overall privacy risk value $\Psi$ for Alice and Bob is 0.75 and 0.92 respectively. The privacy risk values 0.75 and 0.92 represent the reluctance level of each individual with respect to the revelation of his or her personal information. In reality, some people value their personal information more than others; say a celebrity's cell phone number versus a student's cell phone number. Therefore, the computation performed by the database agent captures the different valuations that people might assign to their private data. This value will be used by the payoff agent to place a reward amount that reflects the level of risk perceived by each individual.

**Table 2: Context-dependent weights for Alice and Bob**

| Weight | | Category (contact) | Category (hobbies) | Category (income) |
|---|---|---|---|---|
| Context j = sports | Bob | W1j = 0.4 | Bob W2j = 0.1 | Bob W3j = 0.5 |
| | Alice | W1j = 0.1 | Alice W2j = 0.2 | Alice W3j = 0.2 |

Next we present the details of the trust and reputation agent.

## 4.4　　　Trust and Reputation Agent

As mentioned earlier in section 4, in order for Alice to assign privacy risk values to her private data categories, she wants to know the reputation and the competency of the online business with respect to privacy and private data handling.

In recent years, trust and reputation systems have emerged as a way to reduce the risk entailed in interactions among total strangers in electronic marketplaces. While there are many definitions of trust, we follow Gambetta (1988), where trust is defined to be "*a particular level of subjective probability with which an agent assesses that another agent will perform a particular action, both before the assessing agent can monitor such an action and in a context in which it affects the assessing agent's own action*." According to Gambetta (1988), trust is the perception of confidence. It helps reduce the complexity of decisions that have to be taken in the presence of many risks. In the context of eCommerce (or any kind of online transaction), perceived privacy is often thought to be another important antecedent of trust (Chen and Barnes 2007) and is defined as the subjective possibility of accordance between consumers' anticipation and their cognition of how their private information is being used (Yang 2005). It is the perception that the online business will adhere to an acceptable set of practices and principles (Lee and Turban 2001).

On the other hand, reputation is based on past behavior where derived reputation scores are assumed to predict likely future behavior. In the context of eCommerce, reputation acts as a signal of the service provider's skill and integrity. Consumers can use reputation information to distinguish dishonest sellers from good ones, thereby overcoming adverse selection. While there are a large number of online commercial reputation services (e.g., ivouch.com, bizrate.com, Mcafee.com, etc.), reputation information about privacy practices is not that obvious. Furthermore, highly reputable websites do not always mean that the service provider has high privacy credential ratings. To understand how the service provider is going to handle the personal information, it is essential to assess its trustworthiness based on privacy credential attributes which the online provider has obtained (e.g., trustmark seals, privacy certificates, contents of the privacy statement, use of encryption mechanisms, etc.).

In our system, we focus on trust attributes (explained next) that are related to privacy and private data handling. For the reputation scores, we rely on commercial reputation services such as ivouch.com, bizrate.com, Mcafee.com, etc. Figure 12 depicts the architecture components of the trust and reputation agent. The main components of the trust and reputation agent are the communication module and the trust manager. Similar to the database agent, the communication module is responsible for sending and receiving messages, while internally interacting with the trust manager. The trust manager consists of the fuzzy rule sets and the fuzzy engine which produce the privacy credential report.

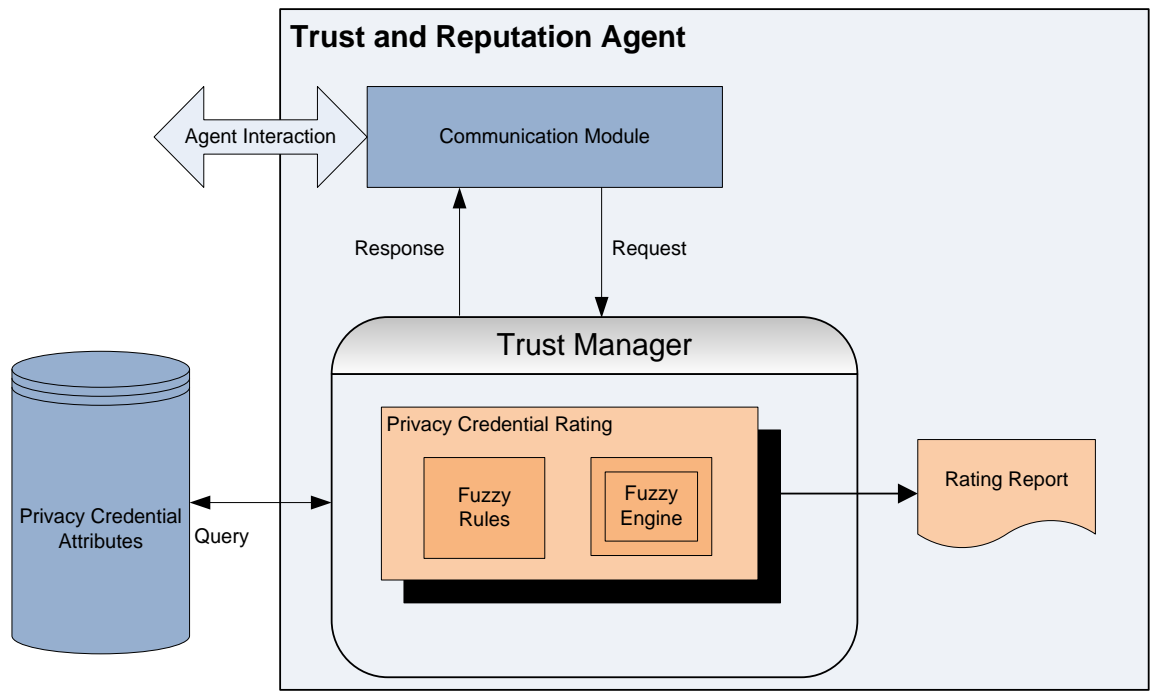

**Figure 12: Architecture of the trust and reputation agent**

Fuzzy logic is a form of multi-valued logic derived from fuzzy set theory to deal with reasoning that is approximate rather than precise (Zadeh et al. 1997). In contrast with "crisp logic," where binary sets have binary logic, the fuzzy logic variables may have a membership value of not only 0 or 1—i.e., the degree of truth of a statement can range between 0 and 1 and is not constrained to the two values only (i.e., 0 or 1). Furthermore, when linguistic variables are used, these degrees may be managed by specific functions. In our case, linguistic variables are often used to describe the importance/significance of a trust attribute on the perceived risk (e.g., the importance/significance of obtaining a security seal from a known trusted third party is HIGH). HIGH is a linguistic variable that needs to be managed by specific functions. In subsection 4.2.2, we are going to provide more details about these functions/rules.

In Figure 12, when the agent prepares the rating report, it is shared with the consumer and logged in a local repository. The fuzzy engine gets its feed from the repository which stores the privacy credentials information about each service provider. Privacy credentials information are attributes that describe the privacy practice of the service provider with respect to handling of personal information. Empirical research on the impact of such attributes on individuals' perceived trust, e.g., that of Ratnasingam and Pavlou (2003), Sha (2009), and Teo and Liu (2003), to name a few, report that the impact varies depending on the type of the attribute. Examples of such attributes and their significance are shown in Table 3.

**Table 3: Examples of trusting attributes and their significance**

| Credential | Type | Significance |
|---|---|---|
| Trustmark seals | e.g., Reliability Seals, Security Seals, Vulnerability Seals, Privacy Seals | Varies depending on the type of the seal and the issuing party (Sha 2009, Zhang 2005) |
| Dispute resolution | Independent resolution mechanisms, e.g., American Arbitration Association | Medium (Chouk et al. 2007) |
| Membership to privacy compliance auditing service | e.g., Pricewaterhousecooper, PrivaTech | High (Cline 2003, Pennington et al. 2003) |
| Privacy statement | Privacy policy generator, e.g., Organization for Economics Cooperation and Development OECD | Varies depending on the contents (Meinert et al. 2006) |
| Authentication-based disclosure of information | e.g., Kerberos-based vs. username/password authentication | Low (Sha 2009, Hu et al. 2003) |
| Allows consumers to opt out | Opt out of data sharing and marketing solicitations | High (Gideon et al. 2006) |
| Use of Encryption Mechanisms | e.g., 128-bit is better than 64-bit encryption scheme | Low (Sha 2009, Gideon et al. 2006) |

This thesis classifies credential attributes into abstract sets that represent their significance (low, medium, and high). The classified sets will be used, as explained in the next paragraph, as a reference to measure the reliability of the online businesses. The classification is based on consumer research analysis (Chouk et al. 2007, Gideon 2006, Hu et al. 2003, Larose 2004, Sha 2009, Kimery and McCord 2002) and industry studies (Cline 2003, Hussain et al. 2007). Most of these studies analyze the impact of different trust attributes on consumers' online purchasing behavior and their perceived trust. For example, based on Sha (2009) and Gideon et al. (2006), the "use of encryption mechanism" and "periodic reporting on privacy matters" can be placed in the set of attributes that have low impact on the perceived trust. In the case of a new attribute, such as the introduction of a new third-party seal, the reference sets are updated based on the perceived impact of the attribute.

### 4.4.1        *Fuzzy Engine*

In order to describe how the fuzzy engine works, let us assume that $\zeta_k$ is the set of $K$ attributes, written as follows: $\zeta_k = \{\zeta_1, \zeta_2, \zeta_3 ..., \zeta_K\}$; $k = 1,..,K$. As mentioned earlier, each attribute has its own impact on the perceived trust as shown in Table 3. The set $\zeta_k$ is classified into three subsets $\zeta_f^h$, $\zeta_g^m$, and $\zeta_w^l$ of high, medium, and low credentials respectively, such that $\zeta_f^h \cup \zeta_g^h \cup \zeta_w^l = \zeta_k$. Subsets $\zeta_f^h$, $\zeta_g^m$ and $\zeta_w^l$ will be used as reference sets against which the online business will be measured. In reality, online service providers obtain different types of attributes, each of which has a different impact on the perceived trust. The trust and reputation agent uses a set of fuzzy logic rules to derive a rating score based on such attributes as are explained next. While it is possible to consider other mathematical computations to determine the rating score, the reason for considering fuzzy logic is that linguistic labels are often used to describe the significance of an attribute (Chouk et. al 2007, Schlager and Purnel 2008, Sha 2009). For example, values are set by using linguistic variables such as "I believe that if the internal protection increases, the increase in the trust level is high". HIGH may be interpreted to represent more than one numerical value. To avoid assigning a specific numeric value to this rather subjective concept, we use fuzzy rules (provided in the next subsection) where we can subdivide a range [0,1] into a number of linguistic labels such as VeryLow, Low, Moderate, High, and VeryHigh. The rules help us determine the final linguistic label which is translated into a single crisp value, which could be a numeric scale if needed. In our case, the final linguistic label is translated into a scale composed of five "Stars" such that one star is VeryLow, two stars is Low, and so on and so forth.

In this thesis, we introduce a factor, called Online Provider Reliability $\Phi$, which helps determine the competency of the online business with respect to privacy and private data handling. The competency here is an indication to what extent the service provider can be trustworthy to respect consumer's private data and adhere to the best practices of respecting consumers' privacy. Let $\varphi$ be the set of credential attributes that the online business has. In reality, $\varphi$ would include attributes of different types (i.e., high, medium, and low). Consider that X, Y, and Z are the number of attributes that belong to $\varphi$ and represent the types low, medium, and high respectively. The higher the number of credential attributes, the more reliable is the online provider to protect consumers' personal data. However, as mentioned earlier, the service provider may have obtained different types of attributes. Therefore, its reliability $\Phi$ is based on such combination. We denote by $\Theta$ to represent the combination of X, Y, and Z based on the set of rules presented in Table 4, then we determine the online provider reliability $\Phi$ as follows:

$\Phi = $ X $\Theta$ Y $\Theta$ Z, such that $X \leq \left| \zeta_f^h \right|$, $Y \leq \left| \zeta_g^m \right|$, and $Z \leq \left| \zeta_w^l \right|$, where $\left| \zeta_f^h \right|, \left| \zeta_g^m \right|$, and $\left| \zeta_w^l \right|$ are the cardinalities of sets $\zeta_f^h, \zeta_g^m$ and $\zeta_w^l$ respectively.

Using fuzzy logic, the reliability of the online business is determined as follows:

**IF** $\Phi$ is high

**THEN** perceived privacy risk is low

The heuristic behind this rule is that online service providers who show a high degree of competency to protect consumers' data are assumed to honor their promises and can be trusted. Figure 13 shows the fuzzy sets for the variable $\Phi$.

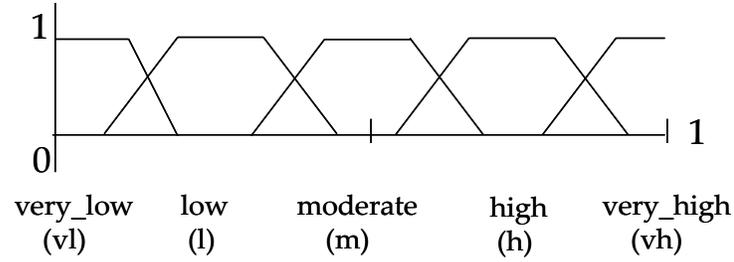

| very_low (vl) | low (l) | moderate (m) | high (h) | very_high (vh) |

**Figure 13: Fuzzy sets for variable $\Phi$**

### 4.4.2 Fuzzy Rules

An important facet of a fuzzy engine is the rules that determine how the engine will produce the result. The fuzzy rules presented in this subsection represent our interpretation of trustworthiness in a certain service provider given the presented credentials. The weights of the attributes represent the impact that one type of credential has on the trusting intuition and vary in the interval [0, +1]. In Table 4, vh, h, m, l, and vl stand for Very high, High, Medium, Low, and Very low respectively. The rules presented in Table 4 tend to ensure the minimum credentials to determine the value of the variable $\Phi$. For example, the first rule imposes a condition that all online service providers which obtain at least half of the required attributes in each category are considered very reliable. Other rules can also be considered as desired.

**Table 4: Example of reliability fuzzy rules**

| Fuzzy Rules |
| --- |
| **IF** $(X \geq \dfrac{\lvert \zeta_f^h \rvert}{2}, Y \geq \dfrac{\lvert \zeta_g^m \rvert}{2}, Z \geq \dfrac{\lvert \zeta_w^l \rvert}{2})$ **THEN** $\Phi$ *is vh* |
| **IF** $(X \geq \dfrac{\lvert \zeta_f^h \rvert}{2}, Y \geq \dfrac{\lvert \zeta_g^m \rvert}{2}, Z \leq \dfrac{\lvert \zeta_w^l \rvert}{2})$ **THEN** $\Phi$ *is vh* |
| **IF** $(X \geq \dfrac{\lvert \zeta_f^h \rvert}{2}, Y \leq \dfrac{\lvert \zeta_g^m \rvert}{2}, Z \geq \dfrac{\lvert \zeta_w^l \rvert}{2})$ **THEN** $\Phi$ *is h* |
| **IF** $(X \geq \dfrac{\lvert \zeta_f^h \rvert}{2}, Y \leq \dfrac{\lvert \zeta_g^m \rvert}{2}, Z \leq \dfrac{\lvert \zeta_w^l \rvert}{2})$ **THEN** $\Phi$ *is h* |
| **IF** $(X < \dfrac{\lvert \zeta_f^h \rvert}{2}, Y \geq \dfrac{\lvert \zeta_g^m \rvert}{2}, Z \geq \dfrac{\lvert \zeta_w^l \rvert}{2})$ **THEN** $\Phi$ *is m* |
| **IF** $(X < \dfrac{\left\|\zeta_f^h\right\|}{2}, Y \geq \dfrac{\left\|\zeta_g^m\right\|}{2}, Z \leq \dfrac{\left\|\zeta_w^l\right\|}{2})$ **THEN** $\Phi$ *is m* |
| **IF** $(X < \dfrac{\left\|\zeta_f^h\right\|}{2}, Y \leq \dfrac{\left\|\zeta_g^m\right\|}{2}, Z \geq \dfrac{\left\|\zeta_w^l\right\|}{2})$ **THEN** $\Phi$ *is m* |
| **IF** $(X < \dfrac{\left\|\zeta_f^h\right\|}{2}, Y \leq \dfrac{\left\|\zeta_g^m\right\|}{2}, Z \leq \dfrac{\left\|\zeta_w^l\right\|}{2})$ **THEN** $\Phi$ *is l* |
| **IF** $(X < \dfrac{\left\|\zeta_f^h\right\|}{2}, Y < \dfrac{\left\|\zeta_g^m\right\|}{2}, Z \geq \dfrac{\left\|\zeta_w^l\right\|}{2})$ **THEN** $\Phi$ *is l* |
| **IF** $(X < \dfrac{\left\|\zeta_f^h\right\|}{2}, Y < \dfrac{\left\|\zeta_g^m\right\|}{2}, Z \leq \dfrac{\left\|\zeta_w^l\right\|}{2})$ **THEN** $\Phi$ *is vl* |
| **IF** $(X < \dfrac{\left\|\zeta_f^h\right\|}{2}, Y < \dfrac{\left\|\zeta_g^m\right\|}{2}, Z < \dfrac{\left\|\zeta_w^l\right\|}{2})$ **THEN** $\Phi$ *is vl* |

### 4.4.3        *Privacy Credential Report*

The privacy credential report emphasizes key items of privacy concerns that are likely to be most interesting to users; for example, information about the service provider's data-sharing practices and information about whether the website allows opt out of data sharing and marketing solicitations. Furthermore, the privacy credential report makes privacy items easier to understand, as the information is communicated to consumers through a user interface. Examples of such items are presented below in Table 5.

**Table 5: Example of items presented in the privacy credential report**

| |
|---|
| Has valid privacy seal certificate (BBB, TRUSTe, …) |
| Implements secure infrastructure to protect consumer data |
| Reports to consumer of any impact on their private data |
| Member of privacy compliance auditing service |
| Allows consumers to opt out from mailing lists |
| Shares consumers' data that identifies them with marketing companies |
| Uses consumers' data that identifies them for advertisement |
| Employed personnel are trained to respect consumers' privacy |

In addition to the items, a privacy credential score about the reliability of the service provider with respect to privacy and private data handling is also presented to the consumer (the implemented report will be presented in Chapter 6).

## 4.5 Payoff Agent

The responsibility of the payoff agent is to determine the payoff which the user should receive once he or she has decided to reveal the personal information. As mentioned earlier in subsection 1.3, consumers need to be far-sighted about the value of their personal information. Hence, they need a mechanism that associates the value of the information to the individual's behavior taking into consideration of the privacy concerns of the owner, i.e., the user to whom the information refers. The responsibility of the payoff agent is to determine the payoff value that the consumers should receive once they decide to reveal their personal information.

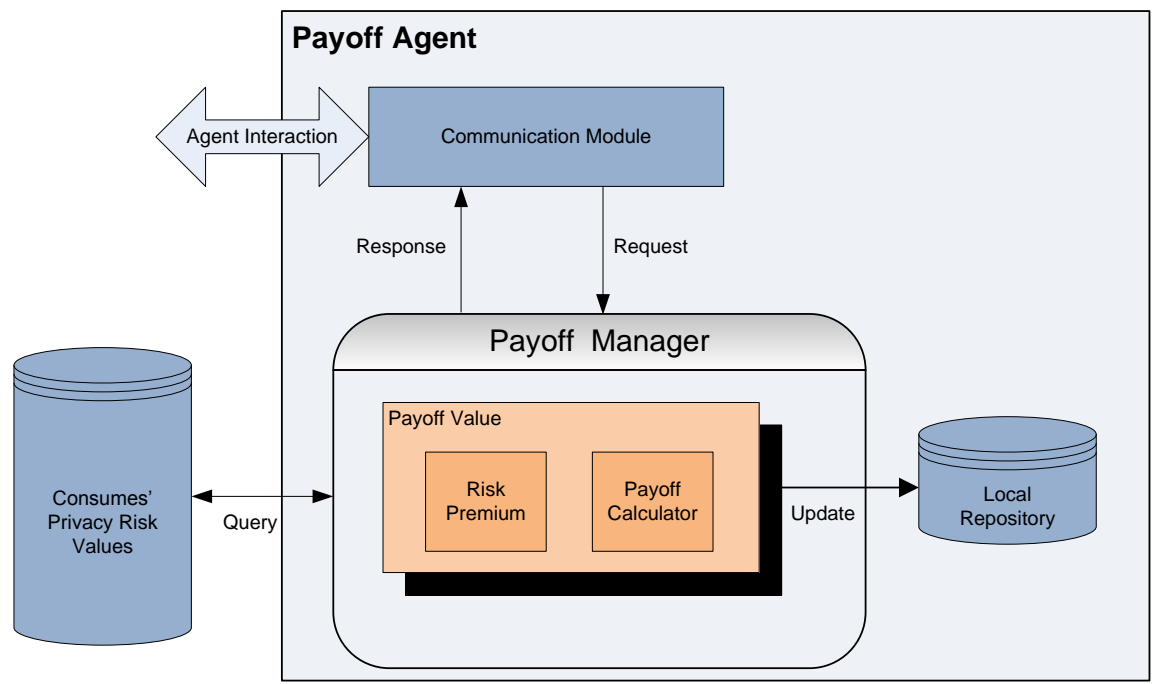

**Figure 14: Architecture components of the payoff agent**

Figure 14 depicts the architecture components of the payoff agent. It consists of a communication module like all the agents in the system. It allows the agent to communicate with the agents in the system and with the internal components as well. The payoff manager is responsible for determining the value of the payoff that the consumer should receive given the privacy risk of the potential transaction and the risk premium. Once the value is determined, it is then communicated with the facilitator agent and logged in the local repository. Below are the details of the payoff calculations.

### 4.5.1       *Payoff Calculation*

According to Cavoukian (2009), consumers incur privacy costs when their information is violated (such as the sale of personal information to a third party). Privacy cost varies according to the level of privacy violation and the type of the consumer. Studies have shown that consumers differ in their concerns for their privacy (He and Jutla 2006).

Previously, in subsection 4.3, we showed how the database agent determines the privacy risk value $\Psi(\Lambda, \beta)$ that the consumer expects when revealing his or her personal information. The payoff agent's responsibility is to determine the payoff value that the consumer should receive given the privacy risks that are involved. Intuitively, we want to associate high benefit/compensation with $\Psi(\Lambda, \beta)$ that allow high identification of $I$ (the identity of the consumer) given $\Lambda$ and high risk $\beta$. One approach for determining this value is to construct a risk premium that is valuated in response to the amount of potential damages that might occur (cost of risk) with respect to the risk exposure. In this manner, the compensation paid to the consumer is justified, at least in part, from the damages that might occur. This approach will help make service providers more conservative when handling users' personal information. This is because privacy risk penalties and reputation consequences on the violators of users' presumed privacy rights are more likely to be costly. This approach is widely used in the financial and auto insurance industry (Danthine and Donalaon, 2002).

Theoretically, the expected payoff of a risky asset in conjunction with expectations of the risk-free return should be used to construct the risk premium (Danthine and Donalaon, 2002). The basic idea is that those who are considered more risky should pay more interest. The standard model (Danthine and Donalaon, 2002) is:

$$Expected\,return = RiskfreeRate + AssetRisk \cdot Market\,Risk\,premium \qquad (4.5)$$

where:

Expected return is the payoff given to the consumer against revealing their private data (i.e., their asset);

RiskfreeRate is the expected return value at zero risk;

Market Risk premium is the expected market risk and it is positive; and

AssetRisk is the calculated risk of the private data (i.e. $\Psi$ calculated from subsection 4.3).

Assuming that the *RiskfreeRate* is zero (this assumption is valid since consumers do not expect to be compensated for private data if the level of risk is zero), this model requires two inputs. The first is the risk value of the private data asset being analyzed, and the second is the appropriate market risk premium(s). Equation (4.5) is now formally written as follows:

$$E\big(U_q\big) = \Psi_q \bullet E(\Re) \qquad (4.6)$$

where:

E $(U_q)$ is the expected return $U$ of revealing private data q;

$E(\Re)$ is the expected market risk premium; and

$\Psi_q$ is the calculated privacy risk of revealing private data q.

Equation (4.6) is simply saying that consumers' demand for revealing their personal information is measured based on the perceived risk and the market risk premium. The market risk premium is a monetary value estimated by looking at historical premiums over long time periods (Danthine and Donalaon, 2002). According to Clinebell et al. (1994) and Bakshi et al. (2008), the general behavior of market risk premiums follows a random variable, random walk, or autoregressive process over time. Here, we assume that the risk premium follows a random process and fluctuates over time following a geometric Brownian motion, where the expected risk premium value will be the underlying mean. The geometric Brownian motion assumption has the advantage that the process depends only on two variables, the drift and the standard deviation. The distribution function of a geometric Brownian motion is lognormal, which has the favorable property that negative future values have zero probability. A stochastic (random) process $\Re$ is said to follow a geometric Brownian motion if it satisfies the following stochastic differential equation (Ross 1995):

$$d\Re_t = \mu \Re_t dt + \sigma \Re_t dW_t \tag{4.7}$$

Where $\{W_t\}$ is a Wiener process or Brownian motion, $\mu$ is the drift, and $\sigma$ is the standard deviation. Using Ito's lemma (Ross, 1995), we can derive the analytical solution of equation (4.7) as follows:

$$d \ln \Re(t) = \frac{1}{\Re(t)} d\Re(t) - \frac{1}{2} \frac{1}{\Re(t)^2} d\Re(t)^2 \tag{4.8}$$

$$= \frac{1}{\Re(t)} \Re(t)[\mu dt + \sigma dW(t)] - \frac{1}{2} \frac{1}{\Re(t)^2} \Re(t)^2 [\sigma^2 dW(t)^2] \tag{4.9}$$

$$= \mu dt + \sigma dW(t) - \frac{1}{2}\sigma^2 dt \qquad (4.10)$$

The integral of (4.10) from *0* to *t is*:

$$\int_0^t d\ln\Re(t) = \int_0^t (\mu dt + \sigma dW(t) - \frac{1}{2}\sigma^2 dt) \qquad (4.11)$$

$$\ln\Re(t) - \ln\Re(0) = (\mu - \frac{1}{2}\sigma^2)t + \sigma W(t) \qquad (4.12)$$

Basic calculus gives us:

$$\Re_t = \Re_o e^{((\mu-\sigma^2/2)t+\sigma W_t)} \qquad (4.13)$$

where $\Re_o$ is the risk premium value at time t = 0; i.e., it is the initial risk premium value.

We now take the expectation $E[\Re_t]$ for (4.13):

$$E[\Re_t] = E[\Re_o e^{((\mu-\sigma^2/2)t+\sigma W_t)}] \qquad (4.14)$$

Applying the law of normal variables with mean $\mu$ and variance $\sigma^2$ knowing that the Brownian motion $\sim N(0,t)$ we get:

$$E[\Re_t] = \Re_o e^{\mu t} \qquad (4.15)$$

The risk premium is assumed to be known at time 0. Some studies such as (Hann 2003) estimate the market privacy risk premium to be between $45 and $57 per personal data record. Equation 4.15 estimates the value of the payoff as it varies over time. In Chapter 6, we provide an experiment in which we study the effect of this variation on the payoff benefit of the consumers as well as the service providers' benefit.

The next subsection describes the functionality of the negotiation agent.

## 4.6     Negotiation Agent

It is a daunting prospect for an individual consumer to bargain with a huge service provider about a desired payoff against revealing personal information. Therefore, having an agent working on behalf of a group of consumers, one would be in a better position to bargain over the revelation of their personal information and get something of value in return. Figure 15 shows the architecture of the negotiation agent. The decision-making module is composed of three components, namely the offer construction component, a library of negotiation strategies, and the offers synthesis graph. The offer construction component performs a comparative evaluation of the offers proposed by the opponent agents; this process ultimately aims at finding the best offer, according to the customer's choices and the information at hand. It actually deploys the agent's reasoning mechanism that implements the behavior of the agent by using appropriate rules; for instance, the agent acts proactively upon the reception of some messages, sent by the opponent agent. The library of the negotiation strategies allows the agent to decide on the best strategy to maximize the user's benefit.  The offer synthesis graph stores a relational graph of all the offers and the strategy of executing them. Below we provide the details of the negotiation agent and negotiation process.

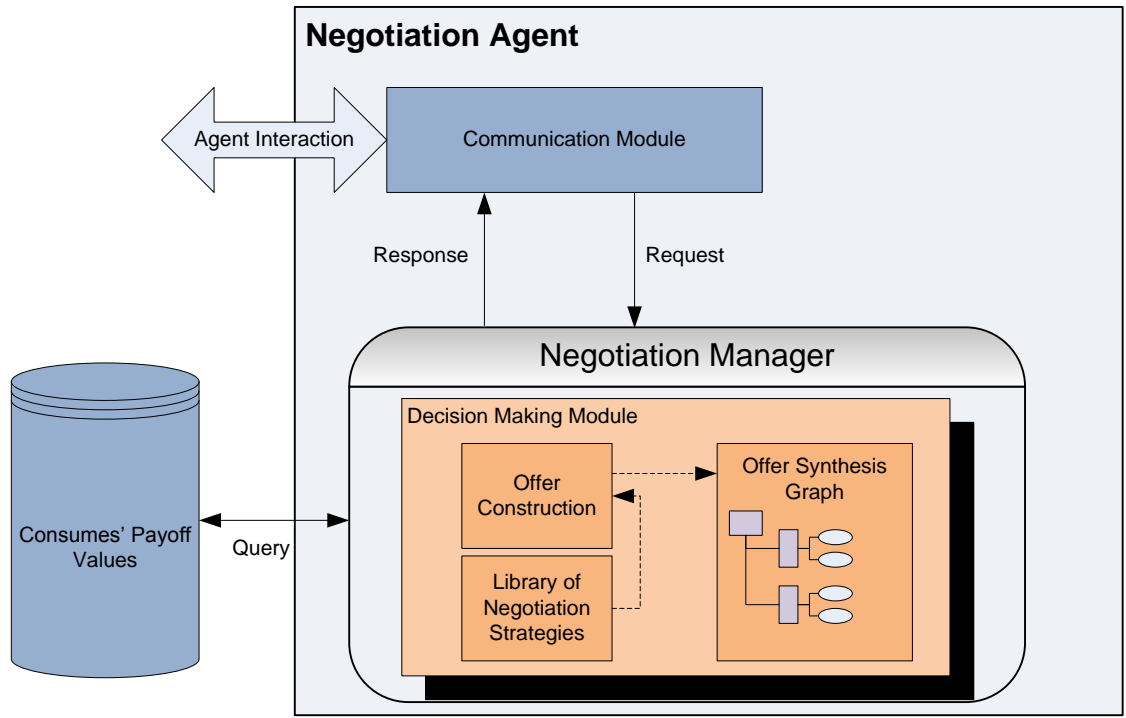

**Figure 15: Architecture of the negotiation agent**

### 4.6.1        *Negotiation Strategy*

In our setting, the intelligence of the negotiation agent is realized by the negotiation strategy. In particular, this consists of the strategy of making concessions and evaluating the incoming offers and of making decisions to interact with the negotiation opponent. Before diving into the details of the negotiation strategy, some rules and assumptions are given, as follows:

- Since a bargaining negotiation is fundamentally time-dependent (Bo et al. 2008), we assume that both the agent representing consumers and the service provider agent utilize a time-dependent strategy while making a concession. This assumption is an appealing one since human traders easily understand deadlines, and it is trivially simple to specify a deadline to a software agent. Thus the consumers' agent has a deadline $T_{deadline}^{CA}$ and the service provider agent has a deadline $T_{deadline}^{SP}$.

- The negotiation is non-cooperative with incomplete information. That is, the consumer agent does not know the reservation price per record or the time deadline of the provider agent and vice versa.

- The agents are negotiating over a single issue, i.e., the payoff that the consumers should receive against the revelation of their personal information

Having studied the rules of the negotiation, we can now present the negotiation strategy. This thesis adopts the work of Bo et al. (2008) in defining a concession strategy. Each agent is assumed to have a different time preference, i.e., its time deadline. In round $0 < t < \min(T_{deadline}^{CA}, T_{deadline}^{SP})$ if the proposal $P_{t-1}^{SP \to CA}$ at round $t - 1$ is not acceptable to the consumers' agent, the agent may make a concession to service provider's agent at round $t$ as reaching an agreement is always better than failing to reach an agreement.

In general, the proposal of agent $A$ to its trading partner at round t $(0 \le t \le T_{deadline}^{A})$ is modeled as a time-dependent function as follows (Bo et al. 2008):

$$P_t^A = IP^A - \phi^A(t) \times (IP^A - RP^A) \tag{4.16}$$

where $t$ is the current trading time, $IP$ is the initial price, and $RP$ is the reservation price and $\phi^A(t)$ is a time-dependent function based on the time deadline preference.

The time-dependent concession strategy is used to decide the amount of concession in each round of the negotiation process. The time-dependent function $\phi^A(t)$ is determined with respect to time preference $\eta^A$ and deadline $T^A_{deadline}$ (where $\eta^A \geq 0$ and $T^A_{deadline} > 0$ is finite) and is given as:

$$\phi^A(t) = \left( \frac{t}{T^A_{deadline}} \right)^{\eta^A} \tag{4.17}$$

The open literature has described a large number of negotiation strategies with respect to the remaining trading time (one for each value of $\eta^A$). In this thesis, we adopt the "sit-and-wait" strategy proposed by Bo et al. (2008). The "sit-and-wait" strategy is used in the case when the negotiation issue does not devaluate over time; it has been proven by Bo et al. (2008) that this strategy is the dominant strategy for an agent using time-dependent negotiation process, regardless of the strategy that its trading partner adopts. Unlike commodities or other services that might be devaluated over time, personal information as a trading object does not have such a problem. Therefore, this strategy is suitable to the type of problem we are facing. For the consumers, even if the offer happens only at some future stage of negotiation, as long as it is anticipated, the consumers' agent who makes this offer has bargaining power. Another important aspect of bargaining power arises from impatience. In order to achieve certain competitive advantages, online businesses have a particular service delivery threshold that makes them more impatient. Therefore, they are more likely to make bigger concessions in order to seal the deal as soon as possible.

The "sit-and-wait" strategy for the consumer agent is as follows: At the time when $t < T^{CA}_{deadline}$, it follows that $(t/T^{CA}_{deadline})^\infty = 0$, and $P^{CA}_t = IP^{CA}$. When $t = T^{CA}_{deadline}$, it follows that $(T^{CA}_{deadline}/T^{CA}_{deadline})^\infty = 1$, and $P^{CA}_t = RP^{CA}$. Let $Q^{CA}_t$ and $Q^{CA}_{T^{CA}_{deadline}}$ be the amounts of concession at $t < T^{CA}_{deadline}$ and $T^{CA}_{deadline}$, respectively. Before the deadline, the agent does not make any concession but "waits" for the service provider agent to concede, since $Q_t^{CA} = P_{t-1}^{CA} - P_t^{CA} = 0$ $(0 \leq t < T_{deadline}^{CA})$. It only concedes at its deadline $Q_{T_{deadline}^{CA}}^{CA} = P_{T_{deadline}^{CA}-1}^{CA} - P_{T_{deadline}^{CA}}^{CA} = IP^{CA} - RP^{CA}$. The service provider agent, on the other hand, concedes to its reservation price at $T_{deadline}^{SP}$. In reality, the time deadline for the consumer agent is set to be a large value so the consumer agent can benefit from the impatience of the online business.

### 4.6.2 Offer Construction

Consider that the negotiation is to acquire consumers' information records which are aggregated together according to a specific interest (e.g., a list of consumers who like soccer or like to travel to England). The consumer and the service provider therefore will have the following objectives in their offers.

*Consumer agent offer:* The consumer agent's objective is to maximize the consumers' payoff as a community; that is, maximizing their social welfare *SW*. Consider that *N* is the number of records in the list; then, at each round of the negotiation the consumer agent constructs an offer as follows:

$$N' = \arg \max_{N' \subseteq N} \sum_{i \in N'} SW_i | U_q \qquad (4.18)$$

such that $N' \subseteq N$ and $U_q$ is the payoff per personal data record q (personal data record refers to the vector of information fields that will be revealed). When the consumer agent receives an offer from the provider agent, it calculates the number of records $N'$ that satisfy this offer. The decision about the number of consumers is taken based on an offer/demand curve according to the consumer's valuation of his or her personal information record:

*At the negotiation manager*
*For each incoming offer*

$$N' \leftarrow \arg\max_{N' \subseteq N} \sum_{i \in N'} SW_i \big| U_q$$

        **repeat** (*offer is within* $< T^{CA}_{deadline}$ )

    **if** (offer $< RP^{CA}$ )
      **then**
        **for** *i=1 to the end of the list*
          *Calculate new proposal* $N'$
          *send ( COUNTER, proposal* $N'$ )
          *wait for new offer*
    **else**
      *state is ACCEPTED*
      *send (CONFIRM)*
    **end**
  **until** ( *a final state is reached* )
  // Final state is time the deadline of the provider is reached or time deadline of
the consumer agent is reached

The above algorithm describes at what time in the negotiation process the consumer agent constructs its offer.

*Provider agent offer*: Assume that the provider's agent has a utility value $V_r$, where $r$ refers to the consumer's record in the list. Let $C_r$ be the cost of acquiring record $r$ (i.e., the payoff offer given to the consumer agent per record), then the provider agent's goal is to maximize the function ($V_r$ - $C_r$). The provider agent constructs its offer so that the value $C_r$ is minimized. The optimal value that maximizes the utility is when $C_r$= 0, but such an offer will end the negotiation process in the first round and the service provider will walk away with nothing. The strategy of the service provider's agent is to start with a small value $C_r$ (for example, a predefined percentage of the utility $V_r$) and then in each round of the negotiation gradually increase the offer at a step rate equal to $\theta$ (an arbitrary bid increment value determined by the provider) until it reaches its *RP* or the deadline expires and in this case concedes to *RP*.

Theoretically, the maximum reservation price which the provider can offer is equal to $V_r$ which yields to zero profit. But practically, this is not acceptable for the service provider. Hence, in practice the upper limit of *RP* is the value that yields minimum acceptable utility. The minimum acceptable utility is that at which the difference between the utility and the reservation price is relatively small, that is, $(V_r - C_r) = l$ where $l$ is a positive small value slightly greater than zero.

### 4.6.3    *Negotiation Protocol*

A basic condition for the automation of the negotiation process among agents is the existence of a negotiation protocol, which encodes the allowed sequences of actions. Although FIPA (Foundation for Intelligent Physical Agents) (FIPA 2002) provides a plethora of protocols, such as FIPA brokering, FIPA English auctions, FIPA Contract net protocol, etc., we found that there is no agreed-upon standard interaction protocol for 1-1 automated negotiation. As a result, we implement the negotiation protocol proposed in (Skylogiannis et al. 2007). This protocol is a finite state machine that must be hard-coded in all agents participating in negotiation. While we understand that hard-coded protocols in agents may lead to inflexibility, the focus of this thesis is not on protocol design but rather a negotiation strategy that maximizes the consumers' benefit, therefore we believe the protocol we use is a good solution for our purpose. Following Skylogiannis et al. (2007), our protocol is a finite state machine with discrete states and transition; see Figure 16.

In Figure 16, the notations Start, S1, S2, S3, S4, S5, and S6 represent the different states of the negotiation and END is the final state in which there is an agreement, or a failure of agreement between the participants. *Send* and *Receive* primitives specify the interactions between the two agents and cause state transitions.

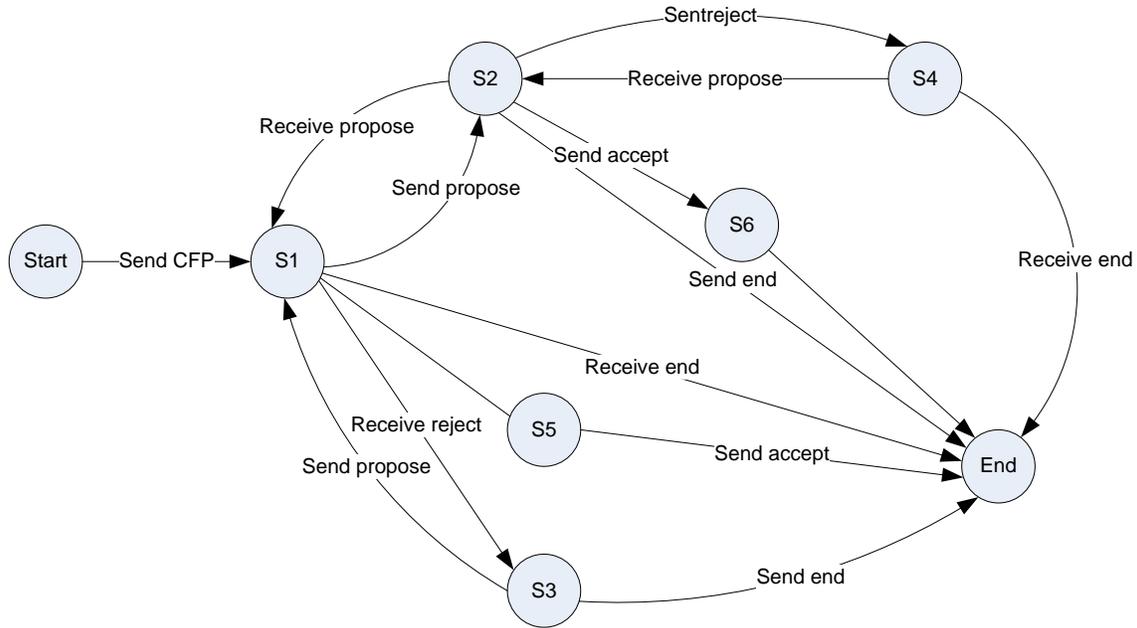

**Figure 16: State machine of 1:1 negotiation protocol**

For example, the sequence of transition START->S1->S2->S6->END can be interpreted as follows: the consumer agent initially sends a call for proposal message (CFP) to the provider agent (START->S1), then it receives a propose message (S1->S2) and after the evaluation it decides to send an accept message (S2->S6). Lastly, it receives an accept message and the negotiation terminates successfully (S6->END). We make the convention that the consumer agent initiates the negotiation by sending a call for proposal.

## 4.7 Facilitator Agent

The facilitator agent manages the interaction of the agents and orchestrates the order of tasks execution and acts as a single point of contact for agents inside and outside our system. The facilitator agent depicted in Figure 17 consists of three main components (namely, the communication module, the task manager, and the decision manager), which run concurrently and intercommunicate by exchanging internal messages.

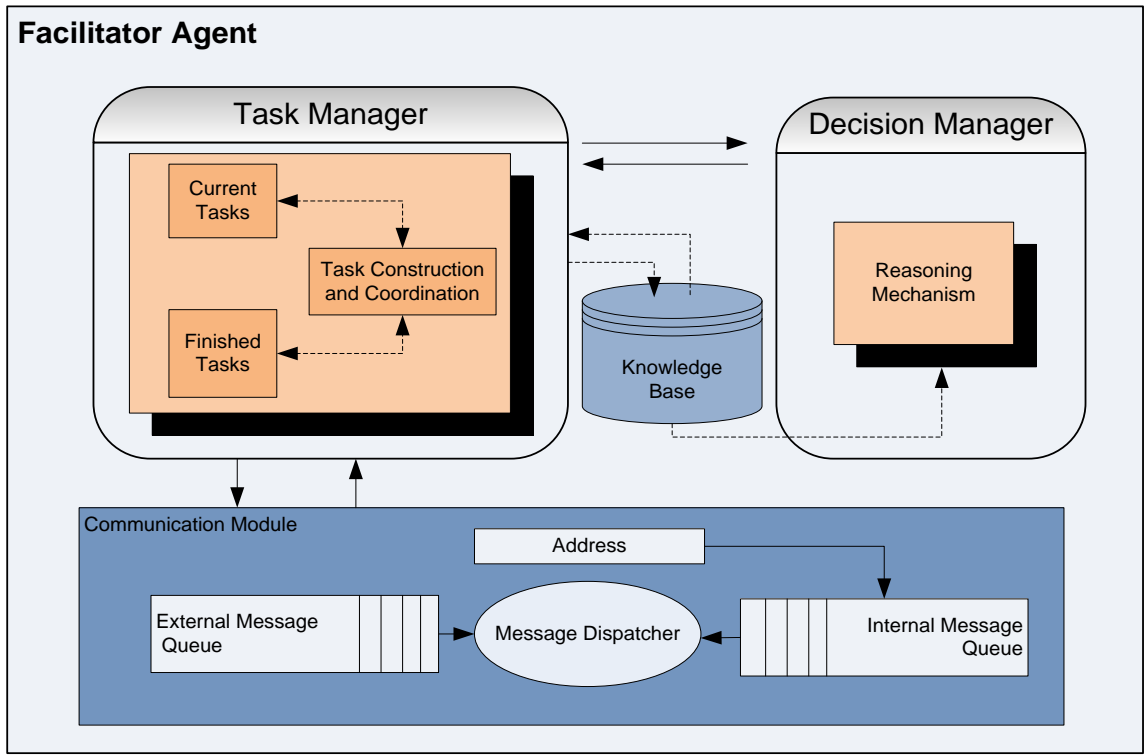

**Figure 17: Architecture of the facilitator agent**

### *4.7.1* *Communication Module*

The communication module of the facilitator agent is responsible for the agent's interaction with its environment; that is, the user and the agents in the system. The communication module sends and receives messages, while internally interacting with the task manager. When a message is received at the communication module, it is then relayed to the task manager using a message queuing mechanism. The communication module handles the interaction as follows: an internal message queue transforms each internally queued message (produced by the task manager) to an inter agent message. If the message is meant for a specific agent, then its address is added in the process. An external message queue handles the opposite case, by transforming each external message received to an internal one and adding it to the internal messages queue. Finally, a message dispatcher looks after the internal and external queues for incoming and outgoing messages, sending the queued messages to their task manager or to another agent.

The cooperation between agents is made possible by means of a common message through which the request for a work will be able to be performed and the result be sent. The communication paradigm is based on asynchronous message passing. Thus, each agent has a message queue and a message parser. In our system, the message is based on ACL (Agent Communication Language) of FIPA-ACL standards. Each message includes the following fields:

- The *sender* of the message

- The list of *receivers*

- The Communication Act, called performative, indicating what the sender intends by sending the message. If performative is REQUEST, the sender wants the receiver to do something; if it is INFORM, the sender wants the receiver to know something

- The *content* containing the actual information to be exchanged by the actual message

### 4.7.2        Task Manager

The task manager constructs and coordinates the tasks carried out in the system. It monitors both the current tasks and the finished tasks. It records the result of the finished tasks in the repository. As shown in Figure 17, the task manger interacts with both the communication module and the decision manager. For instance, each time the decision manager needs to interact with the user, it first sends a message to the task manager, which, in turn, attaches additional information (if required) and forwards it to the communication module. Similarly, the task manger may filter the content of a received message before forwarding the related data to the decision manager.

### 4.7.3　　　　Decision Manager

The decision manager decides on different kinds of queries. A user may send a request for information about a service provider, leading the decision manager to consult the knowledge base for information. The knowledge base stores information about service providers such as their reputation assessment, line of business, preferences, and services. This requires that all service providers register with the system. The decision manager decides about eligible service providers according to past experience, reputation, and agreements.

## 4.8　　　　Summary

The main goal of this chapter was to develop the architectural component of the proposed AAPPeC system that can address the coordination and cooperation challenges for the privacy payoff problem.

The architectural components of the database agent were presented with the analysis of the data categorization and the ontological formulation entities. We have shown the trust attributes and the fuzzy logic mechanism used to produce the privacy credential components of the trust and reputation agent. The details of the payoff agent and the market risk model used by the agent to calculate the benefit of the consumers are provided. With respect to the negotiation process between the consumer agent and the provider agent, we discussed the components of the consumer agent with emphasis on the negotiation strategy analysis and formulation. At the end of this chapter, the facilitator agent components were developed and described.

Now after we have described the architecture, the design specifications of AAPPeC are presented in the next chapter. In particular, the next chapter will present the AAPPeC software design using Gaia and JADE. Gaia is an agent software engineering methodology, while JADE is the implementation platform.

# Chapter 5. AAPPeC design specification

## 5.1    Introduction

Over the last few years, there has been a growth of interest in the potential of agent technology in the context of software engineering. This has led to the proposal of several development environments to build agent systems and agent applications in compliance with the FIPA specifications (e.g., FIPA-OS, JADE [2002], etc.). These development environments and software frameworks demanded that system analysis and design methodologies, languages, and procedures would support them. As a consequence, several methodologies have been proposed for the development of agent-oriented software such as Gaia, AUML, MASE, PASSI, ADELFE, etc. A review of these methodologies can be found in Henderson and Giorgini (2005).

This thesis is not concerned with proposing a new software design methodology for multi-agent systems, but rather a structural design of AAPPeC that satisfies the intended results of the system. Furthermore, we have found that there is no single methodology that answers all the questions regarding the design of MASs. According to Juan et al. (2002) and Henderson and Giorgini (2005), it appears that the most developed agent-oriented software engineering methodology is Gaia. Gaia is a general methodology that supports both the levels of the individual agent structure and the agent society in the MAS development process (Zamobolini et al. 2003). The Gaia methodology is considered quite easy to learn and use in order to analyze and design a multi-agent system (Moratias et al. 2003). For this reason, this thesis adopts Gaia for the design of AAPPeC as described by Moratias et al.

(2003). In the development phase we used the JADE platform. JADE is an off-the-shelf platform that complies with FIPA standards for implementation of software agents.

The organization of this chapter is as follows: an overview of Gaia and JADE is provided in subsections 5.2 and 5.3 respectively. Subsection 5.4 describes the AAPPeC design with Gaia and JADE followed by a summary that closes the chapter in subsection 5.6.

## 5.2 Overview of the Gaia Methodology

The Gaia methodology is an attempt to define a complete and general methodology that is specifically tailored to the analysis and design of MASs. According to Gaia, MASs are viewed as being composed of a number of autonomous interactive agents that live in an organized society in which each agent plays one or more specific roles.

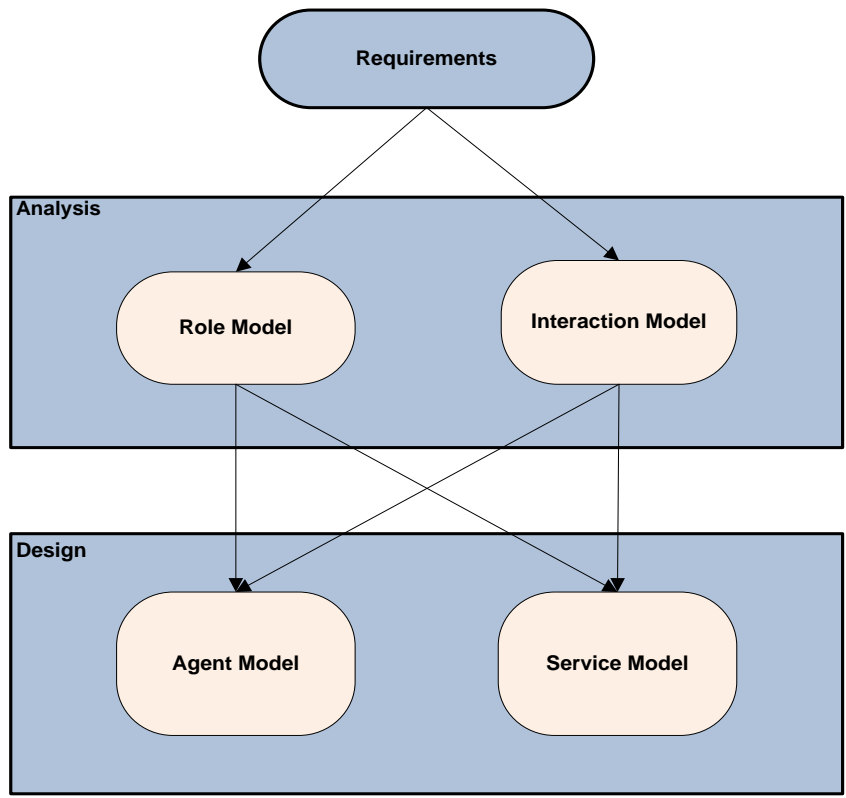

**Figure 18: The Gaia Models (Zamobolini et al. 2003)**

The methodology is applied after the requirements are gathered and specified, and covers the analysis and design phases. Figure 18 shows the artifacts produced by using Gaia. In the analysis phase, the role model and the interaction model are constructed. The two models depict the system as interacting abstract roles. These two models are then used as input to the design stage. These two models are then taken into the detailed design stage. Gaia does not cover detailed development and relies on conventional methodologies for that purpose. Gaia defines the structure of MAS in terms of a role model. The model identifies the roles that agents have to play within the MAS and the interaction protocols between the different roles. The objective of the Gaia analysis process is the identification of the roles and the modeling of interactions between the roles found. Roles consist of four attributes: responsibilities, permissions, activities, and protocols.

Responsibilities are the key attribute related to a role since they determine the functionality. Responsibilities are of two types: liveness properties—the role has to add something good to the system, and safety properties—the role must prevent and disallow that something bad happens to the system. Liveness describes the activities that an agent must fulfill and safety ensures that an acceptable state of affairs is maintained during the execution. Gaia supports the set of liveness expression-formulas presented in Table 6.

In order to realize responsibilities, a role has a set of permissions. Permissions represent what the role is allowed to do and in particular, which information resources it is allowed to access. Activities are tasks or actions a role can take without interacting with other agents, and protocols are tasks or actions a role can take that involve other agents.

**Table 6: Gaia operators for liveness formula**

| Operator | Interpretation |
|----------|----------------|
| x.y | x followed by y |

| x|y | X or y occurs |
| X* | x occurs 0 or more times |
| X+ | x occurs 1 or more times |
| x ω | x occurs infinitely often |
| [x] | X is optional |
| x||y | x and y interleaved |

Protocols are the specific patterns of interaction, e.g., a negotiator role can support different auction protocols. Gaia has formal operators and templates for representing roles and their attributes that can be used for the representation of interactions between the various roles in a system.

## 5.3  JADE Overview

JADE is a software development framework fully implemented in JAVA language aiming at the development of multi-agent systems and applications that comply with FIPA standards for software agents. JADE provides standard agent technologies and offers to the developer a number of features in order to simplify the development process, as follows:

• Distributed agent platform. The agent platform can be distributed on several hosts, each one of which executes one Java Virtual Machine.

• FIPA-compliant agent platform, which includes the Agent Management System the Directory Facilitator and the Agent Communication Channel.

• Efficient transport of ACL messages between agents.

All agent communication is performed through message passing and the FIPA ACL is the language that is used to represent the messages. Each agent is equipped with an incoming message box and message polling can be blocking or non-blocking with an optional timeout. Moreover, JADE provides methods for message filtering.

FIPA specifies a set of standard interaction protocols such as FIPA-request, FIP-Aquery, etc. that can be used as standard templates to build agent conversations. For every conversation among agents, JADE distinguishes the role of the agent that starts the conversation (initiator) and the role of the agent that engages in a conversation started by another agent (responder). JADE provides ready-made behavior classes for roles, following most of the FIPA-specified interaction protocols. In JADE, agent tasks or agent intentions are implemented through the use of behaviors. Behaviors are logical execution threads that can be composed in various ways to achieve complex execution patterns and can be initialized, suspended, and spawned at any given time. The agent core keeps a task list that contains the active behaviors. JADE uses one thread per agent instead of one thread per behavior to limit the number of threads running in the agent platform. A scheduler, hidden to the developer, carries out a round robin policy among all behaviors available in the queue. The behavior can release the execution control with the use of blocking mechanisms, or it can permanently remove itself from the queue at run time. Each behavior performs its designated operation by executing the core method action ().

The behavior in JADE has two children base classes, SimpleBehavior and CompositeBehavior. The classes that descend from SimpleBehavior represent atomic simple tasks that can be executed a number of times specified by the developer. Classes descending from CompositeBehavior support the handling of multiple behaviors. The actual agent tasks that are executed through this behavior are not defined in the behavior itself, but inside its children behaviors.

## 5.4   Using Gaia and JADE for AAPPeC Design

This subsection provides more details on using Gaia and JADE to design AAPPeC. The work in this subsection follows the studies presented by Moraitis and Spanoudakis (2003). However, before getting into the details of using the JADE and Gaia methodologies for the design specifications of AAPPeC, below are some points that need to be highlighted:

1. Gaia does not commit to any special-purpose notations for its models. It suggests the adoption of some simple and intuitive notations without excluding the possibility of coupling these notations with more formally grounded notations. Because of such freedom, we use UML (Unified Modeling Language) notation to describe the diagrams, activities, etc. in AAPPeC.

2. Gaia does not guide developers to take advantage of richer requirements enabled by agent technologies or with goal-oriented approaches to requirements engineering. Hence, we extend the analysis phase of Gaia to include a module that specifies the hierarchical goal structure of AAPPeC.

3. Gaia suggests a sequential approach to software development, proceeding linearly from the analysis to design and implementation (without specifying the implementation structure of the agents). In Gaia a service is a block of tasks which the agent will perform. Such service represents the autonomous nature of the agent in performing the required tasks and also implies that there are internal decision capabilities. For that reason in AAPPeC design we interpreted the service model to describe the tasks that the agent should perform to provide its service and hence replaced the service model by a task specification that describes the activities of the agents.

4. Gaia does not deal with dynamic change in the system such as adding tasks on run time or step back from the current activities and re-think some of the decisions. Our system assumes such behavior, and for that reason the tasks in our system are not added on run time.

5. Gaia does not deal directly with implementation issues. Such a radical choice is mainly driven by conceptual simplicity and generality. In the AAPPeC case, we have added a development phase to describe the agent implementation structure using JADE. At this phase, we make some assumptions and definitions which are related to the liveness part of each role in corresponding to the JADE terminology. For example, a simple or complex behavior representation of each role is discussed in the development phase.

Based on the above discussion, the modified model of the Gaia methodology is shown in Figure 19. In the analysis phase, a goal hierarchy is introduced to specify the goal structure of AAPPeC. The role model and the interaction model (discussed in the next subsection) follow the conventional models of Gaia. In the design phase, the agent model amounts to identifying which agent classes are to be defined to play specific roles. Task specification is used to describe the tasks that the agent should perform to provide its service.

Finally, the structure of the agents and their behavior are defined for implementation. Next, we provide the details of each phase for AAPPeC design using the modified Gaia methodology.

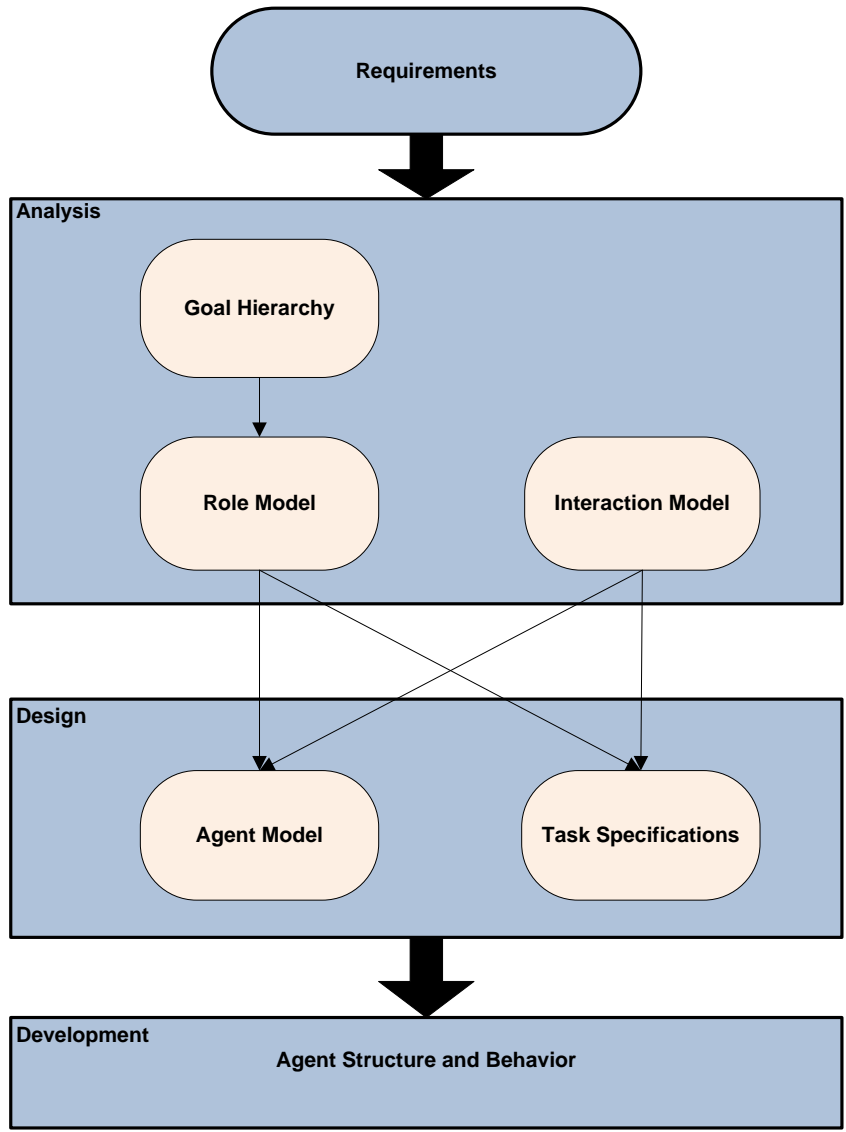

**Figure 19: Modified Gaia Model**

### 5.4.1        Requirements

In this phase, Gaia does not deal with any guidelines for the activity of capturing requirements. In AAPPeC the requirements of the system are captured by an example scenario similar to the one discussed in Chapter 4. The requirements are described by a use case diagram following the approach coming from Burrafato and Cossentino (2002).

In this example scenario, Alice is willing to share her personal data preferences with certain online service providers for a discount value or a reward fee. But Alice wants complete information about the service provider's privacy practices and its trustworthiness. She also wants to determine the level of risk involved in the transaction and wants the system to valuate her data composition risk. The system must also calculate the reward or discount value and negotiate with the service provider to maximize her benefit. Figure 20 shows the use case diagram depicting the analysis of this scenario.

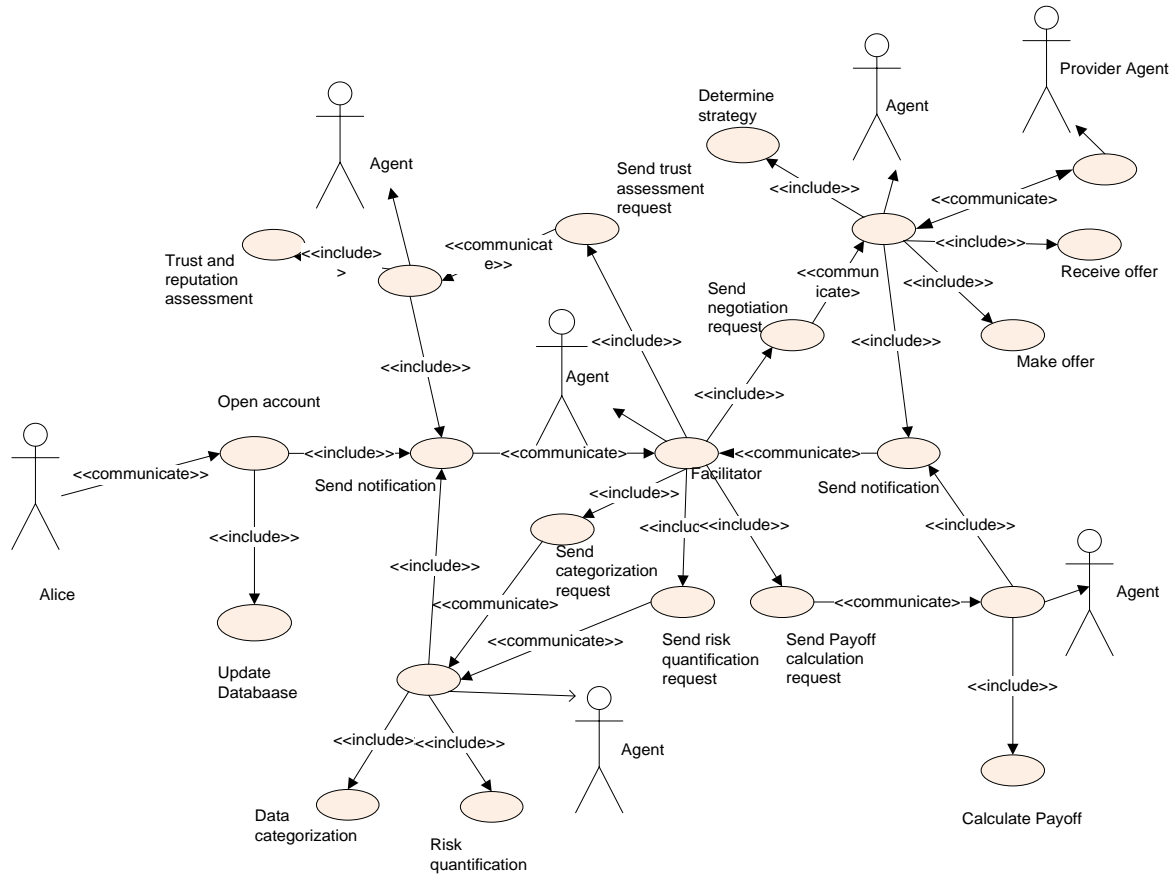

**Figure 20: Use case diagram of a case study example**

The stereotypes used here come from the UML standard, while relationships between use cases of different agents are stereotyped as "communicate." The convention adopted for this diagram is to direct communication relationships between agents from the initiator towards the participant. For the sake of clarity, we omitted the feedback that the agent will receive in this example in order to focus on the use cases of the agents.

### 5.4.2 Analysis

The analysis phase provides a preliminary definition of system goals. Also at this phase the agent's roles and the interaction protocol are defined. Below are the details of this phase.

### 5.4.2.1 Goal Hierarchy

The first step in the modeling process is to take the initial system specification and transform it into a structured set of system goals as depicted in Figure 21, the goal hierarchy diagram.

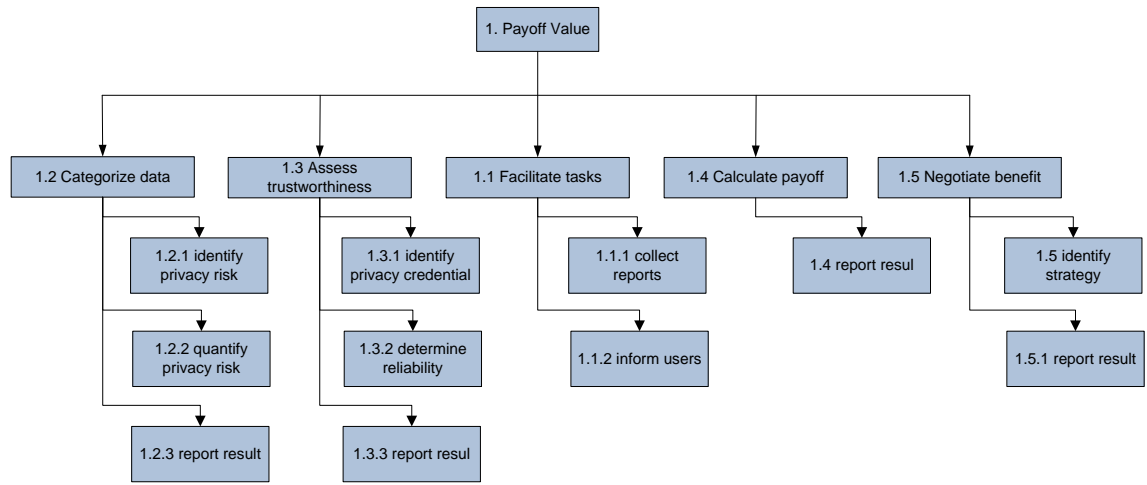

**Figure 21: AAPPeC goal hierarchy diagram**

In the goal hierarchy diagram the goals are organized by importance. Each level of the hierarchy contains goals that are roughly equal in scope and sub-goals are necessary to satisfy parent goals. For example, in AAPPeC, determining the benefit that the users should receive against revealing their personal information is made as the goal at the top of the hierarchy. All other sub-goals underneath are only to satisfy the user's benefit. Eventually, each goal will be associated with roles and agent classes that are responsible for satisfying that goal.

*5.4.2.2        Role Model*

Roles are identified from the system goals. In AAPPeC modeling we make sure that all system goals are accounted for by associating one or more goal with a specific role that is played by at least one agent in the final design. There are studies, such as DeLeaoch et al. 2002, which suggest that each goal should be mapped to one role, but there are situations where it is useful to combine multiple goals in a single role for convenience or efficiency. For example, in AAPPeC the categorization of data, identifying risk, and quantifying risk are goals that can be mapped to one role.

In AAPPeC we have identified six roles (at least one role per agent and a common role by all agents): *Facilitator* handles the collaboration process and events in the system, *DataCategorization* handles personal data classification and quantifies risk, *TrustAssessment* provides trustworthiness assessment to the users, *PayoffCalculator* serves the users by determining the payoff value of their personal information, *Negotiator* negotiates the user's benefit, and, finally, a *SocialInteraction* role that should be taken by all entities. The Gaia roles model for AAPPeC is presented in Table 7.

**Table 7: AAPPeC Roles Model**

| |
|---|
| **Role**: Facilitator<br>**Description**: It monitors and coordinates all the events/requests and whenever a new request is detected it forwards it to the responsible entity according to the nature of the request.<br>**Protocols and Activities**: <u>CheckForNewRequest</u>, <u>ScheduleTasks</u>, RequestDataCategorization, RequestRiskQuantification, RequestTrustAssessment, RequestPayoffValue, RequestStartNegotiation, ReceiveResults, InformUsers<br>**Permissions**: read agents data structure<br>**Responsibilities**:<br>**Liveness**:<br>FACILITATOR =<u>CheckForNewRequest.</u> (InformForNewRequests) \|\| ReceiveResults. (InformUsers) ω<br>INFORMFORNEWREQUESTS=RequestDataCategorization\|\|RequestTrustAssessment.\|<br>RequestRiskQuantification. RequestPayoffValue.RequestStartNegotiation<br>**Safety**: A successful connection with the users and entities in the system. |
| **Role**: TrustAssessment<br>**Description**: It provides assessment on trust and reputation related to private data handling. These assessments are presented to the user. These assessments help the user to understand the privacy risks involved in revealing personal information. |

**Protocols and Activities**: <u>QueryPCRRepository</u>, <u>ApplyFuzzyRules</u>, <u>RateServiceProviders</u>, SendResults, <u>UpdateLocalRepository</u>

**Permissions**: read privacy credential information repository, read write local repository

**Responsibilities**:

**Liveness**:

TRUSTASSESSMENT = <u>QueryPCRRepository</u>. (ServeUsers) ω

SERVEUSER = <u>ApplyFuzzyRules</u>.<u>RateServiceProvider</u>.SendResults || [<u>UpdateLocalReposity</u>]

**Safety**: A successful connection with the privacy credential information repository is established.

---

**Role**: DataCategorization

**Description**: It classifies personal information based on users' defined risk values. It also quantifies the privacy risk of data composition. This process is an important one in order to determine the payoff value that the user should receive based on the potential risk.

**Protocols and Activities**: <u>ObtainUserData</u>, <u>ApplyAttributeOntology</u>, <u>CategorizePersonalData</u>, ReceiveNewRequest, <u>QuantifyPrivacyRisk</u>, SendResults, <u>UpdateLocalRepository</u>

**Permissions**: read personal information repository, read write local repository

**Responsibilities**:

**Liveness**:

DATACATEGORIZATION= <u>ObtainUserData</u>.(ServeUserData) ω || (ReportResults) ω

SERVEUSERDATA = <u>ApplyAttributeOntology</u> . <u>CategorizePersonalData</u> || ReceiveNewRequest

REPORTRESULT = SendResults. [<u>UpdateLocalRepository</u>]

**Safety**: A successful connection with the personal data repository is established.

---

**Role**: PayoffCalculator

**Description**: It applies financial models to calculate the payoff value which the consumer is entitled to receive against revealing private data to the service provider.

**Protocols and Activities**: <u>ObtianIndividualRiskVlaue</u>, <u>CalculateRiskPremium</u>, <u>CalculatePayoffValue</u>, SendResults, <u>UpdateLocalRepository</u>

**Permissions**: read privacy risk value repository, read write local repository

**Responsibilities**:

**Liveness**:

PAYOFFCALCULATOR = <u>ObtianIndividualRiskVlaue</u>. (ServeRequest) ω

SERVEREQUEST = <u>CalculateRiskPremium</u>. <u>CalculatePayoffValue</u>. SendResults || [<u>UpdateLocalRepository</u>]

**Safety**: A successful connection to the privacy risk repository is established.

---

**Role**: Negotiator

**Description**: It strategically engages in the negotiation process with the service providers' agent in order to maximize the consumers' benefit. The negotiator is responsible for selecting the negotiation strategy and prepares offers and counter offers. When the negotiation terminates the final result is sent to the user.

**Protocols and Activities**: <u>SetTimedeadline</u>, <u>SetReservationPrice</u>, <u>DecideNegotiationStrategy</u>, StartNegotiation, Receive Proposal, <u>EvaluateOffer</u>, <u>ProposeCounterOffer</u>, SendResults, UpdateLocalGraphRepository

**Permissions**: read, strategy, read write create local repository

**Responsibilities**:

**Liveness**:

NEGOTIATOR = <u>SetTimedeadline</u>|| <u>SetReservationPrice</u> .( ServeRequest)ω

SERVEREQUEST = <u>DecideNegotiationStrategy</u> . (Engage )ω

ENGAGE = Receive Proposal. <u>EvaluateOffer</u>. <u>ProposeCounterOffer</u> .SendResults || [<u>UpdateLocalGraphRepository</u>]

**Safety**: A successful connection with the GIS is established.

**Role**: SocialInteraction
**Description**: It requests agents that perform specific services from the DF (JADE Directory Facilitator).
**Protocols and Activities**: <u>RegisterDF</u>, <u>QueryDF</u>,
**Permissions**: create, read, update agent's data structure.
**Responsibilities**:
**Liveness**:
SOCIALINTERACTION =.(<u>RegisterDF</u> **.** <u>QueryDF</u>) ω
**Safety**: true

Here it must be noted that another role is involved in the MAS operation. It is the Directory Facilitator (DF) FIPA role that is supported by JADE. However, this role concerns the operational level of the MAS and not the application itself, which is why a Gaia representation is not supplied for this role. Moreover, interactions with it are not presented as protocols, as they are defined in Gaia methodology as activities. Indeed, the activities RegisterDF (denoting the registration to the DF) and QueryDF (querying for agents of specific types or that have registered specific services activities) are DF services provided directly by the JADE framework, provided not as a result of interaction between agents, but as method invocations.

### 5.4.2.3        Interaction Model

The Gaia interaction model denotes which action returns from a request along with the roles that can initiate a request and the corresponding responders. Table 8 holds an example from our model. However, since our intention is to use a standard FIPA protocol we are going to describe the informative used in each interaction message among the agents.  For example, an ACL message "RequestDataCategorization" sends from the facilitator agent to the database agent could use the FIPA informative INFORM with SL language.

**Table 8: Interaction model**

| Protocol | Request Data Categorization | Request Risk Quantification | Request Trust Assessment | Request Payoff Value | Request Start Negotiation |
|---|---|---|---|---|---|
| **Initiator(s)** | Facilitator | | | | |
| **Receiver(s)** | Data Categorization | Data Categorization | Trust Assessment | Payoff Calculator | Negotiator |
| **Informative** | INFORM | INFORM | INFORM | INFORM | INFORM |
| **Responding Action** | Categorize Personal Data Report Result | Quantify Privacy Risk, Report Result | Rate Service Providers Report Result | Calculate Payoff Value Report Result | SendResults |
| **Purpose/ Parameters** | Inform that the personal information needs to be categorized. The required data is stored in the database | Inform that privacy risk needs to be determined for each individual. The calculation is based on the data categorization | Inform to rate the service provider and report the result. The contents is based on the privacy credentials attributes | Inform to calculate the individual payoff and report the result. The content is based on the individual privacy risk assessments | Inform to start the negotiation with the provider agent and to report the result. |

### 5.4.3 Design Phase

During the design phase the agent model is achieved, along with the task specification. It is worth emphasizing that, while the analysis phase is mainly aimed at understanding what the MAS will have to be, the design phase is where decisions have to be taken about the actual characteristics of the MAS. Therefore, even if some of the activities have been specified in the analysis phase, the actual tasks that each agent is going to perform is decided in the design phase that leads to the actual characteristics of the system.

#### 5.4.3.1 Agent Model

The agent model creates agent types by aggregating roles. Each agent type can be represented as a role that combines all the aggregated roles attributes (activities, responsibilities, permissions, protocol).

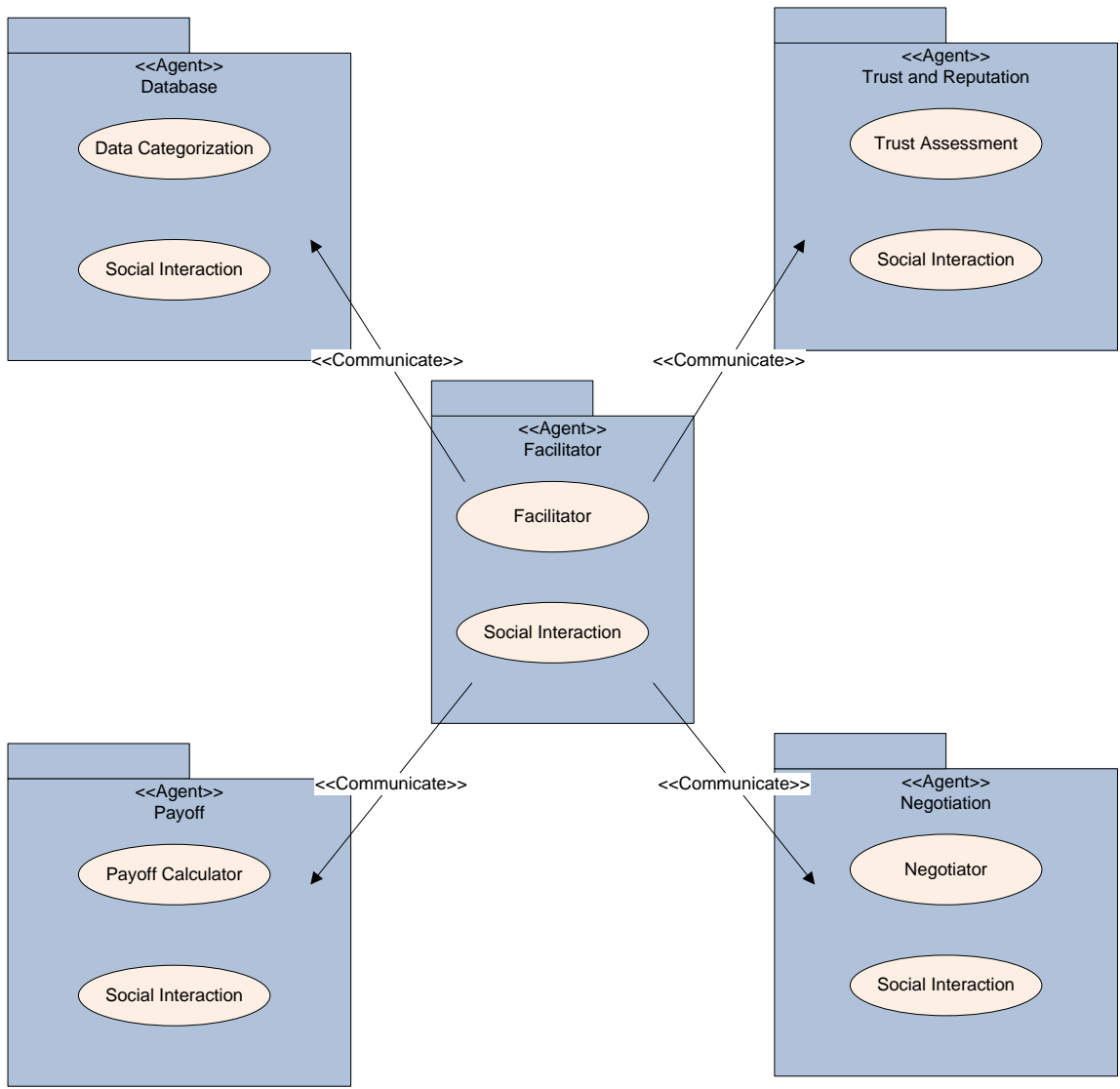

**Figure 22: AAPPeC agent model**

The agents model of the AAPPeC system will include five agent types: the facilitator agent type, which fulfills the Facilitator and SocialInteraction roles, the database agent type, which fulfills the DataCategorization and SocialInteraction roles and the trust, the reputation agent type, which fulfills the TrustAssessment and SocialInteraction role, the payoff agent, which fulfills the PayoffCalculator and SocialInteraction role, and the negotiation agent type, which fulfills the Negotiator and the SocialInteraction roles. The agent model is presented in Figure 22. The roles that the agent is going to assume are grouped into stereotyped packages.

Relationships among the agents are stereotyped as "communicate." The convention adopted for this diagram is to direct communication relationships among agents from the initiator towards the participant.

### 5.4.3.2 Task Specification

The task specification summarizes what the agent is capable of doing. Once roles have been identified, the detail tasks, which define how a role accomplishes its goals, are defined and assigned to specific roles. Relationships between activities signify either messages between tasks and other interacting agents or communication between tasks of the same agent. As mentioned earlier in subsection 5.3, AAPPeC tasks are static and predefined so that the agent performs them upon request. In future work, however, this model needs to be modified so that agents can perform tasks added dynamically at run time.

For every agent in the model, we draw an activity diagram that is made up of two swimlanes. The one from the right-hand side contains a collection of activities symbolizing the agent's tasks, whereas the one from the left-hand side contains the activities representing the other interacting entities or agents.

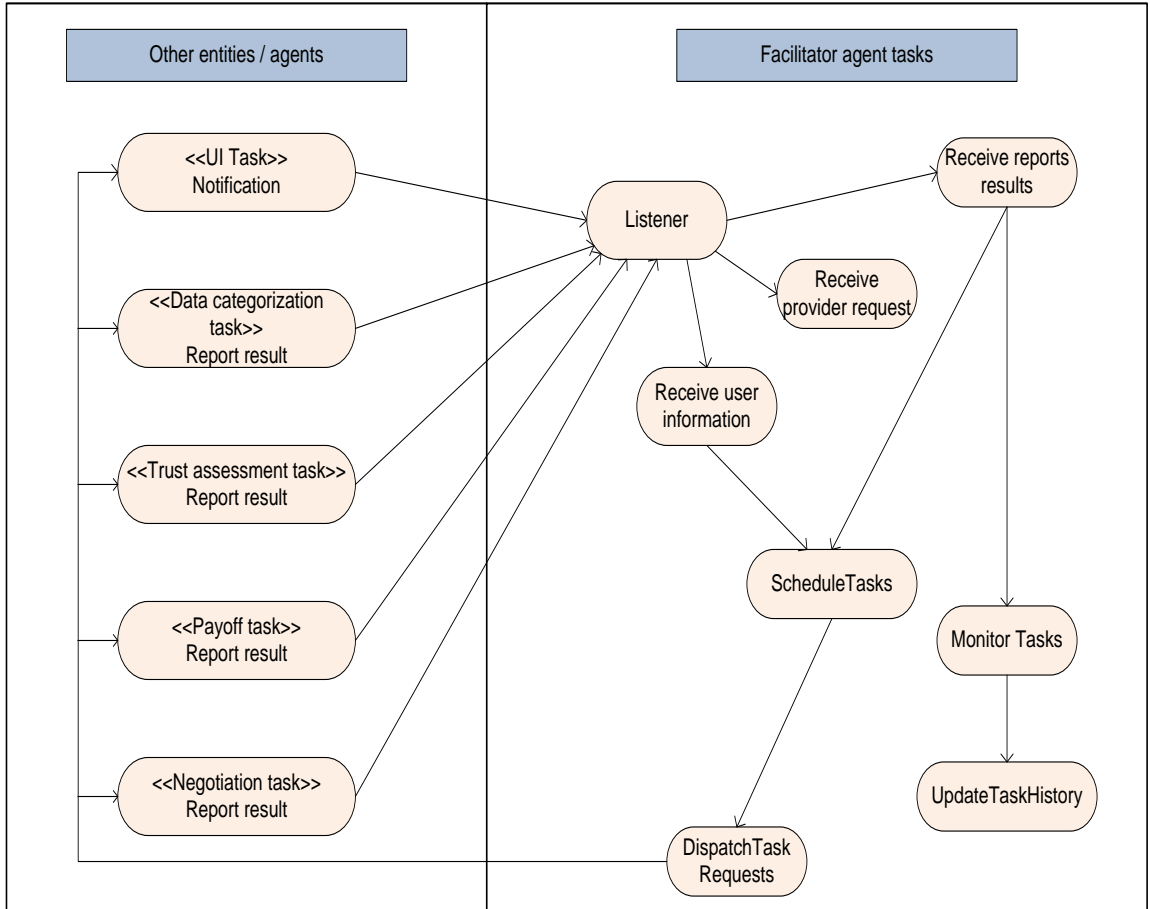

**Figure 23: The tasks of the facilitator agent**

*Facilitator agent*: Figure 23 shows the task diagram of the facilitator agent. A listener task is needed in order to pass incoming communication to the proper task. Further tasks are needed to handle all the incoming messages (ReceiveReportresults, ReceiveProviderRequest, and ReceiveUserInformation) from the user interface and the other agents. Other tasks such as ScheduleTasks, MonitorTasksStatus, and UpdateTasksHistory are introduced for better decomposition of the agent and to make sure that all dispatched activities resulting from the outgoing messages are tracked and monitored.

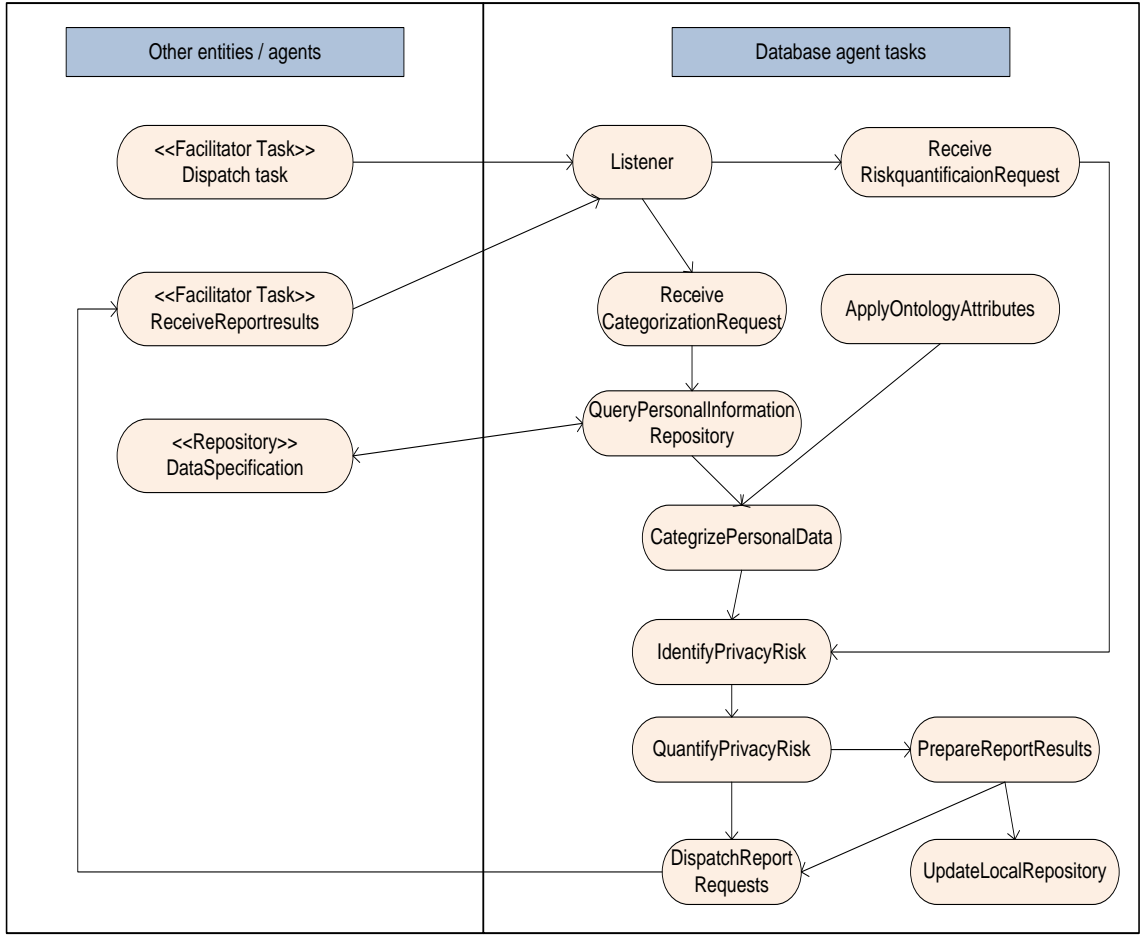

**Figure 24: The tasks of the database agent**

*Database agent*: Figure 24 shows the task diagram of the database agent. The database agent is required to handle two types of requests coming from the facilitator agent, namely the data categorization request and the risk quantification request. The tasks that the database agent is supposed to perform are QueryPersonalInfomation repository, ApplyOntologyAttributes on the data objects, CategorizePersonalData, IdentifyPrivacyRisk, and QuantifyPrivacyRisk. Supporting tasks are also needed to finish off the service provided by the database agent. These tasks are PrepareReportResult, UpdateLocalRepository, and they send the result to the facilitator agent.

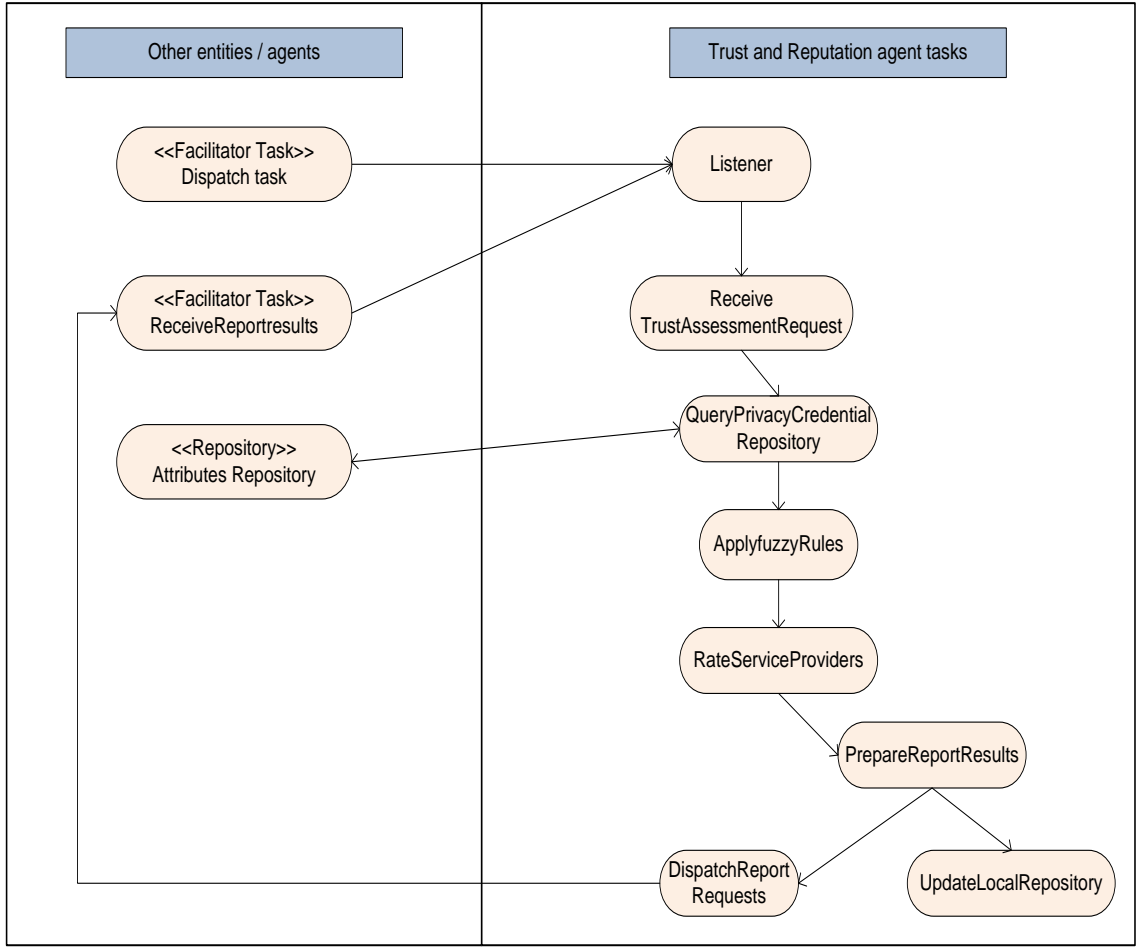

**Figure 25: The tasks of trust and reputation agent**

*Trust and reputation agent*: Figure 25 shows the task diagram of the trust and reputation agent. The trust and reputation agent is required to handle one request coming from the facilitator agent, which is the TrustAssessmentRequest. In order to provide its service the agent is required to access the privacy credentials of the service providers that are stored in the repository and ApplyFuzzyRules to calculate the reliability of the service provider (RateServiceProviders). Once the assessment task is completed, the report is prepared with the associated task PrepareReportResult. Finally, the report result is sent to the facilitator agent.

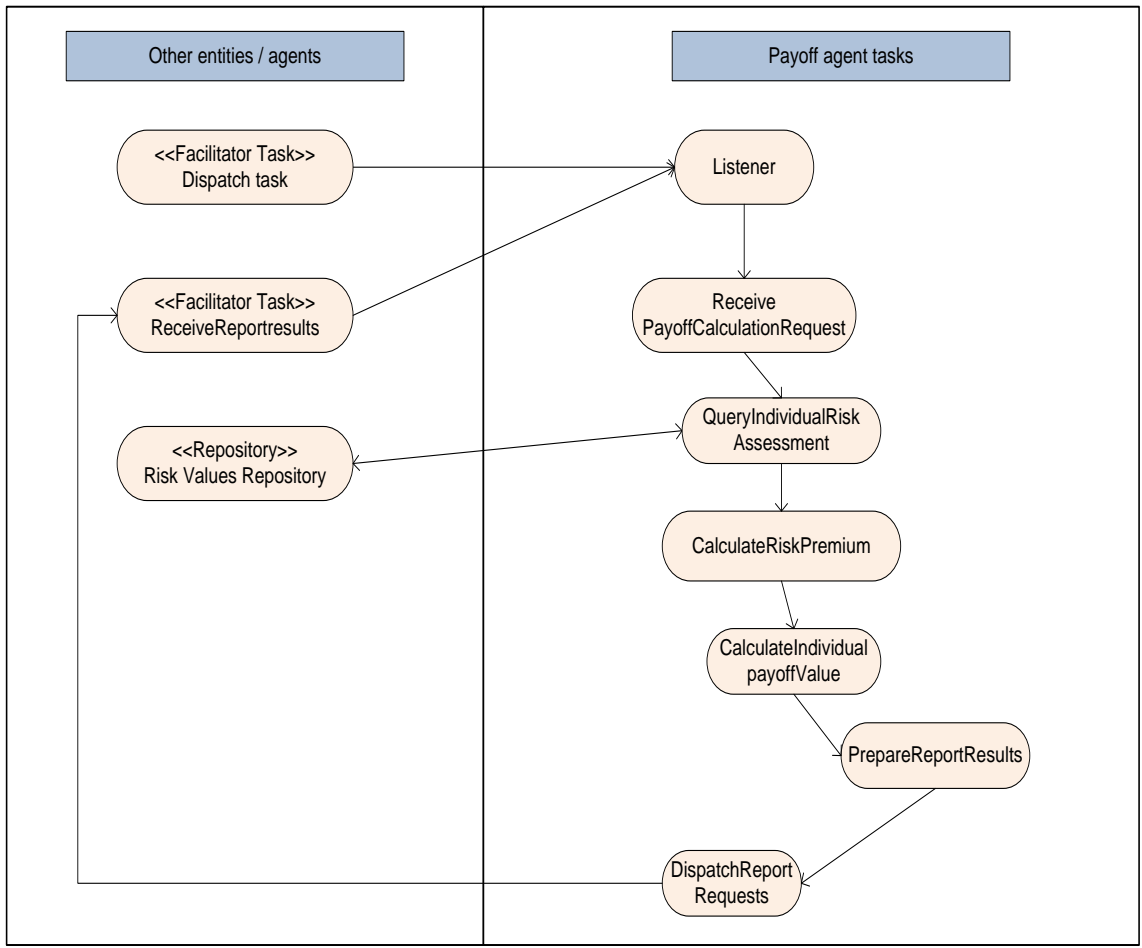

**Figure 26: The tasks of payoff agent**

*Payoff agent*: Figure 26 shows the task diagram of the payoff agent. The payoff agent is required to handle a request coming from the facilitator agent to calculate the payoff. In order to provide its service the agent is required to access the privacy risk value of each consumer that is stored in the repository (QueryIndividualRiskAssessment), determine the risk premium, and calculate the individual payoff value. Finally, the report is prepared and dispatched to the facilitator agent.

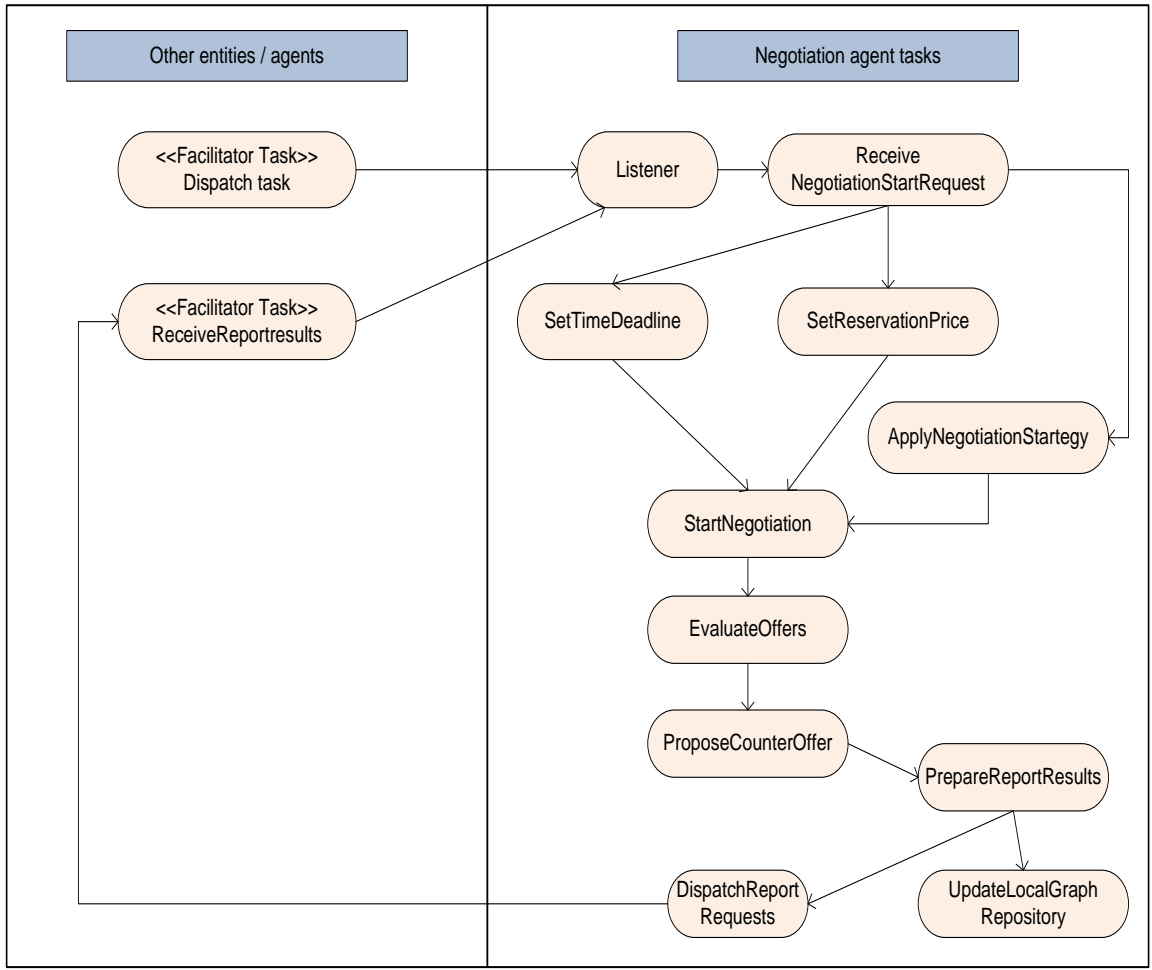

**Figure 27: The tasks of the negotiation agent**

*Negotiation agent*: Figure 27 shows the task diagram of the negotiation agent. The agent is required to start the negotiation with the service provider agent upon receiving the NegotiationStartRequest from the facilitator agent. The agent sets the reservation price, the time deadline, and applies the negotiation strategy. During the negotiation, the agent evaluates the incoming offers and proposes counter ones until the negotiation process terminates. Finally, the report is prepared, the local graph of offers is updated, and the final result sent to the facilitator agent.

### *5.4.4*      *Development*

As mentioned earlier, some assumptions have to be made when moving from the Gaia model to the JADE framework. For example, the liveness part of each role can be considered as a behavior in correspondence with the JADE terminology (Moraitis and Spanoudakis 2003). Hence, we can say that a simple or a complex behavior represents each role. This behavior is considered as the top-level behavior of the role. Each behavior may contain other behaviors, as in the JADE behavior model. The contained behaviors are the children of the top behavior. The Negotiator role in AAPPeC, for instance, has the Negotiator top behavior. This behavior has several lower level behaviors such as tracking the deadline, reservation price, evaluate offers, etc. The ω and ‖ operators on Gaia liveness formulas now have the following meaning. The ω means that a lower level behavior is added by the behavior that contains it in the Gaia liveness formula and is only removed from the agent's scheduler when the behavior that added it is removed itself. If such behaviors are more than one, they are connected with the ‖ symbol which denotes that they execute "concurrently." Concurrency in JADE agent behaviors is simulated using parallel behaviors (JADE allows one thread per agent and the behaviors are executed in a round robin fashion). All Gaia liveness formulas are translated to JADE behaviors. Activities and protocols can be translated to JADE behaviors, to action methods or to simple methods of behaviors. The JADE behaviors that can be useful for the negotiation agent for example are the SimpleBehaviour, ParallelBehavior, and SequentialBehaviors.

The safety properties of the Gaia roles model must be taken into account when designing the JADE behaviors. Some behaviors of the role, in order to execute properly, require the safety conditions to be true. Towards that end, at least one behavior is responsible for monitoring each safety condition of a role. In our case, the safety requirement of the Negotiator role, for example, is the establishment of communication with the provider agent. One SimpleBehavior is responsible for monitoring the validity of this safety requirement. Below, we provide more details about the behaviors used to implement the negotiation agent. The full implementation of this agent is shown in Appendix C. We just note here that the naming convention used in the task diagram might have been changes in the implementation for convenience; however, the functionality is kept in place as required.

• Listener: It is a *SimpleBehaviour* that waits until it receives a specific ACL message.

```
public class sb extends SimpleBehaviour {

    public Boolean finished = false ;
    public void action( ) {
            ACLMessage msg = CA.revceive ();
            if  (msg!=null) {
                        do();
                        inished = true;
            }
            else {
                        block();
            }
    }
    public Boolean done() {
    return finished;
    }
}
```

• **SetTimedeadline** and **SetReservationPrice** are implemented with a SimpleBehavior that gets the time deadline specified by the user and monitors its decrement operation.

```
Consumer_agent.addSubBehaviour(new Behaviour(this) { // this behavior is part of a sequential behavior
        @Override
        public void action() {
           // upon receiving the message
                try {
                   facilitator.invokeProviderAgent(msg.getContent());
                   if (msg!= null){
                        System.out.println(getLocalName() );
                        System.out.println(msg.toString());
                        String content = msg.getContent();
                        System.out.println(content);
```

```
try {
    if(content.contains("PR:")){
        actionPR(content);
        return;
    }
    if(content.startsWith("inbid:")){
        if(deadline == -1){
            myAgent.doDelete();
            return;
        }
        roundTrip++;
        if(deadline  == roundTrip){
            deadline = -1;
            roundTrip=0;
            double pare =
            Double.parseDouble(AIMSViewPanelPhase3.jTxtCAReserPrice.getText().trim());
            invokeProviderAgentForProposal("finalbid:"+(paRe*2));
            WriteXLSheet.data.setConsumerOffer(String.valueOf(paRe*2));
        }
        else{
            String msgStr = content.toString();
            msgStr = msgStr.substring(msgStr.indexOf("inbid:")+6);
            double  cares =
            Double.parseDouble(AIMSViewPanelPhase3.jTxtCAReserPrice.getText().trim());
            WriteXLSheet.data.setProviderOffer(String.valueOf(msgStr));
            if(caRes>Double.parseDouble(msgStr)){
                invokeProviderAgentToReject("regectedbid:");
            }
            else{
                invokeProviderAgentToAccept("accepted:");
            }
        }
    }
}
```

- ServeRequest: a complex behavior consists of SequentialBehavior(s) and Parallelbehavior. The SequentialBehavior implements the three SimpleBehaviors where the time deadline and the reservation price methods are captured in addition to the offer valuation and the responses to the provider agent. Then a ParallelBehavior is used to perform a concurrent operation: the first one is to inform the facilitator agent that the negotiation process is now terminated and the second one is to record the result so it can be used later to construct the graph synthesis of the results. Below is an example of the parallelbehavior with two sequential behaviors.

```
ParallelBehaviour pb = new ParallelBehaviour(this, ParallelBehaviour.WHEN_ALL);
```

```
// do something
SequentialBehaviour s1 = new SequentialBehaviour(this);
     s1.addSubBehaviour(new OneShotBehaviour(this) {
         @Override
         public void action() {
             p1 = true;
             System.out.println("====================>>>First");
             AID r = new AID();
             r.setLocalName("FacilitatorAgent");
             ACLMessage msg = new ACLMessage(ACLMessage.INFORM);
             msg.setSender(getAID());
             msg.addReceiver(r);
             msg.setContent("Negotiation Terminated");
             send(msg);
             System.out.println(getLocalName() +": send id to Facilitator Agent");
         }
     });

     SequentialBehaviour s2 = new SequentialBehaviour(this);
     s2.addSubBehaviour(new OneShotBehaviour(this) {
         @Override
         public void action() {
             p2 = true;
             System.out.println("====================>>>Second");
             WriteXLSheet xlReader = new WriteXLSheet("c:/AIMS/NagotiationResult.xls");
         }
     });

     pb.addSubBehaviour(s1);
     pb.addSubBehaviour(s2);
```

## 5.5   Summary

In this chapter, we discussed the design specification of the proposed agent-based system. We have used Gaia as a general methodology for the design of AAPPeC. In particular, we modified the methodology so that new components are added to satisfy the design of the AAPPeC architecture. These new components were added to describe the tasks that the agent will perform in order to serve the general goal of valuating and negotiating for the privacy payoff value as described in the architecture.

In this chapter, we also discussed the integration of the Gaia methodology with the JADE platform in the development phase. The development phase is not part of the Gaia methodology; however, we introduced it in order to show how the integration will be performed. For that purpose we provided the mapping between the roles as defined by the Gaia methodology and the agents' behavior structures as defined in JADE. Examples of such mapping were provided and a full implementation of one of the agents is referred to in appendix C.

In the next chapter, a proof of concept implementation is further tested with experiments and case scenarios that examine the intended purpose of the AAPPeC design.

# Chapter 6. Proof of concept implementation

## 6.1 Introduction

A prototype of the proposed system has been implemented using JADE (Java Agent Development Environment) platform. JADE provides a multi-agent environment which is composed of the FIPA standard agents and of a set of application-dependent agents realized by the application developer. The communication component is implemented as a set of classes that inherit the jade.Core.Agent and jade.lang.acl.ACL message of existing classes of the JADE platform. These classes provide a means to construct, send, and receive messages via several FIPA communication performatives. The environment of the prototype characterizes that of consumers' decisions regarding their private data and agent negotiation with potential service providers' agents. The main purpose here is to provide a proof of concept implementation for the agent system as well as to show, from a utility perspective, the effect of compensating consumers for the revelation of their private data on the overall well-being of consumers and the service providers' expected benefit. A set of experiments is carried out involving simulated consumers' perceived privacy risks. This means generating a range of random privacy regimes based on the perceived trustworthiness of the service provider, then allowing the consumers' agent to negotiate using the chosen strategy tactic (as described in subsection 4.6) against the service provider's agent.

The organization of this chapter is as follows: the highlights of the prototype are described next in subsection 6.2. Subsection 6.3 describes the set of experiments and their results followed by a summary that closes the chapter in subsection 6.4.

## 6.2    Prototype Highlights

As mentioned earlier, the prototype has been implemented using JADE. A JADE platform is composed of agent containers. Agents live in containers which are the Java process that provides the JADE run-time and the services needed for hosting and executing agents.

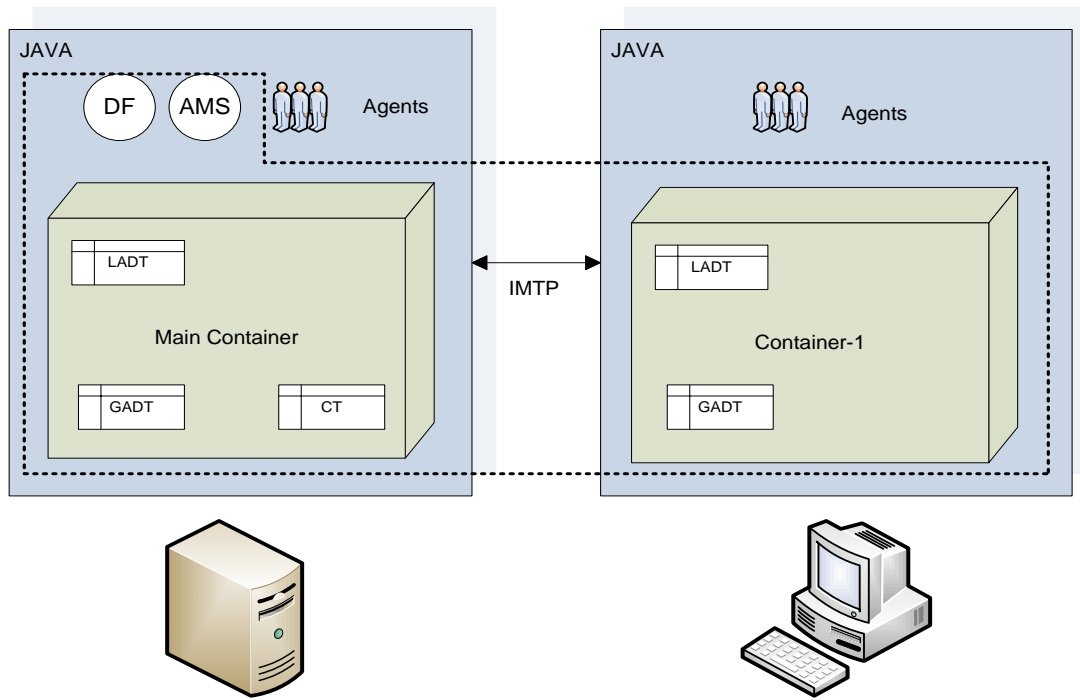

**Figure 28: JADE main architectural components**

Figure 28 shows the main architecture elements of a JADE platform. There is a special container, called the main container, which represents the bootstrap points of the  platform: it is the first container to be launched and all other containers must join the main container by registering with it. As a bootstrap point, the main container manages the container table (CT), which is the registry of the object references and transport addresses of all container nodes. The main container also manages the global agent descriptor table (GADT), which is the registry of all agents present in the platform. The AMS and DF are two special agents that provide the agent management services and the default yellow pages services respectively. Each container has its own local agent descriptor table (LADT) that uses it for as the first place to discover the recipient of a message.

Figure 29 shows a snapshot of AAPPeC agents developed under container-1. Agents developed under container-1 may (but do not have to) exist on a remote machine. For convenience, the agents in our prototype reside on the same host.

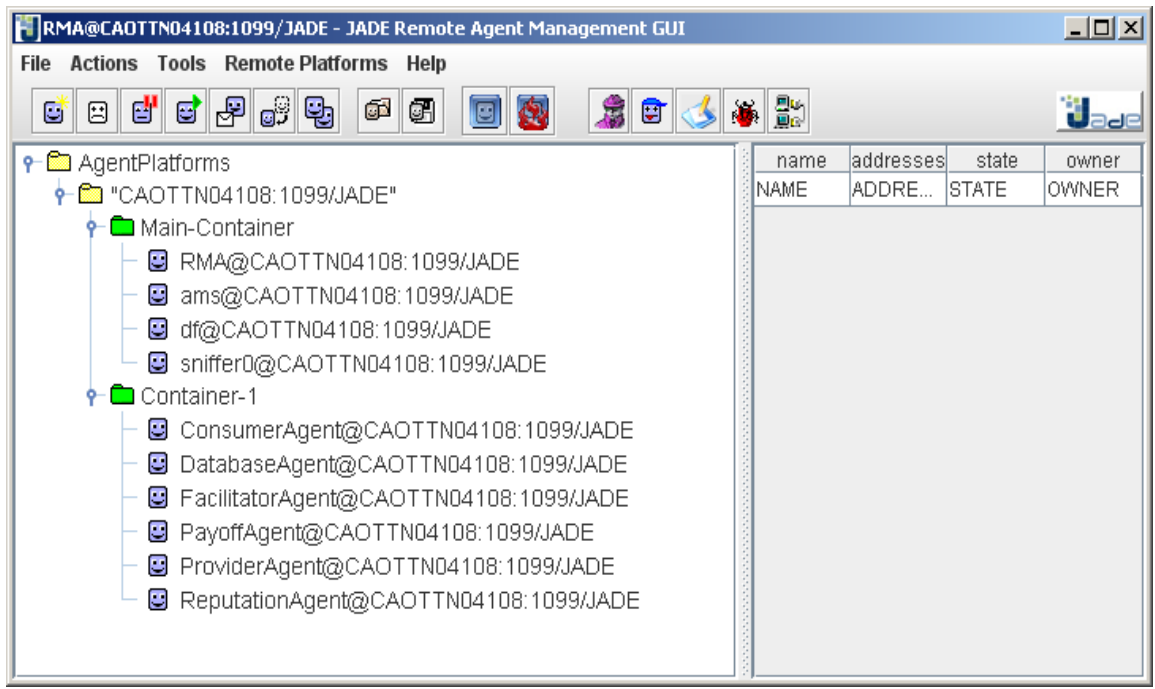

**Figure 29: Screen capture of the Remote Management Agent GUI for JADE set-up showing AAPPeC Agents**

In order for the agents to perform their tasks as described in Chapters 4 and 5, two data repositories were created. One repository is used to hold the consumer information while the other repository is used to hold the privacy trust attributes of service providers.

*Consumers' data*: this is the data that belong to consumers such as their personal information, preferences, tastes, etc. In AAPPeC, consumers record their data through a user interface (as shown in Figure 30). The data is stored in a data repository. The database agent is responsible for accessing the data for categorization purposes as per subsection 4.3.

**Figure 30: User interface for consumers to open an account and record their preferences**

*Service providers credential attributes*: In order to populate the privacy credential repository with privacy attributes and reputation data, 20 online shopping websites have been used. The selected websites are shown in Appendix A. We analyzed the websites' privacy credentials by scrutinizing their privacy policies and checking the validity of their certificates or seals if they post any. Additionally, we used sources such as iVouch.com, SiteAdvisor.com, Bizrate.com, and MacAfee.com to gather reputation scores about these websites (based on consumers' testimonial). The privacy credential attributes are stored in a local repository in order for the trust and reputation agent to access and produce the rating report.

The usage of AAPPeC prototype is as follow: Users open accounts through the interface shown in Figure 30 which are stored in a data repository. The facilitator agent sends a message to the database agent to categorize the personal information as explained in Chapter 4. The information carried in an INFORM message sent by the facilitator agent contains a "Prepare Categorization" request, as shown in the message structure in Figure 31.

```
FacilitatorAgent: Send message to Database Agent for data availability
DatabaseAgent: received the following message :
(INFORM
     :sender  ( agent-identifier :name
FacilitatorAgent@CAOTTN04108:1099/JADE  :addresses (sequence
http://CAOTTN04108.ad3.ad.alcatel.com:7778/acc ))
          :receiver  (set ( agent-identifier :name
DatabaseAgent@CAOTTN04108:1099/JADE ) )
          :content  "Prepare Categorization" )
```

**Figure 31: INFORM message from the facilitator agent to the database agent**

If there is a service provider request to acquire consumers' data, the facilitator agent sends a message to the trust and reputation agent to provide trust assessment (i.e., to generate the privacy credential report). We used an example of a service provider called circuitcity.com for this exercise. Figure 32 shows the trust and reputation agent providing a rating report for the service provider circuitcity.com. Once the report is generated, it is now the consumer's turn to provide their privacy risk weights to their private data categories according to their perceived privacy risk given the privacy credential rating. The privacy credential report emphasizes key items of privacy concerns that are likely to be most interesting to users; for example, information about the provider's data-sharing practices and information about whether the provider allows opt out of data sharing and marketing solicitations. For the reputation score we collected reputation information from iVouch.com
and Bizrate.com.

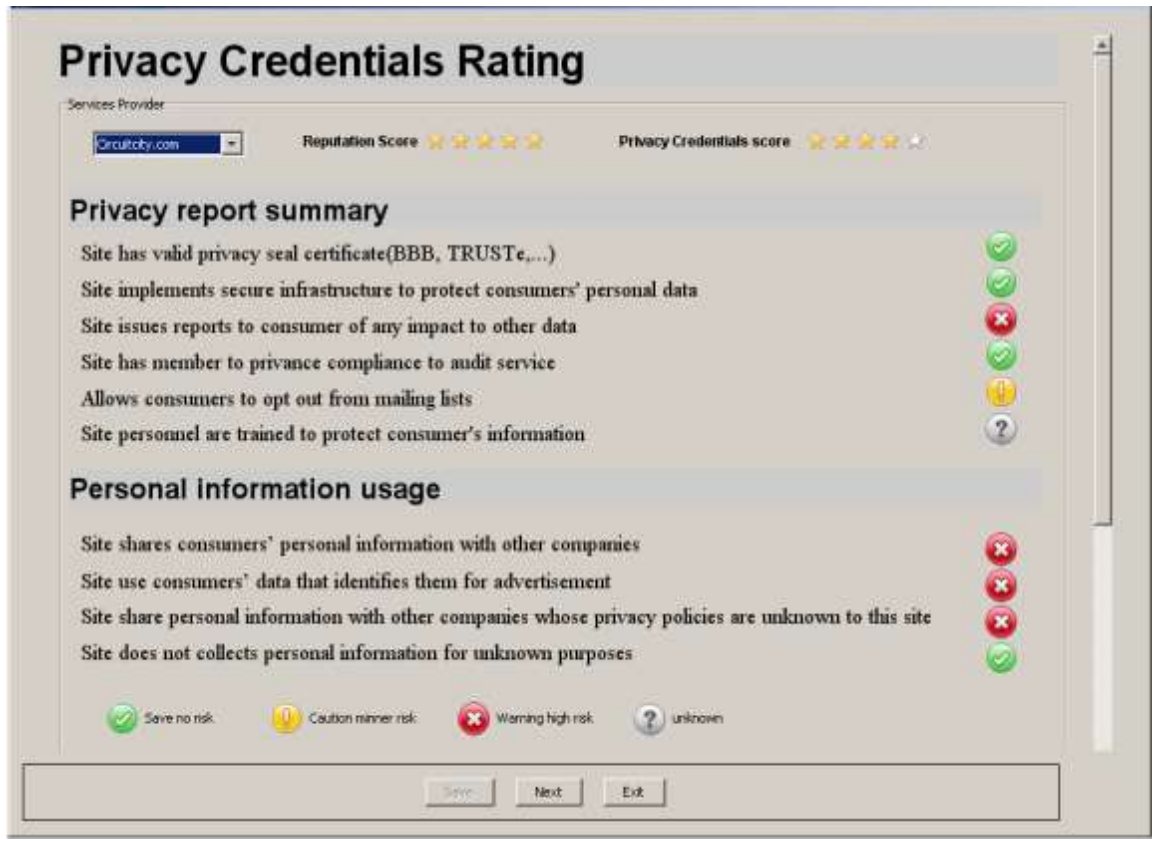

**Figure 32: Privacy credential rating report**

The user uses the credential report to learn about the service provider's privacy
practices and then to assign the level of privacy risk to his personal data categories based on
the perceived privacy risk. Once the user assigns the privacy risk weights, the database agent
quantifies the total privacy risk of the user's personal information. Figure 33 shows a
snapshot of the user interface which the consumers use to assign weights to their private data
categories based on the service provider's privacy credentials and reputation score.

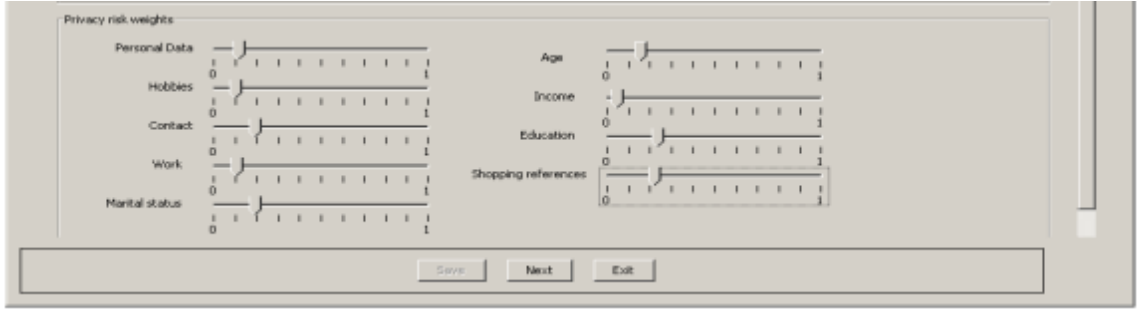

After the risk weights are assigned and the privacy risk is quantified, the payoff agent receives a message to calculate the payoff value for the consumer based on the quantified privacy risk. The message structure received by the payoff agent is shown below in Figure 34.

> *FacilitatorAgent: Send message to Payoff Agent for payoff calculation*
> *PayoffAgent: received the following message:*
> *(INFORM*
> *:sender ( agent-identifier :name FacilitatorAgent@CAOTTN04108:1099/JADE*
> *:addresses (sequence http://CAOTTN04108.ad3.ad.alcatel.com:7778/acc ))*
> *:receiver (set ( agent-identifier :name PayoffAgent@CAOTTN04108:1099/JADE ) )*
> *:content "0.6428572" )*

**Figure 34: INFORM message from the facilitator agent to the payoff agent**

In the above message the privacy risk value calculated by the database agent is 0.6428572 for the user account shown in Figure 30. The payoff agent uses this value to calculate the payoff that the consumer should receive. For this example, we assumed the risk-premium value to be $50, then the consumer should receive $32.142 as a payoff for revealing his personal information record. After the payoff is calculated, the consumer agent (the agent that negotiates on behalf of consumers) receives a message to start the negotiation as shown below in Figure 35.

```
        FacilitatorAgent: Send message to Consumer Agent to start negotiation...
        ConsumerAgent: received the following message :
        (INFORM
          :sender  ( agent-identifier :name
FacilitatorAgent@CAOTTN04108:1099/JADE  :addresses (sequence
http://CAOTTN04108.ad3.ad.alcatel.com:7778/acc ))
          :receiver  (set ( agent-identifier :name
ConsumerAgent@CAOTTN04108:1099/JADE ) )
 :content  "Start Negotiation")
```

**Figure 35: INFORM message from the facilitator agent to the consumer agent**

Figure 36 shows the sniffer agent provided by JADE that keeps track of the messages among the agents.

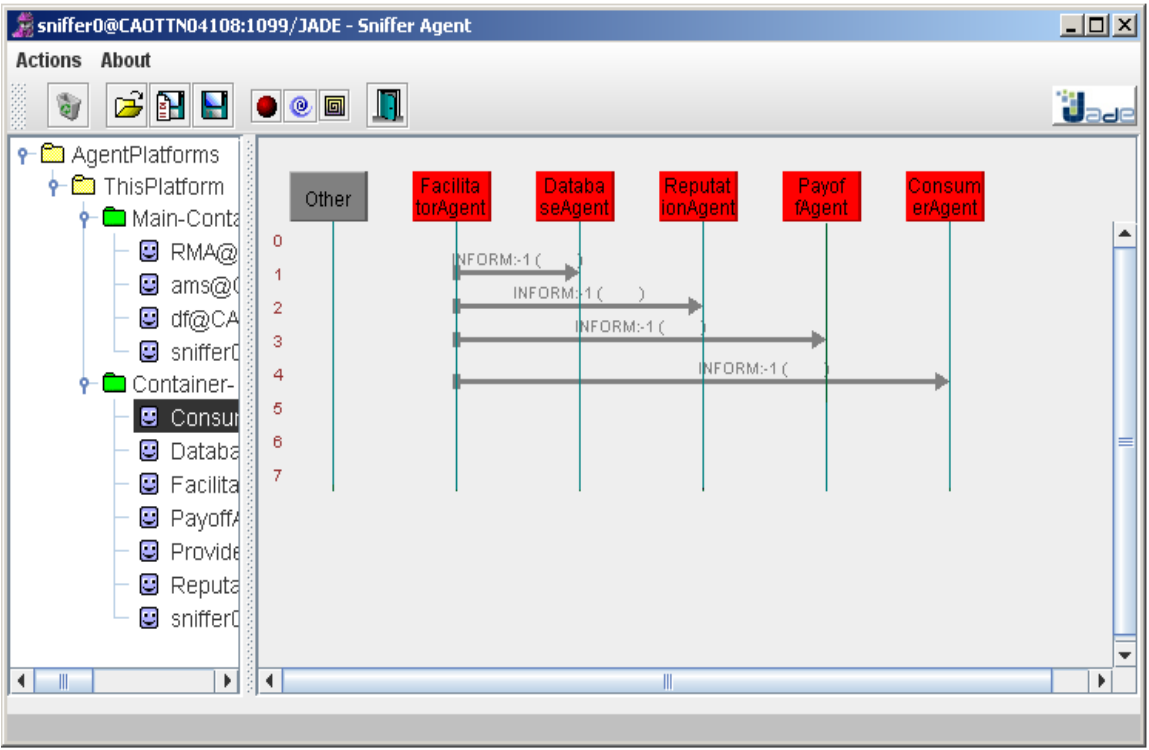

**Figure 36: AAPPeC agents shown in JADE sniffer**

The next subsection provides the negotiation process between the consumer agent and the provider agent.

## 6.3    Negotiation Experiments

As mentioned earlier, it is a daunting prospect for an individual consumer to bargain with a huge service provider about a desired payoff against revealing personal information. Therefore, an agent working on behalf of a group of consumers would be in a better position to bargain over the revelation of their personal information and get something of value in return. The negotiation process in AAPPeC is the final process in which the consumer agent negotiates with the provider agent for the payoff against revealing their personal information. This subsection presents the possible scenarios of negotiation between the two agents and the resulting payoff to the consumers.

It is worth noting here that in order to run meaningful experiments, it is required to have a large number of accounts stored in the database. For this purpose, we simulated 2000 user accounts. The simulated accounts emulate the privacy risk valuation of different types of users (i.e., privacy pragmatic, privacy fundamentalists, etc.). The values are normally distributed between 0 and 1.

Table 9 provides the parameters used in the experiments. The parameters cover some of the possible scenarios of the negotiation process. For example, we have chosen in experiment 1 a time deadline equal to 10 for the consumers' agent and a time deadline equal to 6 for the service provider's agent. In this setup, each time the agent makes an offer and sends it to its negotiation partner it is considered one unit of time. The reservation price is set to $45 and $35 for the consumers' agent and the service providers' agent respectively. Each time the service providers' agent makes an offer it increments its price by the bid increment value. The service providers' agent will concede to its reservation price before the consumers' agent and offer $35 for each record. This is because its deadline will be reached

faster. The utility value represents/measures the relative satisfaction of the service providers' agent. It is also used to determine the initial bid that the service provider will choose to start the negotiation as explained in subsection 4.6.2.

Table 9: Negotiation parameters for both the consumer agent and the provider agent

| Consumer Agent | Provider Agent |
|---|---|
| **Experiment 1- negotiation per record** | |
| Reservation price = 45 | Utility = 70 |
| Time Deadline = 10 | Reservation price = 35 |
| Number of consumer record = 2000 | Time deadline = 6 |
| | Bid Increment = 3 |
| **Experiment 2 – negotiation per record** | |
| Reservation price = 45 | Utility = 70 |
| Time Deadline = 50 | Reservation price = 50 |
| Number of consumer record = 2000 | Time deadline = 25 |
| | Bid Increment = 5 |
| **Experiment 3 – Service providers type** | |
| Reservation price = 45 | Utility = 70 |
| Time Deadline = 100 | Reservation price = 35 |
| Number of consumer record = 2000 | Time deadline = 100 |
| | Bid Increment = 3 |
| | Service providers are Excapes.com, Expedia.com, Travelocity.com, Canadatravel.ca, Itravel2000.com |
| **Experiment 4 – Market risk premium variation** | |
| Reservation price set according to the Payoff valuation and risk premium | Utility = 70 |
| Time Deadline = 15 | Reservation price = 35 |
| Number of consumer record = 2000 | Time deadline = 10 |
| | Bid Increment = 3 |
| | |

The same concept is applied in experiment 2, however, in this experiment the deadline duration of the service provider's agent is assumed to be large so that it reaches its reservation price first. Experiment 1 and experiment 2 are designed to study the effect of the deadline and the reservation price on the payoff value given the concession strategy during the negotiation process.

It should be noted here that in the experiments we chose price values (price per record) that are not necessarily reflecting the real values in real life. However, studies such as the one conducted by Hann (2003) estimates the market privacy risk premium to be between $45 and $57 per personal data record. While these values are based on specific scenarios, as explained by Hann (2003), the actual price that consumers received could be less. However, for the sake of clarity in the simulation we adopt the values provided by Hann (2003).

Experiment 3 is conducted to test the effect of the providers' type (i.e., their privacy regime) on the benefit of both the service provider and the consumers. For this experiment, five service providers were chosen that represent different privacy practice regimes. The five service providers are travel agencies that offer similar services and similar prices but differ in their privacy credential ratings (the details are explained in subsection 6.3.3). Experiments 1, 2 and 3 assume a constant risk premium value of $50. Experiment 4 examines the effect of market risk premium variation on the overall benefit of both the service provider and the consumers. In this experiment, we simulate a variation in the market risk premium by incrementing its value gradually (details are provided in subsection 4.3.4).

The details of the experiments and the results are recorded next.

### 6.3.1 Negotiation Deadline Effect on the Received Privacy Payoff

The trace of the messages exchanged between the two agents is captured via the sniffer agent provided by JADE as shown in Figure 37. The consumer agent sends a call for a proposal (CFP) message to the provider agent. The provider agent sends its proposal to the consumer agent. Each time the consumer agent receives an offer it calculates the number of records $N'$ (as explained in subsection 4.6) and sends it to the provider agent in a propose message. In this scenario, the exchange of messages continues until the provider agent reaches its deadline and concedes to its reservation price.

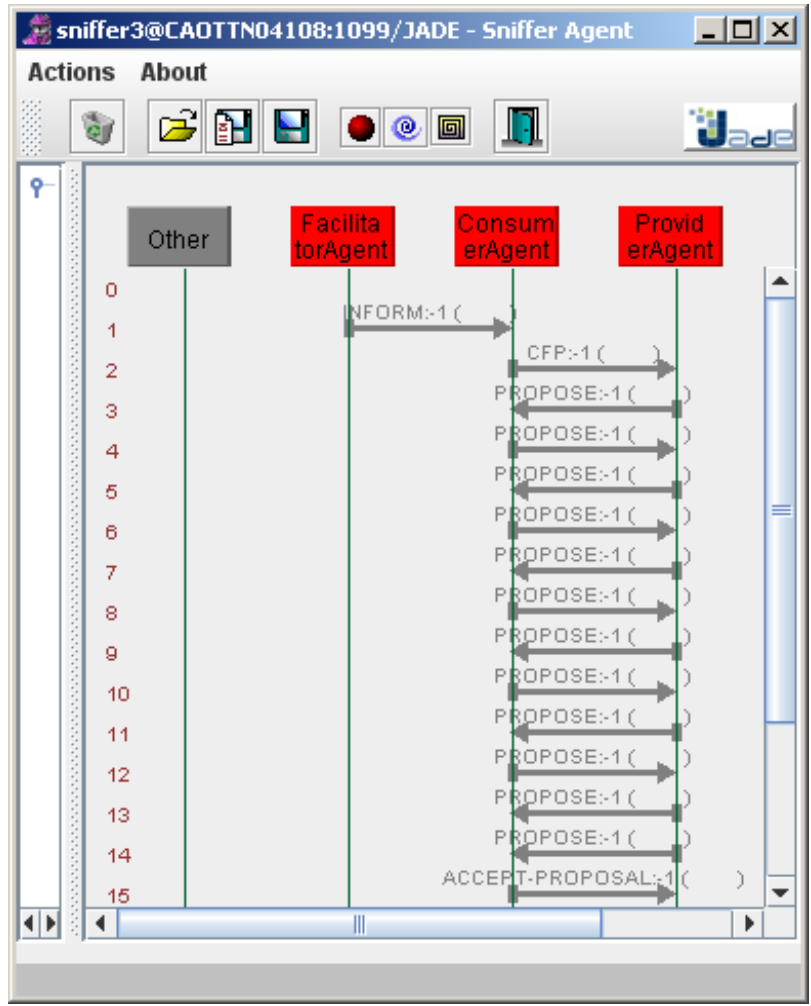

**Figure 37:Traces of per record negotiation between the consumer agent and the provider agent-exp1**

Figure 38 shows the exchange of offers in round 4 (a negotiation round corresponds to the process of sending a propose message and receiving a reply) as well as the final accept message.

The propose message sent by the provider agent contains the provider's offer of $16 per record (shown in the content section of the message). The replied propose message sent by the consumer agent contains an offer equal to 371 consumers' records (shown in the content message). The exchange of messages continues until the provider agent sends its final offer and the consumer agent accepts it. The content of the accept proposal message

includes 1538 records of consumers that potentially will accept to share their personal information for the agreed-upon payoff.

---

**Exchange of messages in round 4:**
(PROPOSE
:sender  (agent-identifier :name ProviderAgent@CAOTTN04108:1099/JADE  :addresses
(sequence http://CAOTTN04108.ad3.ad.alcatel.com:7778/acc))
:receiver  (set ( agent-identifier :name ConsumerAgent@CAOTTN04108:1099/JADE) )
:content  "inbidPR:16.0" )

(PROPOSE
:sender  (agent-identifier :name
ConsumerAgent@CAOTTN04108:1099/JADE  :addresses (sequence
http://CAOTTN04108.ad3.ad.alcatel.com:7778/acc ))
:receiver  (set ( agent-identifier :name ProviderAgent@CAOTTN04108:1099/JADE ) )
:content  "inbidCA:371")

**ACCEPT-PROPOSAL message:**
Provider Agent: received the following message:
(ACCEPT-PROPOSAL
:sender  (agent-identifier :name ConsumerAgent@CAOTTN04108:1099/JADE  :addresses
(sequence http://CAOTTN04108.ad3.ad.alcatel.com:7778/acc ))
:receiver  (set ( agent-identifier :name ProviderAgent@CAOTTN04108:1099/JADE ) )
:content  "inbidCA:1538" )

**Figure 38: Exchange of messages in round 4**

Figure 39 depicts the values of the exchanged offers between the two agents. Each time the provider agent increases its offer, the number of consumers who are willing to share their personal information increases. This is because individuals valuate their perceived privacy risks differently. While some people are willing to accept lower offers to reveal their personal information (this type of individuals according to Spiekermann [2001] are marginally not concerned), others expect offers that are worth the risk of exposure (e.g., privacy pragmatics and privacy fundamentalists).

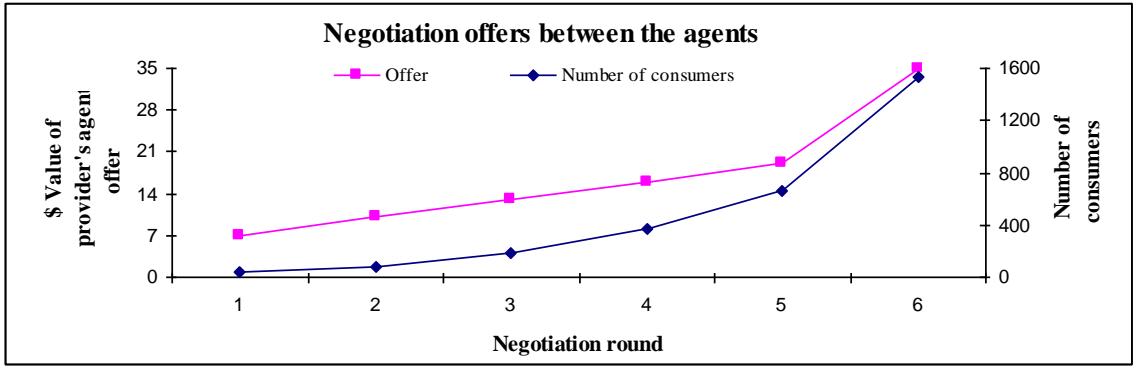

**Figure 39: Negotiation offers between the consumer agent and the provider agent**

In this scenario, the final payoff value is $35 for each record which means that the total gain of the consumers' community is 1538*$35 = $53830. But, as mentioned above, consumers valuate their personal information differently, therefore the distribution of the surplus resulted from the difference in valuation should reflect the contribution of each consumer in the community (Bonnevay et al. 2005).

Let $G_i = \{g_1, ..., g_x\}$ be the gain vector of all consumers according to their initial valuation of their personal information. Let $C$ be the consumers' community in this deal; then the gain surplus $S(C)$ of the community $C$ is:

$$S(C) = v(C) - \sum_{j \in C} g_j \qquad (6.1)$$

The surplus of $C$ is the difference between the community gain $v(C)$ and the sum of the actual gain of each consumer in $C$ according to their initial valuation of their personal information. $S(C)$ is the value that can be divided among consumers of community $C$. The community gain $v(C) = \$53830$. The total initial value gain for each consumer is $28019.99. Then, according to (6.1) the surplus gain is $25810.01. Following Bonnevay et al. (2005), we distribute the surplus among consumers according to the contribution of each consumer's valuation. For example, if two consumers valuated their personal information to be $g_1 = g_2 = $

$25, then both should receive the same surplus gain. However, if the valuation is $g_1 = \$19$ and $g_2 = \$25$, then each one of them should obtain a surplus value reflecting the percentage of contribution in the community, which is in this case $17.50 and $23.03, respectively. Figure 40 depicts the surplus distribution for each consumer in the community and Figure 41 depicts the total gain of each consumer.

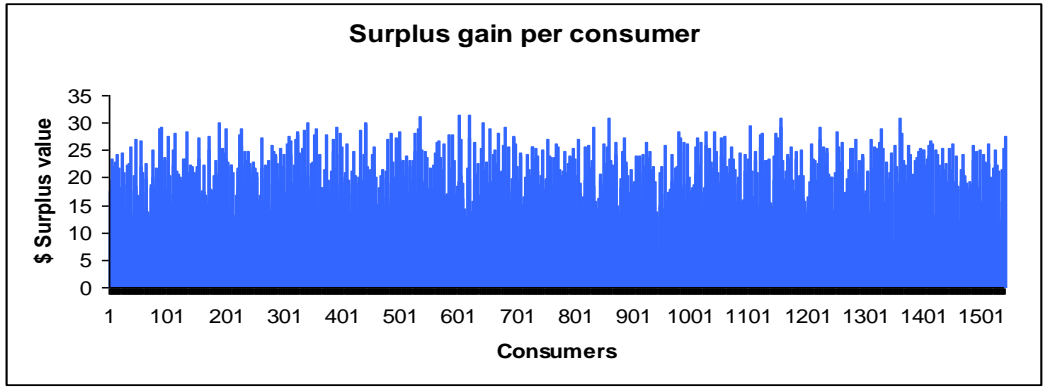

**Figure 40: Surplus distribution among consumers according to the percentage of contribution-exp1**

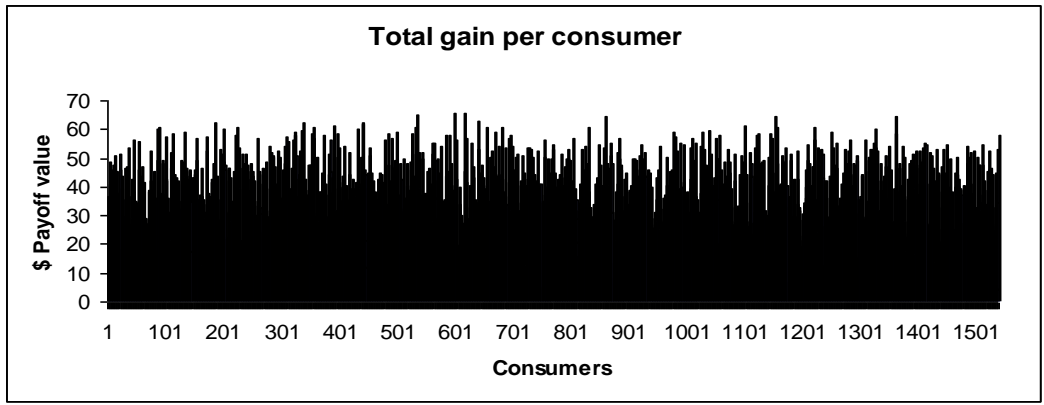

**Figure 41: The total payoff distribution among consumers including the surplus gain-exp1**

### 6.3.2 *Reservation Price Effect on the Received Privacy Payoff*

In experiment 2, the deadline of the provider agent is chosen to be large enough so that during the negotiation the provider agent will reach its reservation price. The trace of the messages exchanged between the two agents is captured via the sniffer agent provided by JADE as shown in Figure 42. The final exchange of messages is shown in Figure 43.

The propose message sent by the provider agent in the last round contains the provider's offer of $47 per record (shown in the content section of the message). The accept message sent by the consumer agent contains an offer equal to 2000 consumers' records. In this experiment, the final offer was more than what the consumer agent has as a reservation price. This is the reason for having all consumer records be included in the deal.

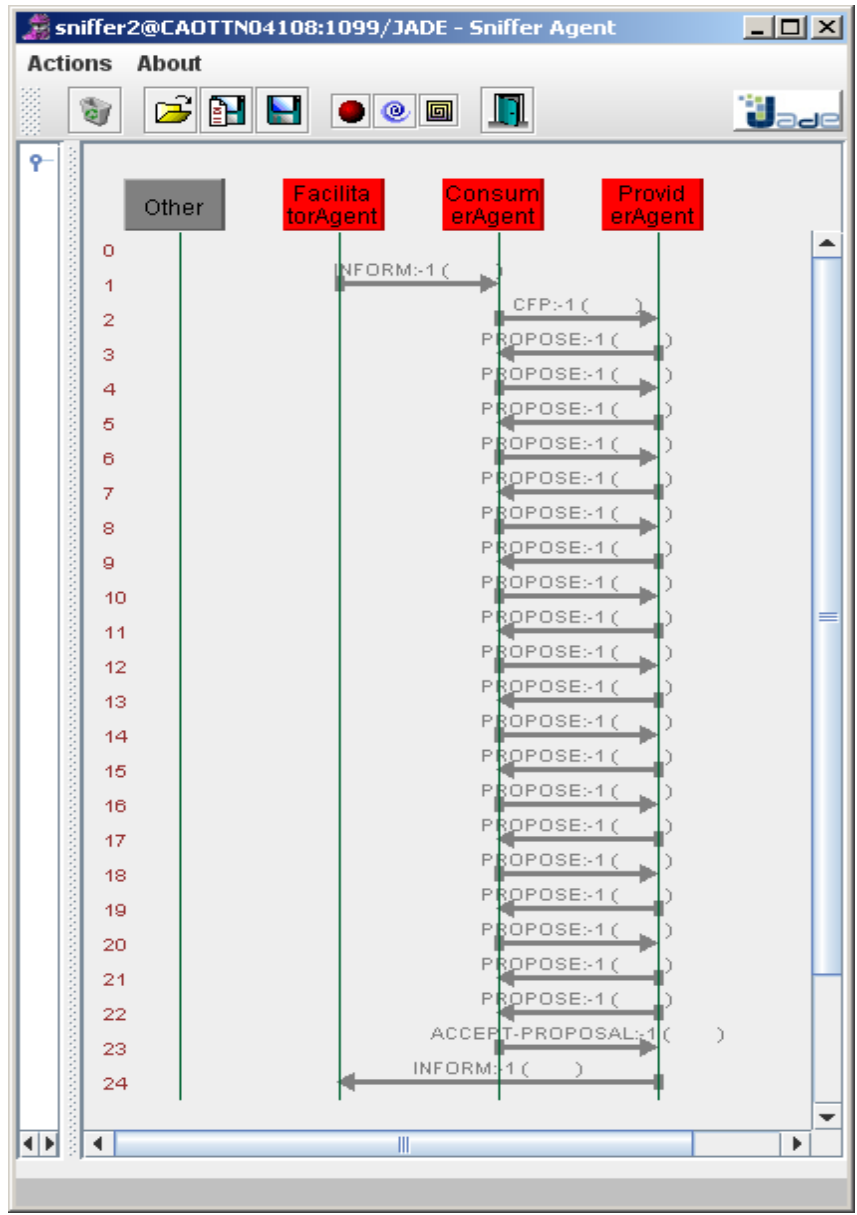

**Figure 42: Traces of per record negotiation between the consumer agent and the provider agent-exp2**

```
Exchange of messages in round 4:
(PROPOSE
:sender   ( agent-identifier :name  ProviderAgent@CAOTTN04108:1099/JADE   :addresses
      (sequence http://CAOTTN04108.ad3.ad.alcatel.com:7778/acc ))
:receiver  (set ( agent-identifier :name ConsumerAgent@CAOTTN04108:1099/JADE ) )
:content  "inbidPR:47.0"

(ACCEPT-PROPOSAL
:sender   (agent-identifier :name  ConsumerAgent@CAOTTN04108:1099/JADE   :addresses
      (sequence http://CAOTTN04108.ad3.ad.alcatel.com:7778/acc ))
:receiver  (set ( agent-identifier :name ProviderAgent@CAOTTN04108:1099/JADE ) )
:content  "inbidCA:2000"
)
```

**Figure 43: Exchange of messages in the last round**

In this scenario, the final payoff value is $47 for each record which means that the total gain of the consumers' community is 2000*$47 = $94000. Based on equation (6.1) in the previous experiment, Figure 44 shows the surplus distribution for each consumer in the community and Figure 45 depicts the total gain for each consumer.

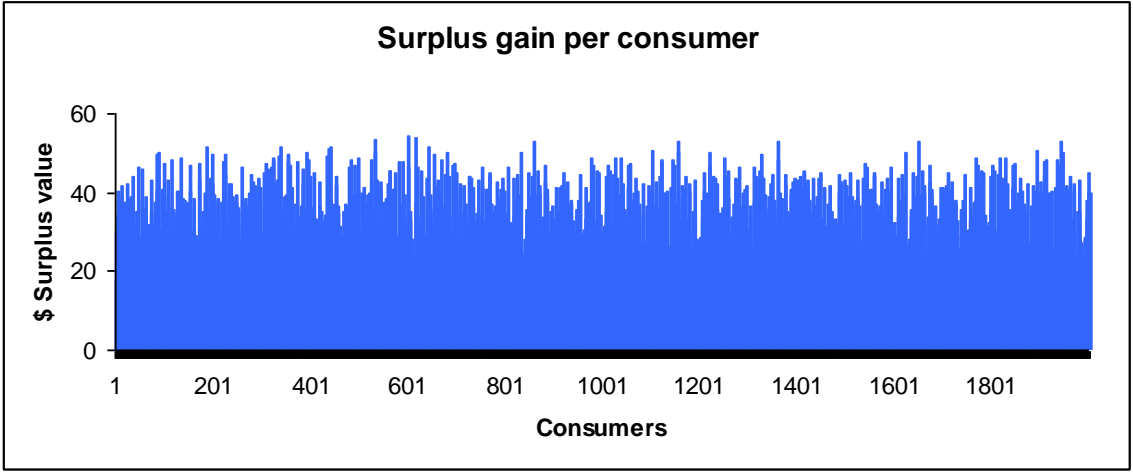

**Figure 44: Surplus distribution among consumers according to the percentage of contribution-exp2**

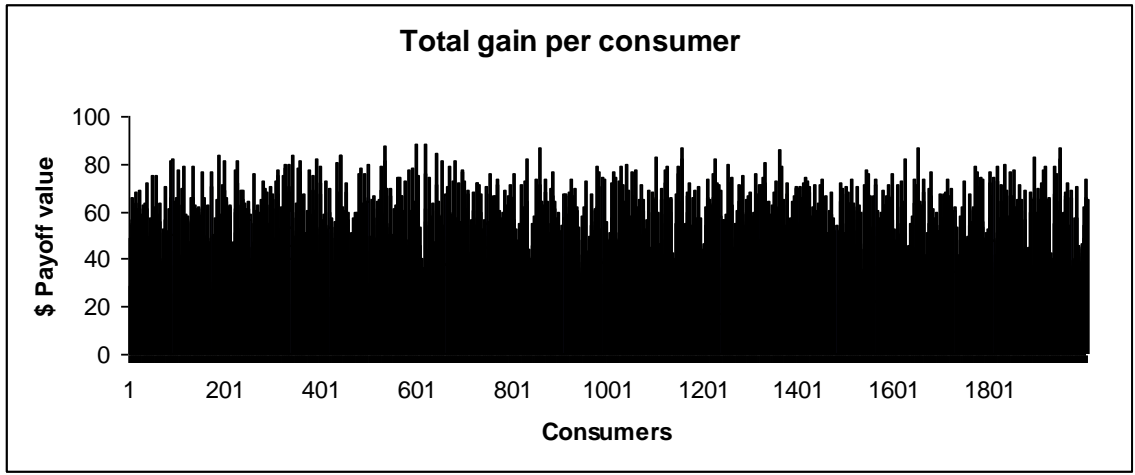

**Figure 45:The total payoff distribution among consumers including the surplus gain-exp2**

### 6.3.3 Provider's Privacy Regime Effect on the Privacy Payoff

This experiment explores how the service provider's type (i.e., their privacy practice regime) affects the benefit for both the service provider and the consumers. For this experiment, five service providers were chosen that represent different privacy practice regimes. The five service providers are travel agencies that offer similar services and similar prices. The service providers are Excapes.com, Expedia.com, Travelocity.com, Canadatravel.ca, Itravel2000.com; for convenience we name them SP_1, SP_2, SP_3, SP_4, SP_5, respectively. The privacy credential ratings of these service providers are as follows: Escapes.com PCR is 1 star, Expedia.com PCR is 1.5 stars, Travelocity.com PCR is 3 stars, Canadatravel.ca PCR is 4 stars, and Itravel2000.com PCR is 4.5 stars.

During the negotiation process, we set a large duration for the negotiation deadline so that all service providers have equal opportunity to obtain consumers' data but differ in their privacy credential score.

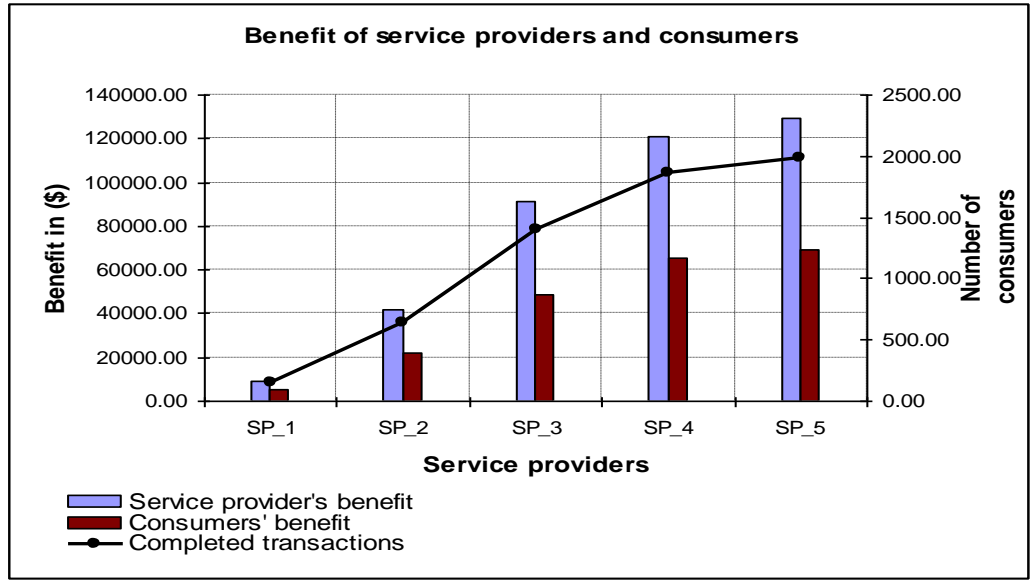

**Figure 46: Benefit of service providers and consumers**

Our first observation from Figure 46 is that there is a difference in intrinsic utility (service provider's benefit) among service providers even though they offer the same price and payoff. We hypothesized that the observed difference in the utilities is caused by the fact that consumers expect higher payoffs when they perceive higher privacy risks. For example, if Alice's privacy risk value $\Psi$ is calculated as 0.9812, 0.8156, 0.7601, 0.6428, and 0.4945 with respect to SP_1, SP_2, SP_3, SP_4, and SP_5 respectively, then her expected benefit is $49.06, $40.78, $38.005, $32.14, and $24.72, respectively. From this we can notice that Alice would not be included in any deal that gives her less than her expected payoff. Since the maximum compensation a service provider can offer is $35 (it is set here as the reservation price per data record), Alice would not be sharing her personal information with SP_1, SP_2, and SP_3. She will only share her personal information with SP_4 and SP_5. This is why we notice a significant difference between service provider SP_5 and SP_1 with respect to the number of consumers who completed the transaction.

As can be seen from the completed transactions in Figure 46, in the case of SP_1, only 143 consumers would accept an offer less than or equal to $35 while in the case of SP_5, 1989 consumers accepted an offer less than or equal to $35. This is because consumers in the case of SP_1 (the service provider with high privacy risk) are more reluctant to reveal their data given the potential risk. Therefore, they asked for higher payoff. However, in the case of SP_5, consumers are more open to share their data given the low level of privacy risk. The result shows that as the number of consumers who completed the transaction increases, the service providers' benefit increases as well. Therefore, allowing consumers to be compensated for the revelation of their personal information is expected to be beneficial for both the consumers and the service providers.

A naïve look at our result may give the impression that service providers are accumulating costs as a result of compensating consumers for the revelation of their data. This is because service providers can obtain this information from their websites for free, simply by tracking the consumers' behavior. However, such practice does not really come free; service providers incur excess cost (junk e-mail, junk advertisement, etc.) in order to attract consumers' attention and offer them services which at best are based on guessing about what the consumers like or dislike (studies show that only 3 to 5 percent of those consumers actually buy the service [Heath 2006]). However, in our experiment, we showed that service providers with high privacy credential rating do not need to make guesses about a certain community of consumers. In particular, SP_4 and SP_5 can target their service to 1863 and 1989 consumers respectively by knowing exactly what they like, thus achieving a particular service delivery threshold and minimizing opportunity costs. Therefore, compensating consumers for the revelation of their personal information is in the service providers' interest.

### 6.3.4    Risk Premium Variation Effect on the Privacy Payoff

This experiment examines the effect of market risk premium variation on the overall benefit of both the service provider and the consumers. In this experiment, we simulate a variation in the market risk premium by incrementing the drift value by 0.1, thus increasing the risk premium gradually. Five different tests were performed, one for each service provider (the same service providers as in experiment 3). Figure 47 shows the outcome of the experiments. It presents the service providers' benefit, the consumers' benefit, and the number of completed transactions. Results from the experiment show that service providers with high privacy credential rating achieve higher utility than service providers with lesser privacy credential rating; see Figure 47.A. In Figure 47.A, when the market risk premium is high (i.e., when the drift value is 2), service provider SP_5 acquired a benefit value equal to $18370 compared to $0, $0, $100, and $2860 for service providers SP_1, SP_2, SP_3, and SP_4, respectively. The reason behind this significant difference is that 334 consumers completed the transaction in the case of SP_5 (shown in Figure 47.F). The 334 consumers were expecting a payoff less than what offered by the service provider. This is based on the valuation of their private data given their perceived privacy risk of service provider SP_5. In case of service provider SP_4, 52 consumers completed the transaction and in case of SP_3 only 2 consumers completed the transaction, as shown in Figure 47.E and Figure 47.D respectively. In both cases, the consumers' perceived risk in SP_4 and SP_3 was low. At high market risk premium, service providers SP_1 and SP_2 could not secure any transactions. Figure 47.B and Figure 47.C show that at high market risk premium,

consumers did not complete any transactions with service providers SP_1 and SP_2. This puts more emphasis on the fact that service providers with higher trust values (i.e., respecting consumers' privacy) will be regarded by consumers despite how the market premiums behave.

Figures 47.B, 47.C, 47.D, 47.E, and 47.F record the performance of each service provider as well as the consumers' benefit. The influence of the market risk premium is significant in each case. When the drift value = (1, 1.1, and 1.2) representing low risk premium, all service providers were able to accumulate a benefit depending on their type, i.e., their privacy credential rating. However, as the risk premium increases to a moderate level (1.4, 1.5, and 1.6), service provider SP_1, for instance, accumulated $0 benefit (shown in Figure 47.B) because all consumers at that level were expecting higher payoff as the privacy risk increases. Such payoff is unaffordable by service provider SP_1. In Figure 47.C, the benefit of service provider SP_2 dropped to $0 at a 1.7 drift value suffering the same dilemma as SP_1 but at a later stage. SP_3 and SP_4 in Figure 47.D and Figure 47.E suffered from a sharp decline of benefit from $5225 and $28600 at moderate market risk premium to $1100 and $2860 at a high market risk premium, respectively. Only SP_5 was able to keep a higher benefit at a high market risk premium.

The outcome of the experiments reveals how the overall perception towards privacy impacts the benefit of both the service provider and the consumers. Consumers who are reluctant to complete transactions because they perceive high privacy risk impact the service providers' benefit as they ignore their services. The most important result is that consumers and service providers are both better off if privacy risk is low.

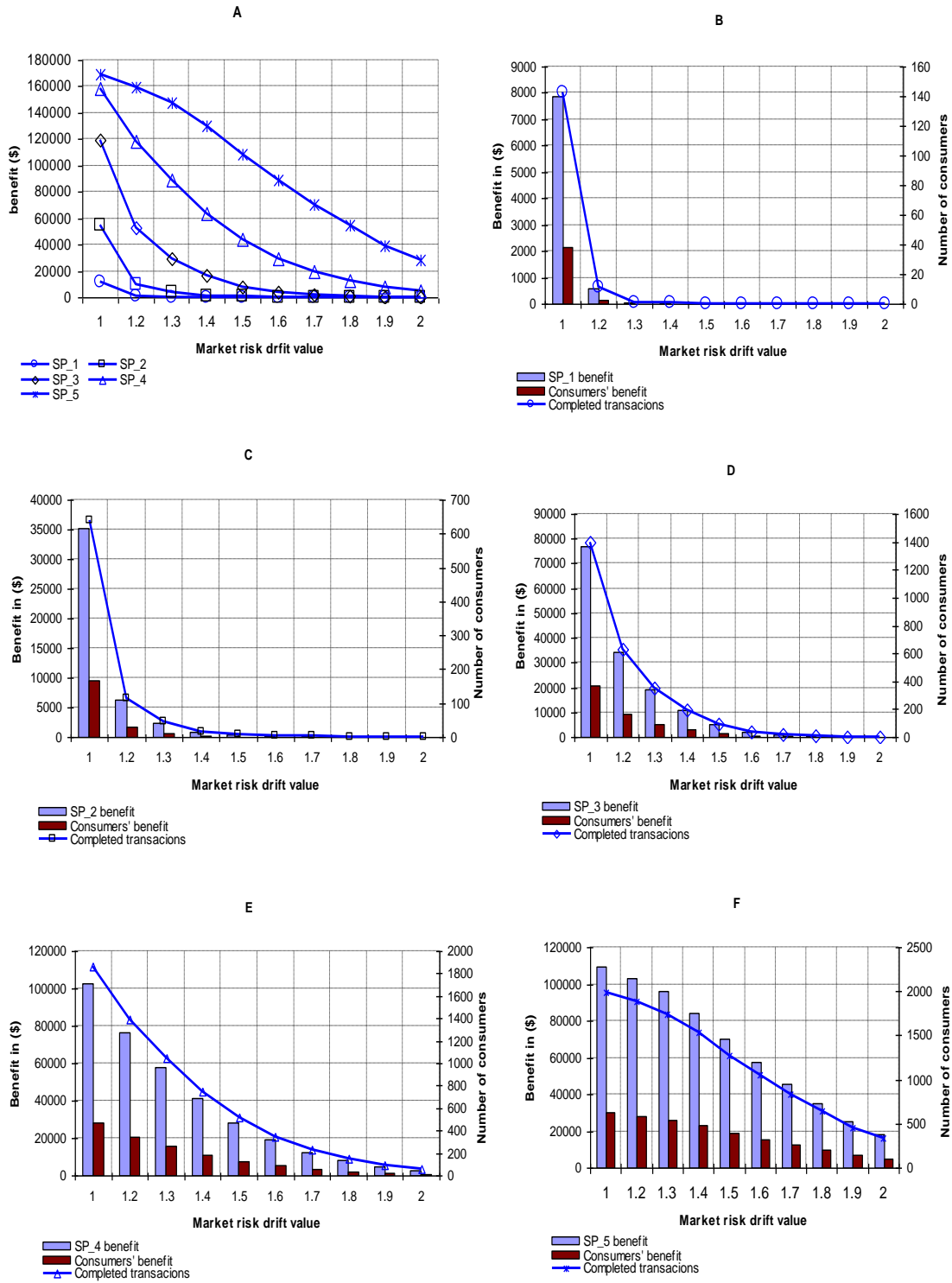

**Figure 47: Benefit of service providers and consumers. (A) Performance of service providers with respect to the market risk premium. (B), (C), (D), (E), and (F) detail the corresponding consumers' benefit and the completed transactions with respect to the market risk.**

It should be mentioned here that in order to validate the results of the simulation experiments provided in subsections 6.3.1 to 6.3.4, we need a reference system for benchmarking. To the best of our knowledge, however, this thesis is the first study that proposes a system for privacy payoff in eCommerce. As such, AAPPeC is a pioneer system in this area of research and hence validation of the results would not be possible without the existence of a comparable system.

## 6.4 Summary

This chapter provided a proof of concept implementation for AAPPeC. The prototype shows that the system is technically feasible and could be deployed using agent technology. Through experiments, we ran a few scenarios to examine the effect of influencing parameters in the payoff value as well as the benefit of service providers. In particular, the first and the second experiments test the negotiation process under deadline constraint. We have shown that a strategically placed consumers' agent can help consumers attain the maximum gain during the negotiation process. In the third and the fourth experiments, we examined two scenarios that affect the privacy payoff and the service providers' benefit. The first scenario is related to the service providers' type (expressed here as in the service provider's privacy regime) and the second scenario is related to the market risk premium fluctuation. In the former, we have shown that service providers with high PCR have the advantages of achieving a particular service delivery threshold and minimizing opportunity costs. In the latter, we have shown that the overall privacy risk perception affects the benefit of consumers and service providers. However, service providers with high PCR were less likely to suffer from revenue loss than those service providers with low PCR.

As we have mentioned in Chapter 4, the privacy payoff valuation is context aware and associated to the privacy risk. But risk valuation is linked to trust evaluation. Therefore, in the next chapter we are going to discuss a field study performed to examine the effect of the privacy credential ratings on consumers' perceived trust, attitude, and awareness.

# Chapter 7. Evaluation

## 7.1    Introduction

The success of electronic commerce is determined in part by whether consumers trust sellers that are invisible to them (Hu et al. 2003). Without trust, it is difficult to imagine that a transactional relationship could be developed or maintained. Many internet users are concerned about the misuse of their personal information when visiting online stores (Acquisti and Grossklags 2005, Gopal et al. 2006, Hussain et al. 2007). Privacy concerns continue to hinder consumers from making online purchases. To gain consumer trust, e-businesses must assure prospective consumers that personal information obtained through e-commerce transactions will not be misused. To this end, they have employed a variety of privacy assurance mechanisms to increase their perceived trustworthiness. The most common solution is to post a privacy policy on the website to ease consumers' minds. Another recognized solution is obtaining seals of approval, third-party certifications, membership to privacy-auditing services, etc. (Pennington et al. 2003, Suh and Han 2003). Online businesses hope that featuring trust attributes in their websites will attract consumers to their storefronts. However, research indicates that it is difficult for many consumers to infer trust and to evaluate trust attributes related to privacy and private data handling (Hu et al. 2003, Hussain et al. 2007). This is because privacy information is invisible to internet users and sometimes difficult to comprehend (Metzger and Docter 2003, Tang et al. 2008, Hu et al. 2003, Hussain et al. 2007, Hoofnagle et al. 2008). For example, it has been shown by many studies such as those of Metzger and Docter (2003) and Tang et al. (2008) that only a small percentage of people actually understand privacy policies, because the policies tend to be long and written in legal terms that require a certain level of legal knowledge to decipher.

In order to help consumers see "good signs" and "bad signs" while checking an online website, we have developed in Chapter 4 a mechanism that examines the competency of the online businesses with respect to privacy and private data handling. However, in order to understand how consumers will react to the rating, we designed an experiment in which 56 consumers were investigated. The experiment was designed to test the effect of the rating score on the participants' attitude, awareness, and perceived trust.

The remainder of this chapter is organized as follows: in the next section, we discuss the hypotheses of the experiment. In section 7.3, the experiment approach is described, followed by the screening survey in section 7.4. In section 7.5 we discuss our findings. Finally, in section 7.6 we conclude the chapter.

## 7.2    Hypotheses

Lack of consumer trust is a critical obstacle to the success of online businesses. Knowledge is one important factor influencing the level of trust as well as consumers' attitude towards online shopping (Wang et al. 2009). A decade ago or before, only those users most familiar with the Internet were concerned about the trustworthiness of online transactions. The rest of Internet users were oblivious, still trying to figure out how to access a webpage. Now that time has passed; more users are very familiar with the Internet and well aware of the potential danger of transactions over the Internet, but are not adept enough to find the way out of traps. Previous studies such as those of Gefen et al. (2003) and Hoffman et al. (1999) have indicated that there is a relationship between knowledge and trust. Knowledge reduces social uncertainty through increased understanding of what is likely to happen. This reduction of uncertainty will contribute to increasing trust.

Information about privacy practices of online businesses is one form of knowledge that contributes to the consumers' confidence and positive attitude towards online shopping. Wang et al. (2009) showed that knowledge about online stores and consumers' attitude are interrelated. According to Jones et al. (2009), attitude is defined as the individual's positive or negative evaluation towards a behavior. It is composed of an individual's salient beliefs regarding the perceived outcome of performing a behavior (Wang et al. 2009).

The hypotheses proposed in this study are designed to examine how participants will respond to questions that are related to the knowledge presented to them about service providers' privacy practices (hypothesis 1). Another experiment is designed to test the effect of the privacy information on participants' decisions regarding sharing sensitive personal information or the decisions to purchase a product that raises privacy concerns (hypothesis 2 and 3). These hypotheses were designed to gauge consumers' reaction in the presence of the privacy information. Specifically, we are looking to see to what extent this information creates awareness that contributes in the decision-making process. Finally, hypothesis 4 aims at testing consumers' perception of trust, given the privacy credential rating. The details of each hypothesis are presented below:

***Hypothesis 1:*** *The presence of the PCR information is positively associated with online shopping decision.*

Knowledge of how online businesses adhere to respectable privacy practice is very important for consumers to participate in online shopping. Hypothesis 1 focuses on the effect of such knowledge (i.e., the presence of the PCR information) on the participants' attitude towards websites. While the consumers' behavior is expected to be obvious (i.e., they will most likely choose websites with high privacy credential ratings), it is rather important to understand their attitude (framed here as purchasing decision) towards certain websites in the presence of privacy credential ratings. This is especially so in the case of famous websites which have low privacy credential rating (as we will see later in this section). Thus, it is essential to realize if participants will consider such knowledge as a factor in their decision making.

In a study of consumers' attitudes towards websites, Gideon et al. (2006) suggest that attitude towards a website's personal data protection transfers to attitudes towards consumers' willingness to share private information with the online business. Similarly, Tsai et al. (2007) found that providing consumers with privacy information about service providers creates a more favorable attitude towards an individual's willingness to buy even if they are required to pay a premium. Both studies suggested that online privacy is tightly related with the perceived risk. The less the risk consumers perceive, the more they are keen to share personal information and to purchase privacy-sensitive products and vice versa. Here, we study the extent to which the perceived risk in the presence of the PCR information is affecting the consumers' decision in scenarios that raise additional privacy concerns. This leads to the following two hypotheses:

**Hypothesis 2:** *The presence of the PCR information is positively associated with the participants' attitude towards sharing sensitive personal information*

**Hypothesis 3:** *The presence of the PCR information is positively associated with the participants' decision to purchase privacy-sensitive products (i.e., products that are likely to raise additional privacy concerns)*

Perceived privacy is often thought to be another important antecedent of trust (Chen and Barnes 2007) and is defined as the subjective possibility of accordance between consumers' anticipation and their cognition of how their private information is being used (Yang 2005). It is the perception that the online business will adhere to an acceptable set of practices and principles (Lee and Turban 2001). The PCR information provides consumers with knowledge about online businesses' practices with respect to privacy and private data handling. Based on this discussion, the following hypothesis is introduced:

**Hypothesis 4**: *Participants' perceived trust in a website increases in the presence of PCR information*

Next, we present the experiment approach used to test the above hypotheses.

## 7.3    Experiment Approach

To examine the above hypotheses, we designed an experiment in which 56 active consumers were investigated. Active respondents tended to be experienced internet users and experienced in online shopping; 93.5% are frequent online shoppers (more than once a month), while 6.4% not as frequent. Since privacy in the online world is less obvious, we targeted only online shoppers in this experiment. Thus, we excluded any consumer with no experience in online shopping. The ages of the participants ranged from 20 to 60 years old. The sample consisted of students and employed people.

Part of the experiment was analyzing the 20 online shopping websites shown in Appendix A. We analyzed their privacy credentials by scrutinizing their privacy policies and checking the validity of their certificates or seals if they post any. Additionally, we used sources such as iVouch.com, SiteAdvisor.com, Bizrate.com, and MacAfee.com to gather reputation scores about these websites (based on customers' testimonial). The aim of this analysis is twofold: first, to apply the proposed mechanism on real data; second, to compare the results of the experiments with some of the available commercial trust systems in the field.

As part of analyzing these websites, we used a scale ranging from 0 to 5 stars to produce the final ratings. The result of the analysis led us to consider four types of websites for our experiment. The first types are the highly reputable websites with low privacy credentials. It should be clarified that being highly reputable is not the same as having a high privacy credential rating (PCR), the former meaning simply that the website is more famous. Examples of this type are: Amazon.com, Target.com, and pogo.com. The second types of website are the newly opened websites that do not have customer testimonials or reputation scores yet, but their privacy credentials are ranked very high. An example of this type is dinodirect.com. The third type of websites are the websites that are ranked somewhere in the middle in terms of reputation, and their privacy credentials are ranked somewhere in the middle too. Examples of this type are: Tigerdirect.com, Travelocity.com, and Bestbuy.com. The fourth type is the highly reputable websites with high privacy credentials. Examples of this type are: Geeks.com, Exceldiamond.com and newegg.com. We also made up four website names (representing the four types) so that the participants would not recognize any of them (see Appendix B). This was done to prevent any prior knowledge from influencing the participant's decision.

The experiment was divided into two phases. Phase 1 includes two separate questionnaires. The first questionnaire sheet includes reputation information of the four websites we made up without the PCR information. The second questionnaire sheet includes the four websites with their reputation information plus the PCR information. The population of the participants was randomly divided into two equal groups of 28. Questionnaire sheets were distributed via emails. In Phase 2, the same questionnaire sheets as in Phase 1 were used except the real names of the websites were revealed to the participants.

In both phases the following conditions were used to analyze the respondents' reaction to the questionnaires:

*Condition 1*, reputation only: Participants were given the four websites' names and their reputation ratings. They were asked a number of questions about their privacy concerns, purchasing decisions, and overall trust (see Appendix B). In phase 2, the same group of participants was given the real names of the websites and asked the same questions as in phase 1.

*Condition 2*, reputation plus PCR: Participants in this condition were given the four websites' names along with reputation ratings and the PCR information. Participants in this condition were asked the same questions as in condition 1 regarding their privacy concerns; purchasing decisions and overall trust (see Appendix B). In phase 2, the same group of participants was given the real names of the websites and asked the same questions as in phase 1.

We structured the experiment so that participants were presented with questions related to their purchasing decision, their willingness to share personal information, their willingness to buy products that might raise some sort of privacy concerns, and, finally, we asked them to rate their perceived trust on a given website based on the information presented to them.

## 7.4 Screening Survey

Before starting the experiments, following Tsai, et al. (2007), we used a screening survey to learn about participants' privacy concerns. The purpose of this screening is to investigate what items of privacy concern have a higher impact on the participants' online purchasing decision. The results in Table 10 show the mean as well as the significance of each item. The scores are based on a 5-point Likert scale with a maximum value of 4.0, ranging from Not Concerned at All to Extremely Concerned. The significance was calculated using the well-known student's t-test comparing the responses to a neutral baseline of 2.

The results show that all participants are statistically equivalent in caring very strongly about these aspects of privacy items. It shows that participants are significantly concerned about their personal information usage ($p < 0.0001$), sharing it with other companies ($p < 0.0001$), and using it for telemarketing and advertisements ($p < 0.0001$). However, as mentioned earlier in this chapter, these items as well as other privacy-related information are often not trivial to discover when visiting a website, especially for an average consumer.

Table 10: The t-test result of the screening survey

| Item of concern | Mean | *t* Value | *p* Value |
|---|---|---|---|
| An online business uses information that does not personally identify you to determine habits, interests, or other characteristics | 1.14 | -6.8 | <0.0001 |
| An online business shares information with other companies that does not personally identify you | 1.71 | -2.16 | 0.034 |
| An online business contacts you via email about other services | 2.41 | 2.52 | 0.01 |
| An online business uses personally identifying information to determine your habits, interests, or other characteristics for advertisement purposes | 3.16 | 8.64 | <0.0001 |
| An online business uses information about you to tailor services and contact you via telephone | 3.21 | 8.59 | <0.0001 |
| | 3.39 | 11 | <0.0001 |

An online business shares information that identifies you
with advertisement companies

An online business does not allow you to opt out from   3.21      7.29      <0.0001
mailing lists

## 7.5 Results

As mentioned earlier, our goal is to determine whether the privacy rating information will

impact consumers' attitude, awareness, and perceived trust. Our findings indicate that the

PCR information was an important part of participants' decisions in condition 2 (reputation plus PCR). Participants were more likely to consider websites offering high levels of privacy protection regardless of their names. Such behavior cannot be attributed to an interest in purchasing from websites labeled with attractive reputation if the privacy information rating was low. Participants in condition 1 (reputation only), however, were generally willing to purchase from websites of high reputation. Next, we present the details of our findings.

### 7.5.1          Shopping Decision

**Hypothesis 1:** *The presence of the PCR information is positively associated with online shopping*
*decision – **Supported**.*

Participants were asked, if they were purchasing a service, which website(s) would they

consider. Overall, we found that there are statistically significant results in this area. Table

11 corresponds to the results done in phase 1 (i.e., with the fake website names); it shows

that there were greater percentages of selecting websites with PCR information in condition

2 (reputation plus PCR). For example, 64.29% of participants in condition 2 selected a

website (got-it-all.com) with no history of reputation (i.e., new in the market) compared to

28.57% of participants in condition 1 (Fisher's exact p < 0.05). This shows that if a website has no reputation available or customer reviews, consumers can rely on the PCR mechanism to make a purchasing decision from a website.

**Table 11:Results of testing hypothesis1 in phase1**

|                       | Condition 1 | Condition 2 | Fisher's Exact p |
|-----------------------|-------------|-------------|------------------|
| amazing-products.com  | 78.57%      | 39.29%      | <0.01            |
| got-it-all.com        | 28.57%      | 64.29%      | <0.05            |
| best-service.com      | 46.43%      | 67.86%      | 0.17             |
| service-place.com     | 50.00%      | 85.71%      | <0.01            |

In the case when the website's reputation is 4 stars and the PCR is 5 stars, 85.71% of participants in condition 2 selected service-place.com compared to 50% in condition 1; the difference is significant (Fishers' exact $p < 0.01$). However, in the case when the PCR is 0 stars, participants tend to avoid websites that do not respect their privacy. This is the case of website amazing-products.com; only 39.29% of the participants in condition 2 selected this website compared to 78.57% of the participants in condition 1. But, we have learned from the screening survey that all our participants showed a high level of privacy concern. This means that the 78.57% of the participants in condition 1 were at risk because the privacy information was not available to them to make an informed decision. We observed from this experiment that there do not seem to be significant differences when the website has a medium reputation rating (3 stars) and medium PCR information. In condition 2, 67.86% of the participants selected best-sevice.com compared to 46% in condition 1 (Fisher's exact $p = 0.17$).

In phase 2 we showed the real names of the websites to the participants. Figure 48 captures the difference in the percentages between the two phases.

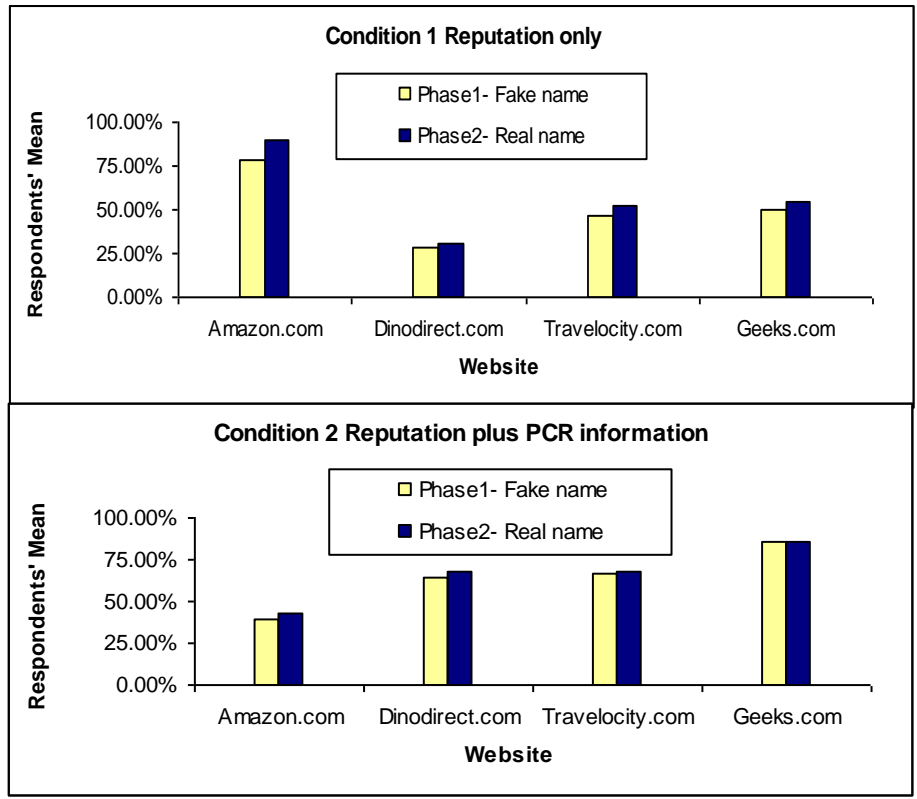

**Figure 48: Percentages of purchasing decision for participants in phase 1 and phase 2**

We have noticed that participants in condition 2 made their decision mainly based on the PCR information regardless of how famous the website is. This is clear when we compare their responses in phase 1 and phase 2 for the website Amazon.com. The percentage of participants who selected Amazon.com in phase 2 was 42% compared to 39.29% in phase 1, less than a 3% increase. Their responses were also in the same manner for the other websites. For example, the percentage increase in case of dinodirect.com was 3.57% while it was 1% and 0% for websites Travelcity.com and Geeks.com respectively. The result means that participants in condition 2 relied on the PCR information to make their decisions, and not on the reputation of the service provider, as we were expecting.

With respect to participants in condition 1, in the absence of the PCR information the name of the website was indeed important. This was noticeable in the case of Amazon.com

where the percentage increase from phase 1 was close to 12%. However, for the case when the website name was not known (dinodirect.com is a new website), the percentage increase was less than 2%. In the other two websites, Travelocity.com and Geeks.com, the percentage increase was 5.57% and 4% respectively.

The above test shows that the PCR mechanism is making a difference in the users' purchasing decision. Participants who were shown the privacy information were well aware in their decisions and chose websites that protect their information without paying much attention to the website's reputable names. This leads to the conclusion that the stated hypothesis is supported.

### 7.5.2        Privacy-sensitive Information

This test is based on a 5-point Likert scale with a maximum value of 4.0, ranging from Not Willing to Consider to Extremely Willing to Consider. We studied the effect of the PCR information on the participants' decision to share sensitive personal information and the participants' decision to purchase privacy sensitive items.

**Hypothesis 2:**  *The presence of the PCR information is positively associated with the participants' attitude towards sharing sensitive personal information– **Supported**.*

In phase 1 we asked the participants which website they would consider if the purchased service requires the submission of information such as credit card number, salary, health condition, social insurance number, marital status, and age. Based on the t-test, we found that individuals who were shown the PCR information were significantly more likely to share their sensitive private information with websites that better protect their personal information (see Table 12).

**Table 12: Results of testing hypothesis 2**

|  | Condition 1 | Condition 2 | t Value | p Value |
|---|---|---|---|---|
| amazing-products.com | 1.36 | 0.93 | 1.54 | 0.13 |
| got-it-all.com | 0.64 | 1.57 | -3.25 | <0.005 |
| best-service.com | 0.75 | 1.36 | -2.34 | <0.05 |
| service-place.com | 1.18 | 2.86 | -5.76 | <0.0001 |

In the case of websites got-it-all.com and service-place.com, the result was significant ($p < 0.005$) and ($p < 0.0001$) respectively. In the case of website best-service.com, the difference was also notable from the mean value of the responses; that is, 0.75 compared to 1.36 for condition 1 and condition 2 respectively ($p < 0.05$).

With respect to participants in the first condition (reputation only), they were in general reluctant to share their sensitive personal information even with a high reputation website; the mean value of the respondents in case of amazing-products.com and service-place.com were 1.36 and 1.18 respectively (below the baseline mean of 2).

The result of phase 1 shows that participants who were shown the PCR information were more likely to share their sensitive personal information with websites that protect their privacy compared to those who did have this information.

Figure 49 shows the participants' mean responses in phase 2. In condition 1 we noticed that the mean response increased from 1.36 to 2.25 in the case of Amazon.com. This result shows that participants' response was driven by the name of the website. The same applies for the case of Travelocity.com and Geeks.com where the mean values increased from 0.75 to 1.14 and from 1.18 to 1.89 for both websites respectively. However, for the website Dinodirect.com, the mean value decreased from 0.64 to 0.61, which indicates that participants are avoiding any website that they do not know.

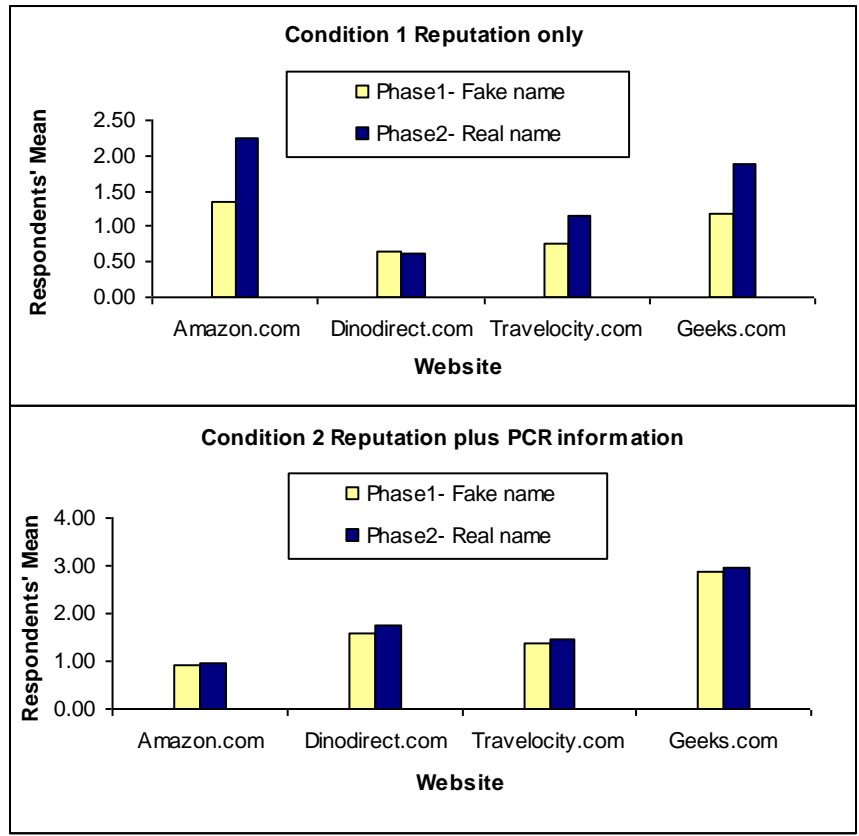

**Figure 49: Responses' mean for participants in phase 1 and phase 2 – hypothesis 2**

As for participants in condition 2, their responses stayed almost the same regardless of the website name. In the case of Amazon.com, the mean value in phase 1 was 0.93 and in phase 2 it was 0.96. As we can see it is only a small change. This means that, in the presence of PCR, participants were not influenced by the fact that Amazon.com is a famous website. Because of the PCR information, they become aware of the privacy practices of Amazon.com. The same behavior is also noticed for their responses in the case of the other websites as shown in Figure 49.

The above test shows that the presence of the PCR information was important for participants to make an informed decision and hence the hypothesis is supported.

**Hypothesis 3:** *The presence of the PCR information is positively associated with the participants' decision to purchase privacy-sensitive products (i.e., products that are likely to raise additional privacy concerns) – **Supported.***

**Table 13: Results of testing hypothesis 3**

|                    | Condition 1 | Condition 2 | t Value | p Value |
|--------------------|-------------|-------------|---------|---------|
| amazing-products.com | 2.04 | 0.86 | 3.29 | <0.005 |
| got-it-all.com | 0.46 | 1.61 | -3.88 | <0.001 |
| best-service.com | 1.04 | 1.54 | -1.66 | 0.109 |
| service-place.com | 1.75 | 2.89 | -4.08 | <0.001 |

In this test, we asked participants in both conditions which website they would consider for the purchase of a privacy-sensitive product. That is, a product that would likely raise additional privacy concerns because participants might feel uncomfortable receiving marketing solicitation or having other people know that they had purchased that item. We selected a box of condoms, sex book, etc. as examples of such products that are likely to raise privacy concerns. Based on the t-test, we found that individuals who were shown the PCR information were significantly more likely to consider websites with higher privacy protection (see Table 13). For example, participants in condition 2 considered go-it-all.com more than participants in condition 1 with a significant difference (p < 0.0001). Despite the fact that go-it-all.com is a new website, participants felt that this website is adequate to protect their privacy. However, participants who did not have the PCR information avoided this website. In the case of service-place.com, the statistical result was significant (p < 0.0001) in favor of condition 2. This is due to the high perception of privacy protection as well as high reputation score. With respect to website best-service.com, the result was not significant (p = 0.109); most participants did not assess that this website is adequate to protect their privacy.

The results of the respondents regarding the first website, amazing-products.com, were statistically significant in favor of condition 1 (p < 0.01); that is, the case when no PCR information is present. But as explained before, this might be misleading for privacy-concerned individuals, which is the case of our participants in this experiment. If we consider all individuals to be statistically equivalent in terms of caring about their privacy, only then can we rationalize why participants who were being shown the PCR information negatively changed their purchasing decision about the website amazing-products.com.

Figure 50 shows the participants' mean responses in phase 2. In condition 1, we have noticed a small increase in the participants' mean response from 2.04 to 2.07 with respect to Amazon.com. The overall mean value was relatively high which is influenced by the website name.

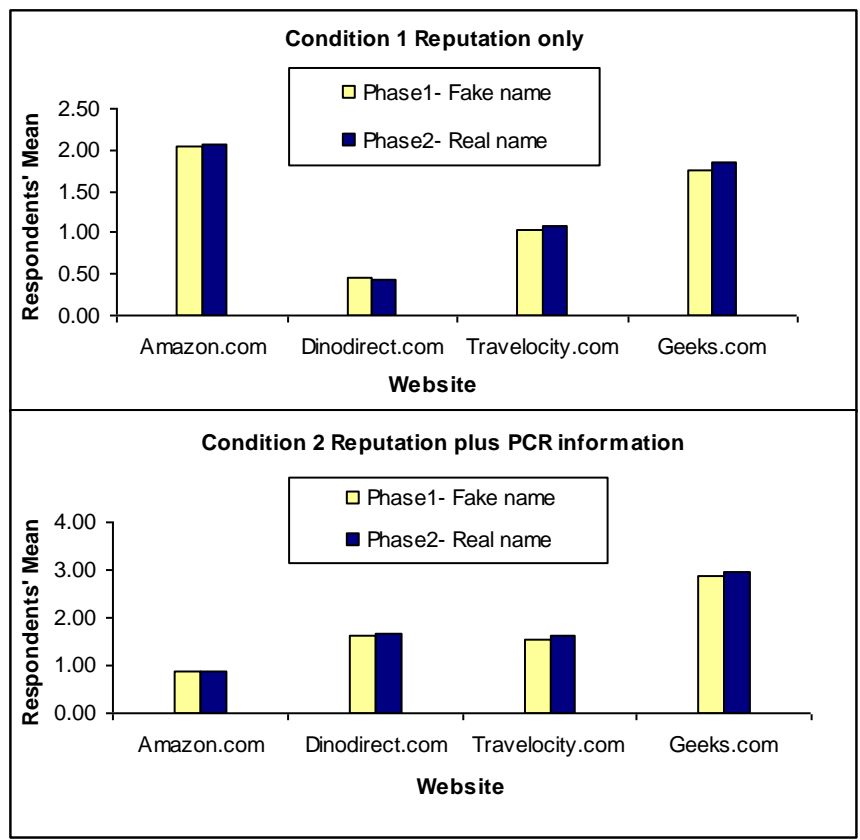

**Figure 50: Responses' mean for participants in phase 1 and phase 2 – hypothesis 3**

In the case of websites Travelocity.com and Geeks.com, the mean response increased from 1.04 to 1.06 and from 1.75 to 1.86 respectively. We have noticed that in the case of Amazon.com the mean response was higher than the case of the previous hypothesis (sharing personal information) as well as the change between phase 1 and phase 2 in terms of mean response was small. It is most likely because of the participants' perception of privacy sensitive product as less risky compared to sharing sensitive privacy information. Future studies may be needed to carefully analyze such behavior.

With respect to participants in condition 2, the mean response in phase 1 for Amazon.com was 0.86 while in phase 2 it was 0.88. In both cases participants were making decisions based on how they see the privacy practices of the website. The same behavior was also noticed for their responses in the case of the other websites, as shown in Figure 50. As expected, participants in condition 2 did not rely on the name and reputation of the websites but rather on the privacy information presented to them about the websites. The awareness created by the PCR information for participants in condition 2 suggests that the hypothesis is supported.

### 7.5.3 Trust

This subsection evaluates the perceived trust by participants who have access to the PCR information. Perceived privacy infuses the consumers' trust in online businesses. In the presence of the PCR information, participants were asked to rate their perceived trust in each website. Scores for each website were based on a 5-point Likert scale from No Trust At All to Strong Trust.

**Hypothesis 4:** *Participants' perceived trust in a website increases in the presence of PCR* information – ***Supported.***

The perceived trust recorded by participants in both conditions shows a significant result in favor of condition 2. In Table 14, the overall mean of trust for the websites in condition 2 ranges from 0.68 for PCR information with 0 stars to 2.86 for PCR information with 5 stars. However, the overall mean of trust for the websites in condition 1 ranges from 0.57 for reputation score of 0 stars to 1.07 for a reputation score of 5 stars. All possible pairwise comparisons between the conditions were significant except for the first website.

**Table 14: Results of testing hypothesis 4 – Phase1**

|                    | Condition 1 | Condition 2 | t Value | p Value    |
|--------------------|-------------|-------------|---------|------------|
| amazing-products.com | 1.07      | 0.68        | 1.46    | 0.15       |
| got-it-all.com     | 0.57        | 1.71        | -4.08   | < 0.001    |
| best-service.com   | 1.04        | 1.54        | -2.15   | < 0.05     |
| service-place.com  | 0.96        | 2.86        | -6.91   | < 0.0001   |

The findings show that respondents were significantly more likely to trust websites with strong PCR information (i.e., scores of 4 or 5 stars), as shown in Table 7. This is the case for websites got-it-all.com and service-place.com; the p values are p < 0.001 and p < 0.0001 respectively. With respect to moderate PCR information level, the pairwise comparisons were significant (p < 0.05). Respondents were more likely to trust best-service.com with a mean rating of 1.54 in condition 2 compared to 1.04 of those in condition 1. The presence of the PCR information, although ranked as medium level, had a noticeable effect on the individuals' perceived trust.

Perhaps the most interesting result was the case of the first website, amazing-product.com. The mean level of the respondents' perceived trust in condition 1 was 0.68 compared to 1.07 to those in condition 1. Although most participants in condition 1 previously showed a relatively strong opinion about the website when they were asked questions related to purchasing decision and sharing privacy information, their perceived trust somehow did not reflect the trend of their previous choices. But when we revealed the real name of the website (Amazon.com) to participants in condition 1 of phase 2, the mean response increased from 1.07 to 1.32 (shown in Figure 51), which is a relatively large change, but not as large as expected compared to the behavior we saw in the previous tests.

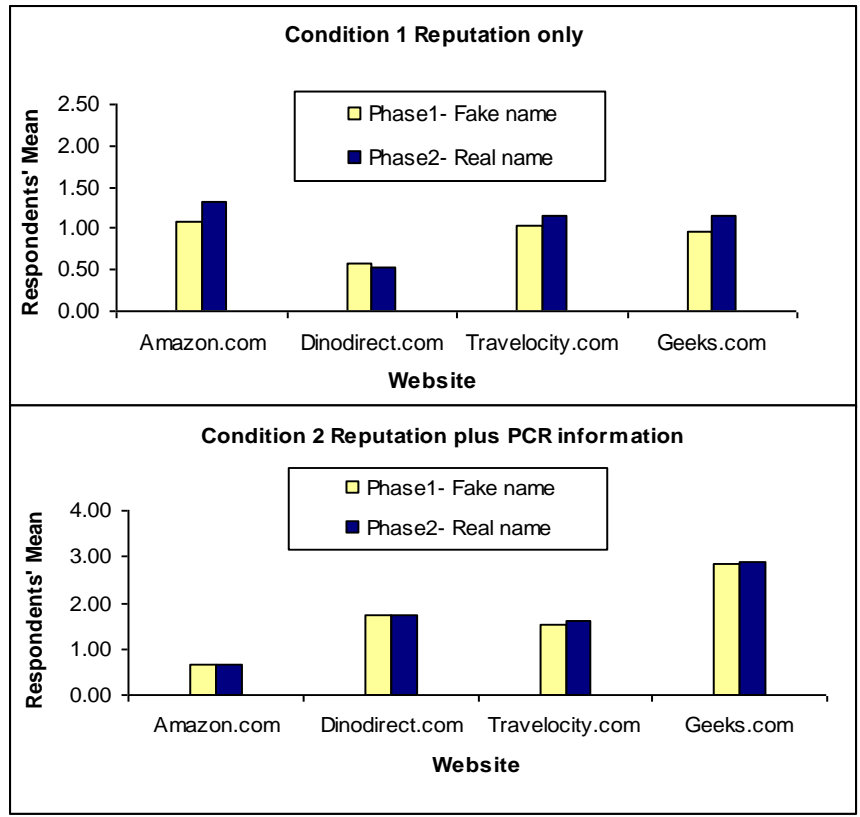

**Figure 51: Responses' mean for participants in phase 1 and phase 2 – hypothesis 4**

A possible explanation of such behavior is linked to the fundamental definition of trust itself as explained by Meinert et al. (2006), in which trust is defined as the consumers' willingness to fully rely on the seller and take action in circumstances where such action makes the consumer vulnerable to the seller. This explanation implies that unless individuals have a definite assurance of the risks involved, their trust in a prospective seller has certain limits. For many individuals, trust is something bigger than a choice of buying a service, but a relationship in which they can feel safe all the time. Such a definition gives a logical explanation as to why individuals being shown the PCR information perceived trust differently compared to those who did not have access to such information.

With respect to the dinodirect.com website, the perceived trust of participants in condition 1 was low. The mean response decreased from 0.57 to 0.54. However, if we compare the mean responses of participants in condition 2 for the same website (the fake name and the real name), we notice a different perception compared to participants in condition 1. The results indicate that the PCR information was a major factor in their perception of trust. The mean response for participants in condition 2 was 1.71 and 1.75 in phase 1 and phase 2 respectively. As for the other websites Travelocity.com and Geeks.com, participants in condition 1 showed a moderate range of percentage increase from 1.04 in phase 1 to 1.14 in phase 2 and from 0.96 in phase 1 to 1.14 in phase 2 respectively. The increase was mainly motivated by the name of the websites. However, the mean response of participants in condition 2 stayed at a small level. For websites Travelocity.com and Geeks.com, the mean response increases from 1.54 in phase 1 to 1.58 in phase 2 and from 2.86 in phase 1 to 2.89 in phase 2, respectively.

We have noticed that participants' perceived trust in website Geeks.com was very high in the two phases for condition 2. This is because participants' perceived privacy risk for this website is low, which induced a higher level of trust. The PCR information is definitely making a difference in the way participants perceived privacy in each website. Thus, this result confirms our expectation on the effect of the PCR information and hence the stated hypothesis is supported.

## 7.6    Summary

One of the key contributions of this research is the introduction of the PCR mechanism. Although previous works have studied several determinants of online trust, very few have focused on the role of the privacy credential ratings of a provider.

In this chapter, we have empirically evaluated the proposed privacy credential rating. The evaluation is based on consumers' responses to a set of hypotheses designed to gauge the participants' reaction to the privacy credential information presented to them. Hypothesis 1, for example, examines the relation between knowledge of privacy information and participants' decision to purchase from online service providers. Our result indicates that the presence of privacy information knowledge was indeed significant for participants who were shown the information. Participants tended to choose service providers that better protect their privacy regardless of how famous they are. Other participants who were shown reputation information only made their decision with incomplete information, which we showed might be misleading.

In hypothesis 2 and 3, we examined the effect of privacy information on consumers' decisions regarding sharing their sensitive personal information or the decisions to purchase a product that raises privacy concerns. In both experiments, the results indicate that consumers' attitude towards sharing sensitive information enhanced significantly with the presence of privacy information compared to those who did not have access to such information.

Hypothesis 4 examines participants' perceived trust in the presence of the privacy information. The findings show that respondents were significantly more likely to trust websites with strong PCR information. Unlike traditional reputation systems where testimonials are subjective and sometimes misleading, trust inferred from privacy credentials is based on attributes that can be verified. That is, if the website claims to be a member of a privacy-auditing organization, usually it is possible to check such claims. We have noticed in this test that consumers who were shown the PCR information did not care about the names of the service providers such as Amazon.com; they were more likely to trust service providers that are not known to them as long their privacy credentials are high.

In sum, our results indicate that the presence of PCR information helped respondents make informed decisions. As for online businesses, it is in their interest to make their privacy credential available to consumers along with their services' and products' qualities to attract consumers to buy from their stores. This is especially true for privacy credentials that the business has paid significant fees to obtain, such as WebTrust seals, security mechanism, etc. (Boulianne and Cho, 2008). The PCR mechanism helps online businesses make consumers aware of such attributes in an easy-to-read report and without additional costs.

The next chapter concludes this thesis with a review of the AAPPeC design goal and objectives as well as the contributions made in this work. We also provide plans for future work.

# Chapter 8. Conclusion and Future Work

In this chapter we conclude the work of this dissertation and provide plans for future work.

## 8.1    Conclusion

The goal of this dissertation was the development of an architecture that allows users to participate in the information market and benefit from sharing their personal information. In Chapter 4, we provided the architectural component of such a system and presented the details of the agents that are responsible for performing the tasks that most consumers would not be able to do on their own. This novel architecture is based on a detailed analysis of existing systems and the privacy challenges that consumers are facing in the emerging information market (as explained in Chapter 3).

A number of contributions were made in this thesis. One of the main contributions is the proposal of a new technique to categorize and valuate privacy risks derived from users' private data set based on attribute ontology—namely, the semantic equivalence and the substitution rate of personal information which was used in the categorization process. This new technique of personal information categorization takes into account the granularity of private data which is achieved by associating the contextual usage risk to the personal data objects. Our system allows users to assign these contextual privacy risks to their private data so that the software agent can quantify the privacy costs for each user. The assignment of these credit values is based on the users' perceived risk, trust, and the level of protection. As such, this thesis contributed to the process of making privacy risk a measuring principle for the quantification of the privacy payoff value. This was one of the main design requirements for AAPPeC; specifically, determining reliable decision criteria for sharing personal information based on trust, risk, and protection level of the private data. Also essential was determining the risk value of a given private data object by means of credit assigning and weighting concept given the risk of potential damages resulting from the misuse of personal information.

One of the challenges that users face in the online world is how service providers use their personal information. In this aspect, this dissertation advances the state of the art by proposing a new assessment scheme to determine the trustworthiness of service providers based on privacy credentials. This is achieved through the design and development of a new mechanism that allows consumers to see "good" signs and "bad" signs when visiting online stores. The trust and reputation agent provides a privacy credential score and a report detailing the competency of the service provider with respect to privacy and private data handling. The score is designed based on a fuzzy logic technique that takes into consideration the significance of each trust attribute that the service provider has obtained. The rating report also includes privacy-related items that concern most consumers. It serves as a source of knowledge that allows consumers to learn how their personal information will be handled by a potential service provider. The development of such a report is one of the design requirements for AAPPeC; in particular, determining a mechanism based on trust that examines service providers' promises regarding privacy and private data handling.

Yet another contribution was the development of a model by which we can valuate the payoff for the consumers against the revelation of their personal information. The reward amount which the consumer should receive must be based on the perceived privacy risk of each data object as well as the risk resulting from the composition of each data object. In this thesis, we were the first to use a financial model to construct the amount of the payoff based on the risk exposure. This model is a well-known one in the financial and insurance industry. Financial models (such as the one we used) are designed so that those who are considered more risky should pay more interest. Since the value of the private data is associated with the amount of privacy risk, financial models were very helpful to determine the fair value that users should obtain. By so doing, we met one of the major requirements of AAPPeC architecture design, which is valuating the payoff value that the consumers should receive against the revelation of personal information.

Negotiating with a big service provider is proven to be a daunting prospect for an individual consumer. Therefore, this thesis contributed in the development of a negotiation strategy that allows for benefit maximization. To achieve this aim, we have extended an agent negotiation model developed by Skylogiannis et al. (2007). Namely, we extended the "sit-and-wait" strategy in order to use it in the negotiation process. We provided the details of this strategy in Chapter 4 and proved through experiments that it maximizes the users' utility. Negotiation with service providers on behalf of consumers is also a requirement for AAPPeC design. We have achieved this requirement by designing a software agent that takes upon the responsibility of performing this task.

In order to provide the detail specifications of AAPPeC design, in Chapter 5 we presented these specifications based on Gaia software agent design methodology and the JADE platform. To meet some of the design requirements, we extended the Gaia model so that new components were added. These new components were added to describe the goal hierarchy of the system as well as the tasks that the agents will perform in order to serve the general goal of valuating and negotiating for the privacy payoff value. We also discussed the integration of the Gaia methodology with the JADE platform in the development phase. In the development phase we presented the mappings between the roles as defined by the Gaia methodology and agent behavior structures as defined in JADE.

In Chapter 6, a proof of concept implementation was presented. The main contributions of this implementation are twofold: first, to show that our system is feasible and second to show how our system can be used by consumers. Furthermore, we performed a set of experiments to examine the variation of the payoff under different scenarios. These scenarios are related to the service provider type (i.e., their privacy regime), the variation of the risk premium, and the negotiation deadline and reservation prices. Through these experiments, we have shown that a strategically placed agent can help consumers attain the maximum gain during the negotiation process. We have shown also that service providers with high privacy credentials have the advantages of achieving a particular service delivery threshold and minimizing opportunity costs. This is because the results of the experiments indicate that service providers with high privacy credentials are less likely to suffer from revenue loss compared to those with low privacy credentials.

Finally, Chapter 7 focuses on the evaluation of the privacy credential rating report. Therefore, our last contribution in this thesis is the design of a field study by which we empirically evaluated the privacy credential rating report. The evaluation is based on consumers' responses to a set of hypotheses designed to gauge the participants' reaction to the privacy credential information presented to them. Hypothesis 1, for example, examines the relation between knowledge of privacy information and participants' decision to purchase from online service providers. Hypothesis 2 and 3 examined the effect of privacy information on consumers' decisions regarding sharing their sensitive personal information or the decisions to purchase a product that raises privacy concerns. Hypothesis 4 examines participants' perceived trust in the presence of the privacy information.

Our result indicates that the presence of privacy information was indeed significant for participants who were shown the information. Participants tended to choose service providers that better protect their privacy regardless of how famous they are (hypothesis 1). Also, the results indicate that consumers' attitude towards sharing sensitive information enhanced significantly with the presence of privacy information compared to those who did not have access to such information (hypotheses 2 and 3). With respect to trust, the findings show that respondents were significantly more likely to trust websites with strong PCR information. The experiment shows that consumers were more likely to trust service providers that are not known to them as long their privacy credentials are high (hypothesis 4).

In conclusion, it is worth discussing few arguments that we often receive from reviewers. Specifically, what are the possibilities of deploying such system and how will it be used by companies, how different privacy laws in different countries could affect the system, and what are the potential disclosure intensification of personal data that may be resulting from the use of this system. Our view on these arguments is as follows:

First, from a pure economic perspective, attention in the information market is the main reason behind the collecting of personal information in eCommerce, if not the most important reason (Taylor 2004). Service providers have a financial stake in seeking ways for accurate information. As profiles become more accurate by participation, targeted services can be delivered to the right consumers, thus achieving a particular delivery threshold and minimizing opportunity costs. In other words, service providers essentially incur excess costs (junk e-mails, junk advertisements, etc.) to attract consumers' attention to their products and services. However, if they knew precisely what the consumer wanted, they could make a much better decision about whether or not to provide the consumer with information about their services. Currently, there are companies such as ChoicePoint where service providers can obtain information and learn about consumers (Focus 2010). According to Focus research group (Focus 2010), ChoicePoint owns the largest database of consumers' records (an estimate of 17 billion records) in the United States with information about 220 million consumers mostly without the consent of those consumers. However, such database may contain inaccurate information about what consumers like and dislike. This is because this information is randomly collected.  As such, service providers have to make guesses about their targeted advertisements when using ChoicePoint as their source of information. However, information provided by a system like the one we propose in this thesis reflects true information about users' taste, preferences, demographic, etc. because it came directly from the users with their consent. Companies and consumers have a mutual interest to use AAPPeC. Companies seek accurate information and consumers seek a payoff from sharing their personal information. From this perspective, a system like AAPPeC will out perform any other system or company that collects information randomly about consumers.

Second, with respect to the effect of different privacy laws in different countries, the proposed system like any other system that uses the Internet which is inherently global. The question is which privacy law should apply on the global Internet so that all the systems that use the Internet are consistent in the applicability and content to privacy laws across jurisdictions. Fortunately, there are few jurisdictional approaches to determine the applicability of privacy laws around the world. Examples of such jurisdictions are: Article 4(1) (a) of the EU Data Protection Directive which looks at the place of origin of the company that makes the decisions about the uses of the data and determines the applicability of the law on that basis (Fleischer 2010). The same approach is used in Canada under the Federal Personal Information Protection and Electronic Documents Act (PIPEDA). The Canadian act controls the collection, use and disclosure of personal information in the course of the commercial activities (Treasury Board of Canada). Under PIPEDA, Canadian entities transferring data outside the country must have provisions in place to ensure a comparable level of protection to that granted by the law. There are other similar approaches from the EU, USA, Australia, and New Zealand that are related to the location of the people whose data is being used and the place where the actual processing happens, more details about these approaches can be found in (Fleischer 2010). However, the most obvious way of resolving the conflict created by the different regulatory regimes would be to have just one global privacy regime. The initiative which was approved in Madrid during the International Privacy Commissioners' Conference in 2009 is a step in that direction.

Third, concerning the potential disclosure intensification of personal data that may be resulting from the use of this system. Our view on this point is that it is unlikely to happen. This is because the payoff or the compensation given to the consumers is for personal information that they usually provide for free anyway while online businesses rack up lucrative amounts of money from personal information profiling. Furthermore, the payoffs are risk-base premiums. Therefore, online businesses are expected to be more conservative when handling consumers' personal information as privacy risk penalties (risk-base premiums) and reputation consequences on violators of consumers presumed privacy rights are more likely to be costly. Also, from a purely practical perspective, a negative reaction could cause consumers to turn away from the service and the service provider, thus counteracting any marketing improvements to the service. In that regard, knowledge of consumer and societal perceptions of privacy invasion are as important as knowledge of the consumer's behavior and habits. With that knowledge, any measure taken by service providers to compensate consumers for their information dissemination in an appropriate manner will potentially be regarded by consumers and maximize their loyalty.

## 8.2 Further Work

This dissertation should serve as a ground for extended research in the future. There are many interesting issues worthy of future work. Below we provide the direction for future work.

The first direction is to consider the effect of future learning in the collected personal information. The architecture presented in this thesis assumes that the possession of information today does not influence the possession of information in the future, and therefore it assumes a linear view of the information valuation with respect to the risk at the time of revealing the personal information. However, in reality, a future learning effect might impose different usage of personal information at different time intervals and therefore a different level of risk at each time. This means that when the agent computes compensation prices, a learning premium should be considered in the belief that what personal information the consumer reveals today will allow online businesses to learn more about this same consumer in the future. In this case, when we generalized the concept of compensation prices of risk to private data, the linear correlation assumption becomes inadequate as the distribution of risk becomes non-normal.

The second direction is to explore other models of payoff determination that use risk as their underlying measurement. This is especially appropriate in the case of risk resulting from future learning effects. In this thesis, we only examined one financial model; however, it would be more interesting to compare this model with other existing compensation models.

The third direction is leveraging machine learning algorithms and techniques for risk prediction. The system may employ trained data gathered from different scenarios about the consumer's reaction to privacy risks and automatically suggest risk prediction metrics for the data object. Such algorithms would be very helpful for consumers in situations where the composition of private data objects spans over different contexts (such as a mix between sports and health condition).

The fourth direction is the consideration of social and cultural background of individuals and the effect on the received payoff value. In AAPPeC, users can specify a subjective weight value to their private data categories based on their privacy risk perception. This value is used to determine the amount of the payoff that the consumer will receive after revealing his/her personal information. However, in reality there are other factors that involve rational analysis of cost and benefit resulted from social and cultural background of individuals. In other words, the perceived risk that each individual foresee in a certain situation does not take into consideration the cost and benefit analysis of the situation which is different from one culture to another. Therefore, a future model is needed to capture the multi-dimensional aspect of the weight values and their integration in the system for fair compensation.

Throughout the thesis, it has been assumed that the agents reside in the same host. However, in dynamic environments such as cloud computing the processing of data is dynamically changing from one location to another. In such environment, agents are required to be loaded at end points (hosts of interest in the computing cloud) to perform specific tasks, provide specific service related to private data valuation, or collect information about service providers' privacy practices. It also requires that agents interact with each other and report their finding to the system. Hence, in order for the system to co-exist with cloud computing, agents' mobility must be considered in future work.

**Appendix A: Credential analysis of service providers**

| Service Provider | Reputation Score | Site share personal information with other companies whose privacy policies are unknown to this site | Site collect personal information for unknown purposes | shares consumers' personal infoamtion with other companies | Site use consumers' data that identifies them for advertisements | Has valid privacy seal certificate (BBB, TRUSTe, ...) | Issues reports to consumers about changes that impact their personal information | Site has membership to privacy compliance and Audit service | Implement secure infrastructure to protect consumers' data | Allows consumers to opt out from mailing lists and marketing solicitations | Site personnel are trained to protect consumers' information |
|---|---|---|---|---|---|---|---|---|---|---|---|
| Amazon.com | 4.5 | 1 | 1 | 1 | 1 | 0 | 0 | 0 | 0 | 0 | 0 |
| Flickr.com | 4 | 1 | 1 | 1 | 1 | 0 | 0 | 0 | 0 | 0 | 0 |
| Photobucket.com | 4.5 | 1 | 1 | 1 | 1 | 0 | 1 | 0 | 0 | 1 | 0 |
| Statcounter.com | 5 | 1 | 0 | 0 | 0 | 0 | 0 | 0 | 0 | 0 | 0 |
| Bestbuy.com | 3.5 | 1 | 1 | 1 | 1 | 1 | 0 | 0 | 0 | 0 | 0 |
| Target.com | 4.5 | 1 | 1 | 1 | 1 | 0 | 0 | 0 | 0 | 0 | 0 |
| Godaddy.com | 3.5 | 0 | 0 | 0 | 0 | 1 | 0 | 1 | 1 | 0 | 1 |
| Tigerdirect.com | 3 | 1 | 1 | 1 | 1 | 0 | 0 | 0 | 0 | 0 | 0 |
| Newegg.com | 4.5 | 0 | 0 | 0 | 0 | 1 | 1 | 1 | 1 | 1 | 1 |
| Aweber.com | 4.5 | 1 | 0 | 0 | 0 | 0 | 0 | 0 | 0 | 0 | 0 |
| Expedia.com | 3.5 | 1 | 1 | 0 | 1 | 1 | 0 | 0 | 1 | 0 | 1 |
| Pogo.com | 4.5 | 0 | 0 | 1 | 1 | 0 | 0 | 0 | 0 | 0 | 0 |
| Orbitz.com | 2.5 | 0 | 0 | 0 | 1 | 1 | 0 | 0 | 1 | 1 | 1 |
| Travelocity.com | 3 | 1 | 0 | 0 | 1 | 1 | 0 | 0 | 1 | 1 | 1 |
| Overstock.com | 4 | 1 | 1 | 1 | 0 | 0 | 0 | 0 | 0 | 0 | 0 |
| Excelldiamond.com | 5 | 0 | 0 | 0 | 0 | 1 | 1 | 1 | 1 | 1 | 1 |
| Dinodirect.com |  | 0 | 0 | 0 | 0 | 1 | 1 | 1 | 1 | 1 | 1 |
| Walmart.com | 4 | 1 | 1 | 1 | 1 | 0 | 0 | 0 | 0 | 0 | 0 |
| Circuitcity.com | 5 | 1 | 0 | 1 | 1 | 1 | 0 | 1 | 1 | 0 | 0 |
| Geeks.com | 4.5 | 0 | 1 | 0 | 0 | 1 | 1 | 1 | 1 | 1 | 1 |

# Appendix B: Questionnaires of the field study

**Phase 1**

**Condition1**

| Website | Reputation |
|---------|------------|
| **amazing-products.com** | ★ ★ ★ ★ ★ |
| **got-it-all.com** | NA (new website) |
| **best-service.com** | ★ ★ ★ ☆ ☆ |
| **service-place.com** | ★ ★ ★ ★ ☆ |

**Condition2**

| Website | Reputation | Privacy Credentials |
|---------|------------|---------------------|
| **amazing-products.com** | ★ ★ ★ ★ ★ | ☆ ☆ ☆ ☆ ☆ |
| **got-it-all.com** | NA (new website) | ★ ★ ★ ★ ☆ |
| **best-service.com** | ★ ★ ★ ☆ ☆ | ★ ★ ★ ☆ ☆ |
| **service-place.com** | ★ ★ ★ ★ ☆ | ★ ★ ★ ★ ☆ |

**In Phase 2 :**
**The names are revealed as follows:**
**Amazing-products.com  is Amazon.com**
**Got-it-all.com is Dinodirect.com**
**Best-service.com is Travelocity.com**
**Service-place.com is Geeks.com**

**Condition2: Privacy credential explanation**

| ☆☆☆☆☆ | ★★★★☆ | ★★☆☆☆ | ★★★★★ |
|---|---|---|---|
| • Shares data that identifies consumers with third party **Yes** <br> • Use data that identifies consumers for advertisements **YES** <br> • Has valid Privacy certificate ( BBB, TRUSTe, WebTrust) **NO** <br> • Issues reports to users of any impact to their data **NO** <br> • Member of privacy compliance and auditing service **NO** <br> • Implement secure infrastructure for authentication **YES** <br> • Allows users to opt out whenever they feel like it **NO** <br> • Trains employees to respect consumers' privacy **NA** | • Shares data that identifies consumers with third party **NO** <br> • Use data that identifies consumers for advertisements **NO** <br> • Has valid Privacy certificate ( BBB, TRUSTe, WebTrust) **YES** <br> • Issues reports to users of any impact to their data **NO** <br> • Member of privacy compliance and auditing service **NO** <br> • Implement secure infrastructure for authentication **YES** <br> • Allows users to opt out whenever they feel like it **YES** <br> • Trains employees to respect consumers' privacy **YES** | • Shares data that identifies consumers with third party **NO** <br> • Use data that identifies consumers for advertisements **NO** <br> • Has valid Privacy certificate ( BBB, TRUSTe, WebTrust) **NO** <br> • Issues reports to users of any impact to their data **NO** <br> • Member of privacy compliance and auditing service **NO** <br> • Implement secure infrastructure for authentication **YES** <br> • Allows users to opt out whenever they feel like it **YES** <br> • Trains employees to respect consumers' privacy **YES** | • Shares data that identifies consumers with third party **NO** <br> • Use data that identifies consumers for advertisements **NO** <br> • Has valid Privacy certificate ( BBB, TRUSTe, WebTrust) **YES** <br> • Issues reports to users of any impact to their data **YES** <br> • Member of privacy compliance and auditing service **YES** <br> • Implement secure infrastructure for authentication **YES** <br> • Allows users to opt out whenever they feel like it **YES** <br> • Trains employees to respect consumers' privacy **YES** |

**Experiment questions for participants on Condition1 and Condition2**

1. Do you shop online?

2. What are your most privacy concerns when shopping online?

3- If you were purchasing a service, which website would you consider?

4- If you were purchasing a service that require submitting your email address and contacting number which website would you consider?

5- If you were purchasing a service that require to submit your email address, contact number, credit card number, profession, family size, and salary which website would you consider?

6- If you were purchasing a privacy-sensitive product (such a box of condom, sex book, etc.) which website would you consider?

7- I trust dealing with this website

# Appendix C: Sample code

```
package aims.agents;

import aims.AIMSViewPanelPhase3;
import aims.util.Constants;
import aims.util.ReadXLSheet;
import aims.util.WriteXLSheet;
import jade.core.AID;
import jade.core.Agent;
import jade.core.behaviours.Behaviour;
import jade.core.behaviours.OneShotBehaviour;
import jade.core.behaviours.ParallelBehaviour;
import jade.core.behaviours.SequentialBehaviour;
import jade.core.behaviours.SimpleBehaviour;
import jade.lang.acl.ACLMessage;
import java.util.logging.Level;
import java.util.logging.Logger;

public class ConsumerAgent extends Agent {
    int deadline = 0;
    int roundTrip = 0;
    boolean f = false;
    boolean f1 = false;
    boolean p1 = false;
    boolean p2 = false;

    FacilitatorAgent facilitator = new FacilitatorAgent();
    public void actionPR(String content){
        try{
            String value = content.substring(content.indexOf("PR:")+3);

            if(content.startsWith("inbidPR:") ||
                content.contains("finalbidPR:")){
                double serviceValue = Double.parseDouble(AIMSViewPanelPhase3.jTxtCAServiceValue.getText().trim());
                int numberOfCostomer=0;
                double[] data = getValuesFromFile();
                for(double row : data){
                    if((row*serviceValue) < Double.parseDouble(value) ){
                        numberOfCostomer++;
                    }
                }
                if(data.length == numberOfCostomer || content.contains("finalbidPR:")){
                    if(!content.contains("1finalbidPR:")){
                        invokeProviderAgentToAccept("acceptedPR:"+numberOfCostomer + "");
                    }
                }
                else{
                    invokeProviderAgentForProposal("inbidPR:"+numberOfCostomer + "");
                }
            }
        }
        catch (Exception ex) {
            ex.printStackTrace();
        }
    }
    public double[] getValuesFromFile(){
        try{
            ReadXLSheet xlReader = new ReadXLSheet();
            double data[] = xlReader.getData("c:/AIMS/supporting_values.xls");
            return data;
        }
        catch(Exception e){
            e.printStackTrace();
        }
        return null;
    }
    public static ConsumerAgent thisObject;
```

```java
@Override
protected void setup() {
    f=false;
    thisObject = this;
    final SequentialBehaviour consumer_agent = new SequentialBehaviour(this);
    final SequentialBehaviour d2 = new SequentialBehaviour(this);

    ParallelBehaviour pb = new ParallelBehaviour(this, ParallelBehaviour.WHEN_ALL);
    consumer_agent.addSubBehaviour(new Behaviour(this) {
        @Override
        public void action() {
            ACLMessage msg = receive();
            p1 = false;
            p2 = false;
            if(msg!=null){
                try {
                    facilitator.invokeProviderAgent(msg.getContent());
                    if (msg!= null){
                        System.out.println(getLocalName() + ": received the following message : ");
                        System.out.println(msg.toString());
                        String content = msg.getContent();
                        System.out.println(content);
                        //finished = true;
                        //To delete Agent
                        //myAgent.doDelete();
                        try {
                            if(content.contains("PR:")){
                                actionPR(content);
                                return;
                            }
                            if(content.startsWith("inbid:")){
                                if(deadline == -1){
                                    myAgent.doDelete();
                                    return;
                                }
                                roundTrip++;
                                if(deadline  == roundTrip){
                                    deadline = -1;
                                    roundTrip=0;
                                    double paRe = Double.parseDouble(AIMSViewPanelPhase3.jTxtCAReserPrice.getText().trim());
                                    invokeProviderAgentForProposal("finalbid:"+(paRe*2));
                                    WriteXLSheet.data.setConsumerOffer(String.valueOf(paRe*2));
                                }
                                else{
                                    String msgStr = content.toString();
                                    msgStr = msgStr.substring(msgStr.indexOf("inbid:")+6);
                                    double caRes = Double.parseDouble(AIMSViewPanelPhase3.jTxtCAReserPrice.getText().trim());
                                    WriteXLSheet.data.setProviderOffer(String.valueOf(msgStr));
                                    if(caRes>Double.parseDouble(msgStr)){
                                        invokeProviderAgentToReject("regectedbid:");
                                    }
                                    else{
                                        invokeProviderAgentToAccept("accepted:");
                                    }
                                }
                            }
                            else if(content.startsWith("regectedbid:") || content.startsWith("accepted:")){
                                AIMSViewPanelPhase3.btnNagotiationResult.setEnabled(true);
                                if(content.startsWith("accepted:"))
                                    WriteXLSheet.data.setResult("Succeeded");
                                if(content.startsWith("regectedbid:"))
                                    WriteXLSheet.data.setResult("Failed");
                                f = true;
                                if(f1==true){
                                    d2.reset();
                                    d2.restart();
                                }
                            }
                        }
                    }
                }
            }
        }
    }
```

```java
                catch (Exception ex) {
                    ex.printStackTrace();
                }
            }
        } catch (Exception ex) {
            Logger.getLogger(ConsumerAgent.class.getName()).log(Level.SEVERE, null, ex);
        }
    }
}

@Override
public boolean done() {
    return f;
}

@Override
public void reset() {
    f = false;
    super.reset();
}
});

SequentialBehaviour s1 = new SequentialBehaviour(this);
s1.addSubBehaviour(new OneShotBehaviour(this) {
    @Override
    public void action() {
        p1 = true;
        System.out.println("====================>>>First");
        AID r = new AID();
        r.setLocalName("FacilitatorAgent");
        ACLMessage msg = new ACLMessage(ACLMessage.INFORM);
        msg.setSender(getAID());
        msg.addReceiver(r);
        msg.setContent("Negotiation Terminated");
        send(msg);
        System.out.println(getLocalName() +": send id to Facilitator Agent");
    }
});

SequentialBehaviour s2 = new SequentialBehaviour(this);
s2.addSubBehaviour(new OneShotBehaviour(this) {
    @Override
    public void action() {
        p2 = true;
        System.out.println("====================>>>Second");
        WriteXLSheet xlReader = new WriteXLSheet("c:/AIMS/NagotiationResult.xls");
    }
});

pb.addSubBehaviour(s1);
pb.addSubBehaviour(s2);

d2.addSubBehaviour(new SimpleBehaviour(){
    @Override
    public void action() {
        f1=true;
        if(f==true){
            consumer_agent.reset();
            consumer_agent.restart();
        }

    }

    @Override
    public boolean done() {
        return f1;
    }
    @Override
    public void reset() {
        System.out.println("LastReset>>>>>>>>"+f1);
```

```java
            f1 = false;
            super.reset();
        }
    });

    final SequentialBehaviour s3 = new SequentialBehaviour();
    s3.addSubBehaviour(pb);
    s3.addSubBehaviour(d2);

    consumer_agent.addSubBehaviour(s3);
    addBehaviour(consumer_agent);
}

public void invokeProviderAgentInit(String message) throws Exception{
    roundTrip = 0;
    deadline = Integer.parseInt(AIMSViewPanelPhase3.jTxtCATimeDeadlines.getText());
    invokeProviderAgent(message);
}
public void invokeProviderAgent(String message) throws Exception{
    AID r = new AID();
    r.setLocalName("ProviderAgent");
    ACLMessage msg = new ACLMessage(ACLMessage.INFORM);
    msg.setSender(getAID());
    msg.addReceiver(r);
    msg.setContent(message);
    send(msg);
    System.out.println(getLocalName() +": send id to Provider Agent");
}
public void invokeProviderAgentForProposal(String message) throws Exception{
    AID r = new AID();
    r.setLocalName("ProviderAgent");
    ACLMessage msg = new ACLMessage(ACLMessage.PROPOSE);
    msg.setSender(getAID());
    msg.addReceiver(r);
    msg.setContent(message);
    send(msg);
    System.out.println(getLocalName() +": send bid to Provider Agent");
}
public void invokeProviderAgentToAccept(String message) throws Exception{
    AID r = new AID();
    r.setLocalName("ProviderAgent");
    ACLMessage msg = new ACLMessage(ACLMessage.ACCEPT_PROPOSAL);
    msg.setSender(getAID());
    msg.addReceiver(r);
    msg.setContent(message);
    send(msg);
    System.out.println(getLocalName() +": accept bid from Provider Agent");
}
public void invokeProviderAgentToReject(String message) throws Exception{
    AID r = new AID();
    r.setLocalName("ProviderAgent");
    ACLMessage msg = new ACLMessage(ACLMessage.REJECT_PROPOSAL);
    msg.setSender(getAID());
    msg.addReceiver(r);
    msg.setContent(message);
    send(msg);
    System.out.println(getLocalName() +": reject bid of Provider Agent");
}
public void invokeFacilitatorAgentToTerminated(String message) throws Exception{
    AID r = new AID();
    r.setLocalName("FacilitatorAgent");
    ACLMessage ACLmsg = new ACLMessage(ACLMessage.INFORM);
    ACLmsg.setSender(getAID());
    ACLmsg.addReceiver(r);
    ACLmsg.setContent(message);
    send(ACLmsg);
    System.out.println(getLocalName() +": send id to Facilitator Agent");
}
}
```